\documentclass[twocolumn]{aastex631}

\usepackage{graphicx}   
\usepackage{subfigure} 
\usepackage{multirow}
\usepackage{makecell} 
\usepackage{amsmath}

\shorttitle{DM and scattering of FRB}
\shortauthors{Mo, Zhu, \& Feng.}
\graphicspath{{./}{figures/}}

\begin{document}
    \title{The dispersion measure and scattering of Fast Radio Bursts:  contributions from multi-components, and clues for the intrinsic properties}
\correspondingauthor{Weishan Zhu}
\email{zhuwshan5@mail.sysu.edu.cn}

\author{Jian-feng, Mo}
\affil{School of Physics and Astronomy, Sun Yat-Sen University, Zhuhai campus, No. 2, Daxue Road \\
Zhuhai, Guangdong, 519082, China}
\affil{CSST Science Center for the Guangdong-Hong Kong-Macau Greater Bay Area, Daxue Road 2, 519082, Zhuhai, China}

\author[0000-0002-1189-2855]{Weishan, Zhu}
\affil{School of Physics and Astronomy, Sun Yat-Sen University, Zhuhai campus, No. 2, Daxue Road \\
Zhuhai, Guangdong, 519082, China}
\affil{CSST Science Center for the Guangdong-Hong Kong-Macau Greater Bay Area, Daxue Road 2, 519082, Zhuhai, China}

\author{Long-Long, Feng}
\affil{School of Physics and Astronomy, Sun Yat-Sen University, Zhuhai campus, No. 2, Daxue Road \\
Zhuhai, Guangdong, 519082, China}
\affil{CSST Science Center for the Guangdong-Hong Kong-Macau Greater Bay Area, Daxue Road 2, 519082, Zhuhai, China}



\begin{abstract}
Fast radio bursts (FRBs) are luminous, millisecond-duration transients that offer great potential for probing the universe, yet their physical origins remain unclear. The dispersion measure (DM) and scattering time ($\tau$) distributions provide key insights into FRBs' properties, including source population, redshift, and energy distribution. We use a simplified model of FRB source population and intrinsic Schechter function-like energy distribution, coupled with a thorough assessment of various contributors to dispersion and scattering, to replicate the joint distribution of DM and $\tau$ in the CHIME/FRB catalog. A mixed FRB source population, including both young and old progenitors, is considered. Contributions to the DM and $\tau$ from interstellar medium (ISM), circumgalactic medium (CGM) within host and foreground halos are informed by the IllustrisTNG simulation, while contributions from the Milky Way, intergalactic medium (IGM), and local environmental are estimated by updated models. Using MCMC simulations, we identify optimal model that well reproduce the DM distribution and broadly reproduce the $\tau$ distribution in the CHIME/FRB catalog. Our model suggests that the fraction of FRBs tracing star-formation rate is $\rm{f_{PSFR}=0.58^{+0.16}_{-0.27}}$,
while $\rm{log_{10}E_*[erg]=42.27^{+1.17}_{-1.18}}$ and $\gamma=-1.60^{+0.11}_{-0.13}$ in the energy distribution function. Scattering predominantly arises from the circumburst medium or the ISM and CGM of hosts, which cause a DM of $\sim 10\, \rm{pc\,cm^{-3}}$.
Using our optimal model, we estimate FRB redshifts with two methods: DM-only and combined DM-$\tau$. Evaluation with 68 localized FRBs reveals an RMS error $0.11-0.12$, and incorporation of $\tau$ has a minor effect. We further argue that the host galaxy properties of localized FRBs could be a potential tool to validate our model in the future.

\end{abstract}



\section{Introduction} \label{sec:intro}

FRBs represent a captivating enigma within the realm of astrophysics, characterized by their brief, millisecond duration and luminous radio emissions. Initially reported in 2007 by \cite{2007Sci...318..777L} with a single event, their intrigue deepened with the discovery of four additional bursts in 2013 \citep{2013Sci...341...53T}. Since then, the detection rate has surged, surpassing 800 events over recent years. This increment owes much to the concerted efforts of various research teams utilizing a suite of telescopes, including Parkes, Arecibo, ASKAP, UTMOST, CHIME, and FAST (e.g., \citealt{2014ApJ...790..101S,2018ApJ...863...48C,2018Natur.562..386S,2021ApJS..257...59C,2021ApJ...909L...8N}). With the exception of FRB 20200428 (\citealt{2020Natur.587...59B, 2020Natur.587...54C}), all identified events are extragalactic, discerned by their dispersion measures (DMs) exceeding the anticipated contributions from the Milky Way and its gaseous halo. The increase in dectected events has driven intensive efforts to decipher the underlying mechanisms and origins of these enigmatic cosmic phenomena.

The physical origin of FRBs is still unknown. Many theoretical models have emerged since the discovery of the first event (for a comprehensive overview, see reviews by \citealt{2019ARA&A..57..417C,2019A&ARv..27....4P,2019PhR...821....1P, 2020Natur.587...45Z} and references therein). 
For instance, FRB 20200428 is associated with young magnetar; however, it appears increasingly evident that magnetars alone cannot account for all the observed FRB, as evidenced by distinct properties exhibited in certain events (e.g., \citealt{2021ApJ...908L..12T,2022Natur.602..585K}). 
Moreover, the diverse possible channels for magnetar formation, which are capable of inducing FRBs, further complicate the exploration of the origins of FRBs \citep{2019ApJ...886..110M}.
To unravel the true nature of FRBs, critical insights into their distances/redshifts, host galaxies, and local environments are indispensable. However, the effort has been hampered by the limited localization of approximately 71 FRBs to date. Estimating the distances of unlocalized sources typically relies on the relation between the DM induced by the IGM, denoted as $\rm{DM_{IGM}}$, and redshift (e.g., \citealt{2014ApJ...783L..35D, 2020Natur.581..391M}). Nevertheless, such estimations are plagued by considerable scatter along different lines of sight due to the inherently inhomogeneous nature of the IGM (e.g., \citealt{2003ApJ...598L..79I, 2014ApJ...780L..33M, 2015MNRAS.451.4277D, 2018ApJ...865..147Z, 2019MNRAS.484.1637J, 2021ApJ...906...49Z, 2021ApJ...906...95Z,2023arXiv230507022B}).

Furthermore, it is a significant challenge to obtain a reliable $\rm{DM_{IGM}}$ for unlocalized FRBs. The DM of FRBs is caused by multiple components, encompassing electrons within the Milky Way and its gaseous halo, the diffuse IGM, potentially intervening halos and galaxies, and within host galaxies and their halos. Although the contribution from the Milky Way is relatively tractable to assess, often derived using models, such as NE2001 \citep{2002astro.ph..7156C} or YMW16 \citep{2017ApJ...835...29Y}, accurately quantifying contributions from other components for specific events remains elusive. Even the DM attributed to the Milky Way's halo exhibits considerable uncertainty, ranging from 30 to 245 $\rm{pc\,cm^{-3}}$ (\citealt{2019MNRAS.485..648P, 2020ApJ...888..105Y,2023ApJ...946...58C}).
Moreover, substantial uncertainty surrounds the DM contributed by host galaxies, denoted as $\rm{DM_{Host}}$. Theoretical electron density profiles within hosts suggest a log-normal distribution for $\rm{DM_{Host}}$, with a median value of 100 $\rm{pc\,cm^{-3}}$ (\citealt{2015RAA....15.1629X, 2018MNRAS.481.2320L, 2020A&A...638A..37W}). Estimations derived from cosmological simulations of galaxies propose median values of $\rm{DM_{Host}}$ that span 30-180 $\rm{pc\,cm^{-3}}$, depending on the FRB progenitor (\citealt{2020AcA....70...87J, 2020ApJ...900..170Z, 2023MNRAS.518..539M}). Localized events exhibit $\rm{DM_{Host}}$ estimates ranging from 10 to 1121 $\rm{pc\,cm^{-3}}$ (e.g., \citealt{2017ApJ...834L...7T, 2019Natur.572..352R, 2020Natur.577..190M, 2021ApJ...910L..18B, 2021ApJ...922..173C, 2022Natur.602..585K, 2022Natur.606..873N, 2022ApJ...931...87O}), although with considerable uncertainty. For instance, \citealt{2023arXiv230605403L} demonstrated that the $\rm{DM_{Host}}$ of FRB20190520B could range from 340-470 $\rm{pc\,cm^{-3}}$, accounting for the contribution of two foreground galaxy clusters, significantly differing from the previously estimated value of 900-1121 $\rm{pc\,cm^{-3}}$.

The notable uncertainty surrounding the contributions to the DM of FRBs from various components presents a substantial hurdle in inferring the distances of most FRBs and unveiling their physical origins. Meanwhile, accurate estimations of the multiple contributors to the DM of FRB are vital for the broader application of FRBs. Given that the majority of FRBs originate from extragalactic sources and carry information about ionized electron distribution along their propagation paths, they offer invaluable opportunities to probe the universe. FRBs hold promise as tools to investigate fundamental aspects of the universe, including the density of baryonic matter, the reionization history, and the Hubble constant (e.g., \citealt{2014ApJ...783L..35D,2014ApJ...780L..33M,2020Natur.581..391M,2022MNRAS.511..662H}). However, the reliability of these measurements and constraints is largely based on the precision of the estimated relative contributions to the DM of FRBs. Addressing this issue is paramount in advancing the field and using FRBs as potent probes of cosmic phenomena.

One potential avenue for improvement lies in jointly analyzing the contributions to both the DM and scattering of FRBs across various environmental media. \cite{2022ApJ...931...88C} suggest that by assuming the dominance of scattering from the disks of host galaxies or the Milky Way, with a minor role played by the IGM and galaxy halos, the scattering time ($\tau$) of FRBs can be used to estimate the DM contributed by the host galaxies. This combined approach of utilizing scattering and dispersion measurements may yield a more precise redshift estimator. Analyzing 9 events with measured $\tau$ from 14 localized FRB, \cite{2022ApJ...931...88C} found that the combined redshift estimator exhibited a smaller bias by a factor of 4 to 10 and a reduced RMS error compared to the estimator only DM.
However, there are caveats associated with the results in \cite{2022ApJ...931...88C}. First, two of the nine events (FRB 190523 and FRB 190611) exhibit significant offsets from their proposed host galaxies. This discrepancy suggests either that the scattering of these two events originates from a medium outside of their host galaxies or that the proposed host galaxies may not be genuine. In the former scenario, while the diffuse IGM has a minimal impact on the scattering of FRBs (\citealt{2013ApJ...776..125M, 2018ApJ...865..147Z}), intervening foreground galaxies and halos could contribute significantly to scattering for certain events (\citealt{2021ApJ...906...95Z}). Second, the robustness of the conclusion in \cite{2022ApJ...931...88C} is compromised by the limited number of events. To refine and enhance the reliability of the DM-scattering combined redshift estimator, it is imperative to accurately evaluate the contributions to both DM and $\tau$ of FRBs from multiple components concurrently. 

Moreover, the joint distribution of DM and $\tau$ of FRBs encode vital information about their intrinsic properties, including source population models, redshifts, and energy distributions. \cite{2022ApJ...927...35C} conducted a Monte Carlo-based population synthesis study that focused on the joint distribution of DM and $\tau$ of FRBs. Their analysis incorporated theoretical models accounting for contributions from various components, along with several models describing FRB populations and host galaxies. They found that if the CGM of intervening foreground galaxies or the circumburst medium contribute significantly to the scattering of certain FRBs, the joint distribution of DM and $\tau$ of their simulated FRBs within one of nine population models could marginally align with the CHIME/FRB catalog \citep{2021ApJS..257...59C}. Notably, this model favored a population of FRBs offset from the centers of host galaxies. However, several factors indicate the need for further refinement. First, their models do not accurately reproduce the $\tau$ distribution. Secondly, the assumed electron density models in different types of host galaxies that they adopted may inadequately reflect the dispersive and scattering properties of these galaxies. Additionally, adjustments to models regarding the contribution of the IGM and intervening halos are necessary to adequately account for variations in the fraction of baryons in the IGM and CGM of foreground halos as redshift evolves. 

Inspired by recent studies mentioned above, we estimate the contributions of multi-components to the DM and $\tau$ of FRBs based on theoretical models and the properties of galaxies in the IllustrisTNG simulation \citep{2018MNRAS.475..624N}, including the electron, stellar, and star formation distribution. We endeavor to recover the joint distribution of DM and $\tau$ of the FRBs discovered by the CHIME/FRB project through Markov chain Monte Carlo (MCMC) simulations of FRB populations, varying parameters within our models. 
This paper is structured as follows. In Section 2, we introduce our methodology, describing our overall strategy, the observed samples from the CHIME/FRB project, models of mock FRB source populations, and the procedure for calculating DM and $\tau$ caused by different components. Section 3 delves into the MCMC simulations undertaken to identify optimal parameters for FRB source populations, energy distributions, and properties of electron clumps (local environment) within our models. Building on our models and optimal parameters, Section 4 explores the efficacy of combining DM and $\tau$ information to estimate FRB redshifts, in contrast to the DM-only method, analyzed using 32 currently localized FRB events collected from the literature. In Section 5, we discuss the use of fractions of host galaxies with distinct properties to constrain the parameters of FRB populations in our model, along with comparisons with previous findings. Finally, our findings are summarized in Section 6. The cosmological parameters used in this study are: $\Omega_m = 0.3089$, $\Omega_b = 0.0486$ , $\Omega_\Lambda = 0.6911$  and $h = 0.6774$ \citep[][]{2016A&A...594A..13P}. 

\section{Methodology}

Our methodology is outlined in Section 2.1, followed by the criteria employed for event selection from the CHIME/FRB catalog in Section 2.2. Section 2.3 details models of intrinsic redshift distribution, the energy distribution of mock FRBs, and methodology for constructing the mock FRB source population for comparison with selected samples from the CHIME/FRB catalog. Subsequently, Section 2.4 introduces the procedures and models utilized to calculate the contributions to the extragalactic DM and $\tau$ of mock FRBs by multiple components along the lines of sight.

\subsection{Overall strategy}
\label{sec:method_outline}
Firstly, we would like to layout the overall strategy of the methods used in this work.

\begin{enumerate} 
    \item We select samples from the CHIME/FRB catalog with certain criteria. 
    \item We place approximately $10^8$ mock FRB sources at various redshifts, assuming that a fraction of the progenitors trace the cosmic star formation rate (SFR) while others trace the cosmic stellar mass, i.e., a mixed population model.
    \item The intrinsic energy (E) of the FRBs is assumed to follow a Schechter function, from which we sample the energy ($E_i$) for each mock FRB event. 
    \item To account for contributions from the MW and its halo, foregrounds (including galaxies and halos), the IGM, as well as hosts (including galaxies and halos), and local environments, we estimate the extragalactic DM and $\tau$ of each mock source. Detailed calculations are described in Section 2.4.
    \item Subsequently, we select the observable mock sources according to certain criteria and compare their joint distribution of extragalactic DM and $\tau$ with those of selected samples from the CHIME/FRB catalog. Using MCMC simulations, we aim to identify optimal parameters that replicate the joint distribution of the CHIME/FRB sample.
    \item Lastly, we use the estimated relative contributions to extragalactic DM and $\tau$ from multiple components to infer the redshift of FRB events. We further argue to use the properties of host galaxies as an independent indicator to validate our optimal model.
    
\end{enumerate}

\subsection{Observations}
The Canadian Hydrogen Intensity Mapping Experiment Fast Radio Burst (CHIME/FRB) Project \citep[][]{2018ApJ...863...48C} has released the largest sample of FRB sources to date, totaling approximately 500 sources. 
The first catalog (hereafter referred to as `Catalog 1') of CHIME/FRB comprises 536 bursts, including 474 so-far non-repeating bursts and 62 repeating bursts from 18 repeating sources \citep[][]{2021ApJS..257...59C}\footnote{\href{https://www.chime-frb.ca/catalog}{https://www.chime-frb.ca/catalog}}. 
These FRBs were detected across the frequency range of 400 to 800 MHz in a single survey characterized by uniform selection effects. During algorithm testing, the CHIME detection algorithm demonstrated some limitations, notably in missing significant fractions of injected FRBs with low fluences or long scattering times, as well as moderate fractions of injected sources with small or large observed DM and sources with long intrinsic durations. Consequently, it is imperative to account for the selection functions of DM, $\tau$, intrinsic duration, and fluence. We primarily follow the procedures in \cite{2022MNRAS.511.1961H} to select samples.
Samples from the CHIME/FRB catalog that meet the following criteria are considered for analysis.
\begin{itemize} 
    \item $bonsai\_snr>10$, that is the signal-to-noise ratio (S/N) should be greater than 10.
    \item $bonsai\_dm>100 \, \rm{pc\,cm^{-3}} $ and $ bonsai\_dm> 1.5\times \rm{max(DM_{MW,ISM;NE2001}, \, DM_{MW,ISM;YMW16}})$, where $bonsai\_dm$ is the observed dispersion measure, denoted as $\rm{DM_{obs}}$; $\rm{DM_{MW,ISM;NE2001}, \, DM_{MW,ISM;YMW16}}$ are the DM contributed by the ISM in the Milky Way estimated by the NE2001 model \citep[][]{2002astro.ph..7156C} and the YMW16 model \citep[][]{2017ApJ...835...29Y}, respectively. 
    \item $scat\_time<10 \rm{[ms]}$, where $scat\_time$ is the scattering time at 600 MHz, denoted as $\rm{\tau_{obs}}$. Highly scattered bursts are excluded due to significant uncertainties associated with the CHIME detection algorithm \citep[][]{2021ApJS..257...59C, 2022MNRAS.511.1961H, 2023ApJ...944..105S}. For some sources, only the upper limit of $\rm{\tau_{obs}}$ is available. In such cases, we assign $\rm{\tau_{obs}}$ for these events to a random value within the range from 0.1\% to 100\% of the upper limit.
    \item $excluded\_flag=0$, where $excluded\_flag=1$ corresponds to some events that have been excluded from parameter inference due to non-nominal telescope operation \citep[][]{2021ApJS..257...59C}.
    \item The first detected burst, if it is a repeating source.
    \item $fluence>0.4 \rm{[Jy \, ms]}$, which is same as \cite{2022ApJ...927...35C}.
\end{itemize}

Applying these selection criteria, we obtain a sample comprising 352 FRBs from the CHIME/FRB catalog, with 12 repeating sources included. However, if we raise the signal-to-noise ratio threshold to S/N > 12, as adopted in previous studies \citep[][]{2021ApJS..257...59C, 2023ApJ...944..105S}, while maintaining other conditions unchanged, the sample size would decrease further to 259 FRBs.

The selection effects between the observed and intrinsic distributions of measurable properties can be described by \cite{2022MNRAS.511.1961H},
\begin{equation}
    P(\xi) = P_{\rm{obs}}(\xi) \times s(\xi)^{-1},
    \label{eqn:obs and intrinsic}
\end{equation}
where $\xi$ could be dispersion measure, or scattering time, $P(\xi)$ is assumed to follow log-normal distribution, $s(\xi)$ is the selection function. We adopt the selection function derived by \cite{2022MNRAS.511.1961H}, 
\begin{equation}
    s(\rm{DM_{obs}}) = -0.7707(\rm{log DM_{obs}})^2 + 4.5601(\rm{log DM_{obs}}) - 5.6291
    \label{eqn:DM seletion}
\end{equation}
\begin{equation}
    s(\rm{\tau_{obs}}) = -0.2922(\rm{log\tau_{obs}})^2 - 1.0196(\rm{log\tau_{obs}}) + 1.4592 .
    \label{eqn:tau seletion}
\end{equation}
We utilize the selection-corrected data (i.e., intrinsic data) as the observational results against which we compare our simulated results.
More details about the comparison are shown in Section \ref{sec:MCMC}.

\subsection{mock FRB source populations}
\subsubsection{redshift distribution}
The physical origins of FRBs remain an open question. While some FRBs are plausibly associated with magnetars, others may stem from alternative sources. Nevertheless, the progenitors of FRBs can generally be classified into two categories: those associated with young stars (with ages shorter than approximately 50 Myr, corresponding roughly to the main sequence lifetime of a $8\, {M_{\odot}}$ star)
and those linked to aged stars (with ages exceeding $\sim1$ Gyr). The former, referred to as the PSFR population hereafter, exhibits a probability density function (PDF) for the intrinsic FRB redshift distribution that can be correlated with the cosmic star formation history \citep[][]{2022ApJ...927...35C}, i.e.
\begin{equation}
    P(z) \propto \psi(z)  \frac{1}{1+z} \frac{dV_c}{d\Omega dz},
\label{eqn:PSFR z intrinsic distribution}
\end{equation}
where 
\begin{equation}
    \psi(z)=0.015\frac{(1+z)^{2.7}}{1+[(1+z)/2.9]^{5.6} } \rm{M_{\odot} yr^{-1} Mpc^{-3}}
\end{equation} 
is the cosmic star formation rate density as a function of redshift \citep[][]{2014ARA&A..52..415M}, $\frac{dV_c}{d\Omega dz}$ is the comoving volume per solid angle per redshift. 
For the other population that associates with old progenitors, named as the PStar population hereafter, the PDF of the intrinsic FRB redshift distribution could be approximately related to the cosmic stellar mass density history, i.e.
\begin{equation}
    P(z) \propto \rho_{*}(z)  \frac{1}{1+z} \frac{dV_c}{d\Omega dz},
\label{eqn:PStar z intrinsic distribution}
\end{equation}
where $\rho_{*}(z)=(1-R)\int^{\infty}_{z} \psi(z')\frac{dz'}{H(z')(1+z')} $ is the cosmic stellar mass density with a return fraction $R=0.27$ \citep[][]{2014ARA&A..52..415M}. It is worth noting that if most FRB progenitors are middle-aged (around 300-500 Myr), neither the PSFR nor the PStar model can accurately capture them.

\begin{figure}
    \includegraphics[width=1.0\columnwidth]{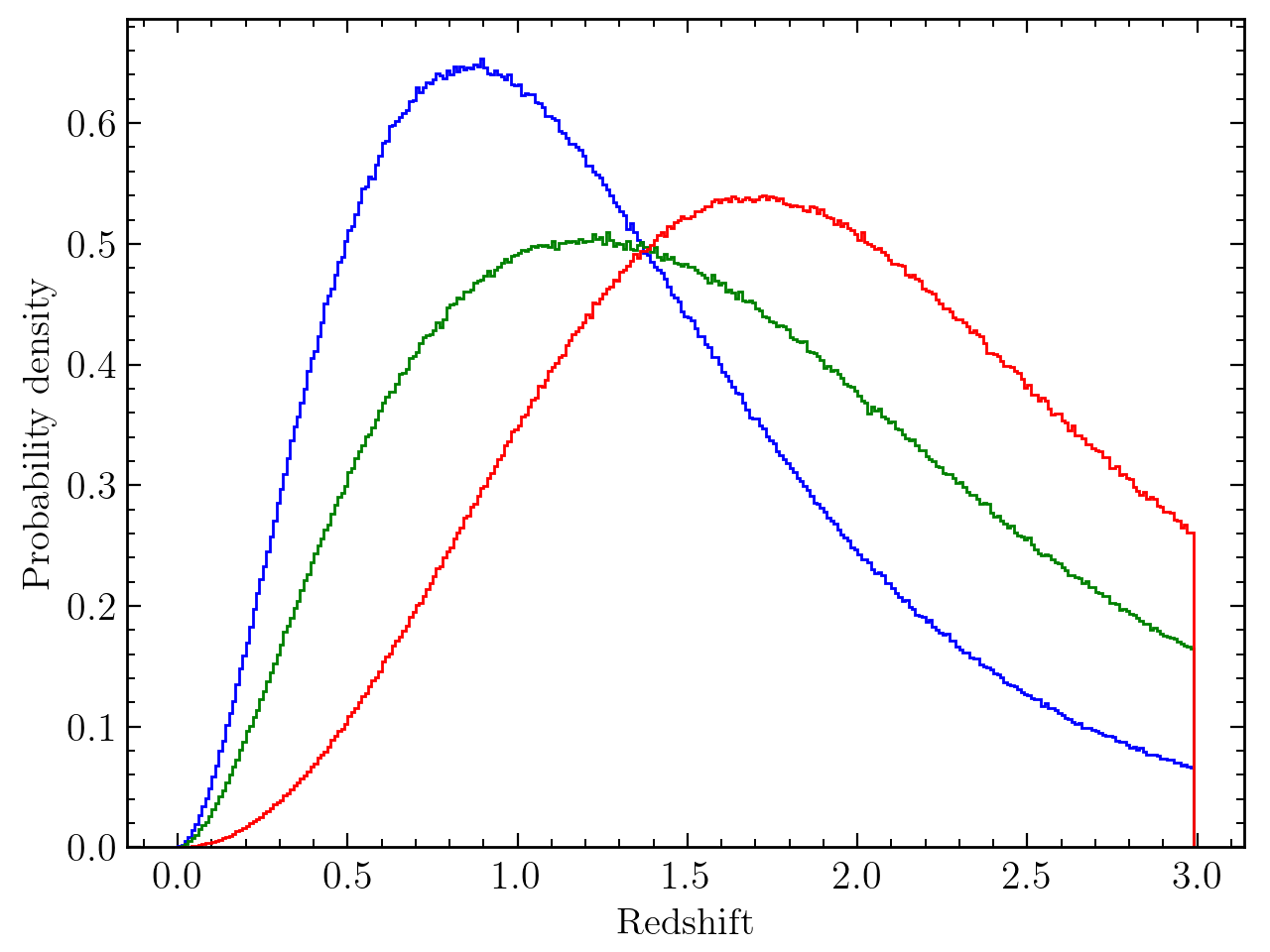}
    \caption{The intrinsic probability density distribution of FRBs between redshift $0.0-3.0$ for the PStar (blue), PSFR (red) and PMix (green) source population models. The blue, green, and red lines indicate the fraction of FRBs  associated with young progenitors (i.e., tracing the cosmic star formation rate density), with $f_{\rm{PSFR}}=0.0, \, 0.5, \, 1.0$, respectively.}
    \label{fig:FRB intrinsic z distribution}
\end{figure}

Given the yet undetermined physical origin of FRBs and the plausibility of young and old progenitors triggering FRB events, it is sensible to adopt a mixed population model that includes both types. This model, termed PMix, assumes the coexistence of young and old progenitors, with the fraction of events induced by young progenitors denoted as $f_{\rm{PSFR}}$. This parameter varies from 0 to 1.0, and its optimal value will be inferred by MCMC analysis in Section \ref{sec:MCMC}. Figure \ref{fig:FRB intrinsic z distribution} illustrates the PDFs of the intrinsic redshift distribution for the PSFR, PStar, and PMix population models. That is, $f_{\rm{PSFR}}$ is set to 1.0, 0, and 0.5, respectively. If $f_{\rm{PSFR}}$ is equal to 1 or 0, the PMix model turns to the PSFR or PStar model, respectively.

\subsubsection{energy distribution}
The intrinsic energy distribution of FRBs is commonly assumed to follow the Schechter function \citep[][]{1976ApJ...203..297S}, 
\begin{equation}
    P(E)dE \propto  \left(\frac{E}{E_{*}} \right)^{\gamma} \rm{exp}\left(-\frac{E}{E_{*}}\right) dE,
\label{eqn:E intrinsic distribution}
\end{equation}
where $\gamma$ is the differential power-law index, and $E_{*}$ is the characteristic energy cut-off. 
Following the procedure employed in many previous works \citep[][]{2022ApJ...927...35C, 2023ApJ...944..105S}, we assume that the intrinsic energy distribution of FRBs does not evolve with redshift. Furthermore, we conservatively set the lower and upper limits of $E_{\rm{min}}=10^{38}$ erg and $E_{\rm{max}}=10^{48}$ erg, respectively, for the intrinsic energy, which is similar to the implementation in \cite{2023ApJ...944..105S}. In the literature, \cite{2022MNRAS.509.4775J, 2022MNRAS.510L..18J} rule out $E_{\rm{min}}=10^{38.5}$ erg at the 90\% confidence level (C.L.) and determine $E_{\rm{max}}=10^{42}$ erg at 68\% C.L., both of which fall within our lower and upper limits. Both $\gamma$ and $E_{*}$ serve as free parameters in our investigation, with their optimal values determined through the MCMC procedure described in Section \ref{sec:MCMC}.

\subsubsection{observable mock FRB sources}
For a mock FRB at redshift z with isotropic energy E, its observed fluence at earth is given by \citep[][]{2018MNRAS.480.4211M,2022ApJ...927...35C},
\begin{equation}
    F=\frac{E(1+z)^{2-\alpha}}{4\pi D^2_L \Delta \nu},
\label{eqn:fluence}
\end{equation}
where $\alpha$ is the spectral index and is set to 0 following \cite{2022ApJ...927...35C}; $D_L$ is the FRB luminosity distance; $\Delta \nu$ is the observing bandwidth and is set to 1 GHz following \cite{2023ApJ...944..105S}. It is worth noting that different values are employed for $\Delta \nu$ in the literature. For instance, it is set to 400 MHz in \cite{2023arXiv230911522M}, to 600 MHz in \cite{2022JCAP...01..040Q}.

In this work, a mock FRB source would be considered observable by CHIME when its fluence is greater than $0.4 \,\rm{Jy\,ms}$, consistent with \cite{2022ApJ...927...35C}.

\subsection{DM and $\tau$ of FRBs}
\label{sec:dm_tau}

The dispersion measure in the rest frame of FRB is defined as
\begin{equation}
    \rm{DM}= \int \it{n_e}  \it{dl},
    \label{eqn:DM}
\end{equation}
where $n_e$ is the electron number density and $\it{dl}$ is the integral element along line-of-sight (L.O.S.). Meanwhile, intervening inhomogeneous plasma induces scattering to the FRB signals. Assuming Kolmogorov turbulence within the plasma, the temporal broadening time, $\delta \tau$, contributed by the intervening medium from redshift $z$ to $z+\delta z$, is related to the effective scattering measure
\begin{equation}
\rm{\Delta SM_{eff}} = \int^{z+\Delta z}_z \frac{C^2_N(z') d_H(z)}{(1+z')^3} \it{dz'},
\label{eqn:SM_eff}
\end{equation}
 where $C^2_N(z')$ scales with the variance of electron density, and $d_H(z)$ is the Hubble radius. The specific relation between $\delta \tau$ and $\rm{\Delta SM_{eff}}$ can be found in Appendix \ref{app:sm and tau}.  

As FRB signals propagate through media in different regions along their path to the observer, the DM and $\tau$ can be described as the sum of contributions from different components, i.e.
\begin{equation}
\begin{split}
    \rm{DM_{obs} = DM_{MW,ISM} + DM_{MW,Halo} + DM_{IGM} }  \\    
  \rm{ + \frac{DM_{Fore}}{1+z_f} + \frac{DM_{Host}}{1+z_h} + \frac{DM_{Local}}{1+z_h}},
    \label{eqn:DMobs sum}
\end{split}
\end{equation}
\begin{equation}
\begin{split}
    \rm{\tau_{obs} = \tau_{MW,ISM}  }   
  \rm{ + \frac{\tau_{Fore}}{(1+z_f)^{x_\tau-1}} + \frac{3\tau_{Host}}{(1+z_h)^{x_\tau-1}}} \\
   +\rm{\frac{6\tau_{Local}}{(1+z_h)^{x_\tau-1}}},
    \label{eqn:tauobs sum}
\end{split}
\end{equation}
where the subscripts `MW,ISM', `MW,Halo', `IGM', `Fore', `Host' and `Local' indicate the contribution from the interstellar medium of Milky Way, the halo medium of Milky Way, the intergalactic medium, the extragalactic foreground halos and galaxies, the host galaxies and halos of FRB, and the local environment of FRB sources, respectively; $z_f, \, z_h$ are the corresponding redshifts of extragalactic foreground and host halos and galaxies, respectively. 
The factor of 3 and 6 preceding $\rm{\tau_{Host}}$ and $\rm{\tau_{Local}}$, respectively, account for the increased broadening resulting from the plane wave approximation for extragalactic sources, as opposed to the spherical wave approximation used for sources within the Milky Way \citep[][]{2022ApJ...927...35C}.
The terms $(1+z)^{-1}$ and $(1+z)^{1-x_\tau}$ in Equations \ref{eqn:DMobs sum} and \ref{eqn:tauobs sum} account for the scaling relations of $\rm{DM}-\nu$ and $\, \tau-\nu$, respectively, i.e., time dilation and photon redshift due to cosmic expansion \citep[][]{2003ApJ...598L..79I}. In this work a power law scattering time-frequency scaling, that is, $\tau(\nu)\propto \nu^{-x_\tau}$, with an index $x_\tau \simeq 4$ \citep[][]{2022ApJ...931...88C} is adopted. Furthermore, the scattering caused by the IGM is negligible (\citealt{2013ApJ...776..125M, 2018ApJ...865..147Z}), and is therefore ignored in this study.  

Next, we present the models and procedures used to estimate the DM and $\tau$ induced by various components in the following subsections. Readers less concerned with the specifics of those models and procedures can simply skip the following subsections and move directly to Section 3. For contributions from the Milky Way, local environment, and the IGM, our estimation will primarily rely on the theoretical models and empirical formulas proposed in the literature, with some updates. 
To estimate the contributions arising from the halo of the Milky Way, extragalactic foreground, and host galaxies and halos, we will take advantage of the cosmological hydrodynamic simulation, TNG100-1 (hereafter referred to as TNG100), from the IllustrisTNG project \citep[][]{2018MNRAS.475..624N} \footnote{\href{https://www.tng-project.org}{https://www.tng-project.org}}. 
The moving mesh code AREPO \citep[][]{2010MNRAS.401..791S} was used to run this simulation in a box size of $(75 \, h^{-1}\, \rm{cMpc})^3$, employing the Planck15 cosmology parameters \citep[][]{2016A&A...594A..13P}. The mass resolutions for dark matter and baryonic matter particles are $6.3 \times 10^6 \rm{M_{\odot}}$ and $1.3 \times 10^6 \rm{M_{\odot}}$, respectively. 

\subsubsection{host galaxies and their halos}
\label{sssec:host}
Based on the TNG100 simulation, the DM contributed by the host galaxies and their halos have been explored in detail in our previous work (\citealt{2023MNRAS.518..539M}, referred as Mo23 hereafter). Note that our model only considers galaxies with stellar masses larger than $10^8 \rm{M_{\odot}}$ as potential hosts, for the same reason as outlined in Mo23. The distributions of $\rm{DM_{Host}}$ for FRB populations with young and old progenitors, namely PSFR and PStar, at various redshift (0.0, 0.1, 0.2, 0.5, 0.7, 1.0, 1.5, 2.0), are available in Mo23. More specifically, we adopt the results of PSFR-VNum and PStar-VNum models for TNG100 simulation in Mo23, where '-VNum' indicates that the number of mock FRBs placed in each galaxy is related to its SFR/stellar mass.
In the PSFR model, the probability of placing a mock FRB in a given cell within the host galaxies is proportional to the SFR of that cell. In contrast, in the PStar model, the probability is proportional to the stellar mass within the cell.
To assign a $\rm{DM_{Host}}$ value to a mock FRB at a specific redshift $z_i$, we proceed as follows: First, we interpolate the cumulative distribution of $\rm{DM_{host}}$ at redshift $z_i$ from the distributions of $\rm{DM_{Host}}$ at redshifts 0.0, 0.5, 1.0, 1.5, 2.0, 3.0 that estimated by the method in Mo23. Then, we randomly sample a value of $\rm{DM_{Host}}$ according to the interpolated distribution and assign it to this mock FRB at $z_i$. 

With respect to determining the values of $\rm{DM_{Host}}$, more effort has been devoted to assigning the values of $\rm{\tau_{Host}}$ to mock FRB events. First, we compute the distributions of $\rm{SM_{Host}}$ and $\rm{\tau_{Host}}$ at specific redshifts (0.0, 0.5, 1.0, 1.5, 2.0, 3.0), following procedures similar to those used in Mo23. Briefly, at a particular redshift $z_{snap}$ that is listed above, we place hundreds of thousands of mock sources (which are different from the mock samples we will use for comparison with observations) into galaxies in the TNG100 simulation with their positions being related to the SFR/stellar mass distribution in the host galaxies. We randomly draw 20 L.O.S. from each mock source to the boundary of the host halo. We cut the path from the mock source to the halo boundary along each L.O.S. into short segments of length $\Delta L = 1\, \rm{kpc}/h$. 

The resolution in TNG100 is limited, with a value of approximately $0.75\, \rm{kpc}/h$ for dark matter and as fine as $0.3\, \rm{kpc}/h$ for baryonic gas. Consequently, the state of the CGM and ISM on scales below this resolution is not resolved in the simulation. 
However, recent observations reveal that the CGM exists in a turbulent multiphase state (see, e.g., \citealt{2017ARA&A..55..389T}). In addition, recent simulations employing a carefully designed refinement strategy suggest that the CGM exhibits significant turbulent behavior, with an injection scale shorter than 1 kpc (e.g. \citealt{2020MNRAS.499..597B,2023ARA&A..61..131F}).
It is also well established that the ISM is turbulent. Therefore, we assume that both the CGM and the ISM in TNG100 can be described by the Kolmogorov turbulence model within a certain scale range below the resolution limit. Additionally, we assume that the outer and inner turbulence scales for the ISM and the CGM are 5 pc and $10^6 $ m, respectively,  based on the consideration in \cite{2021ApJ...906...95Z}. Using Equations \ref{eqn:SM_eff}, \ref{eqn:delta tau for r small} and \ref{eqn:delta tau for r large}, we can evaluate the effective scattering measure and $\tau$ contributions from each short segment of the media. It is important to note that if the actual density distributions below the TNG resolution deviate from our assumptions, the evaluation of scattering by the host and foreground galaxies and their halos would need to be revised accordingly.

\begin{figure}
    \includegraphics[width=1.0\columnwidth]{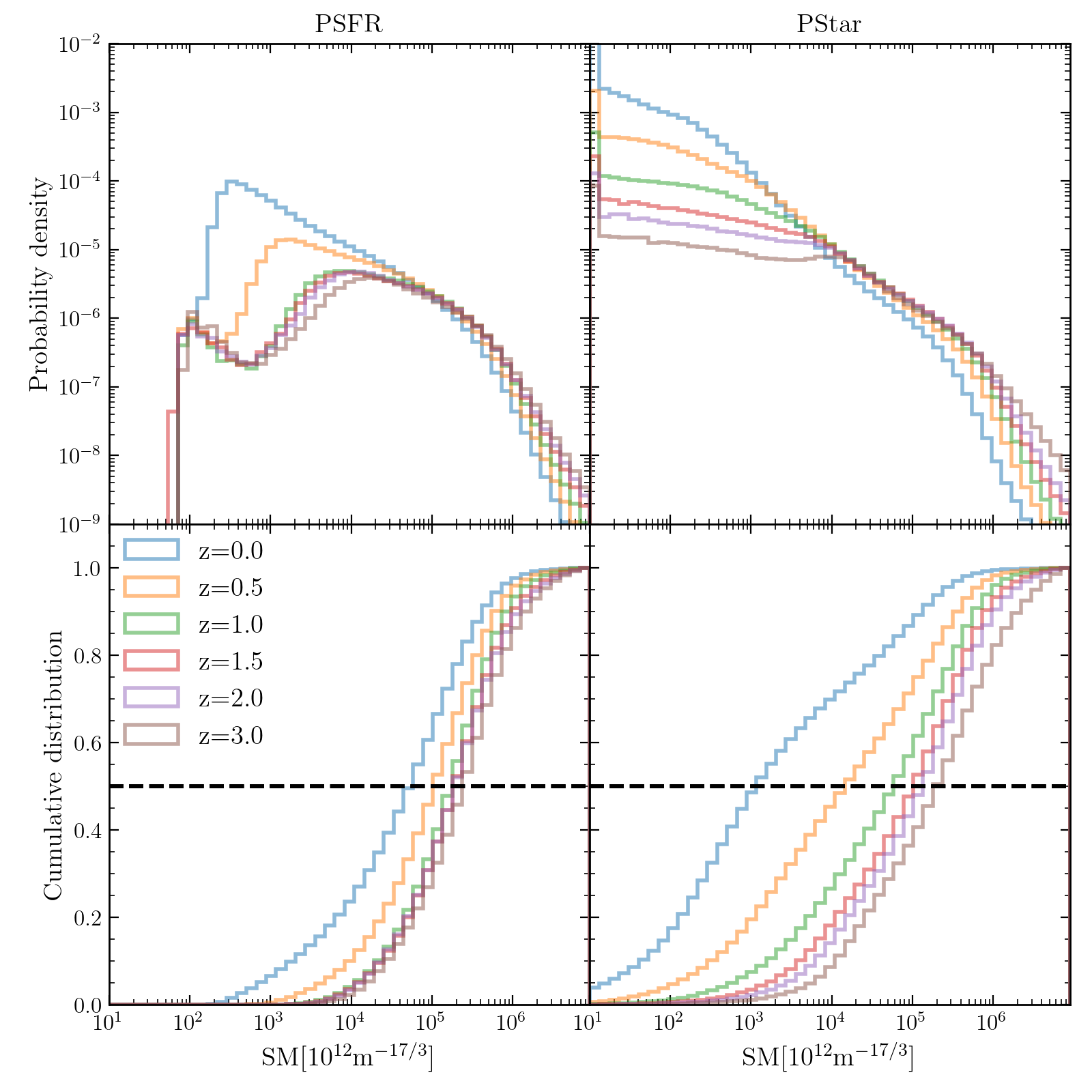}
    \caption{The probability density distribution (top row) and cumulative probability distribution (bottom row) of the scattering measure caused by the host galaxies and their halos, $\rm{SM_{host}}$, for PSFR(left column) and PStar(right column) model at different redshift values based on TNG100 simulation. The blue, orange, green, red, purple, and brown lines are the distribution at z=0.0, 0.5, 1.0, 1.5, 2.0, 3.0, respectively. The black dashed line in the bottom row represents the $50\%$ cumulative probability.}
    \label{fig:smHost}
\end{figure}

\begin{figure}
    \includegraphics[width=1.0\columnwidth]{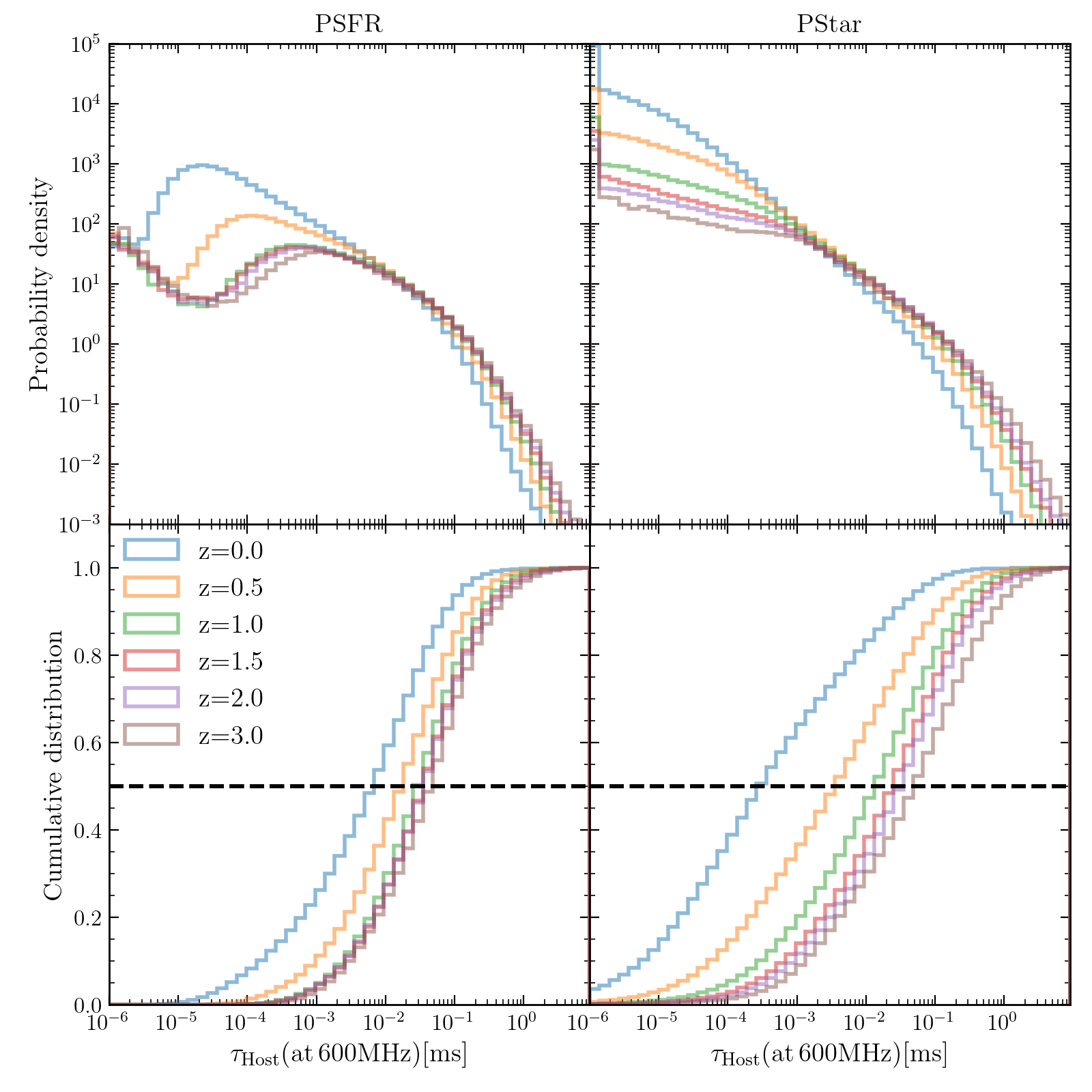}
    \caption{Similar to Figure \ref{fig:smHost}, but for $\rm{\tau_{Host}}$ at 600 MHz.}
    \label{fig:tauHost}
\end{figure}

Summing up the contributions from all segments along each L.O.S., we obtain the cumulative distribution function (CDF) of $\rm{SM_{Host}}$ and $\rm{\tau_{Host}}$ at the redshifts 0.0, 0.5,
1.0, 1.5, 2.0, 3.0, illustrated in Figures \ref{fig:smHost} and \ref{fig:tauHost}, respectively. From these distributions, we can employ interpolation to obtain the distribution of $\rm{SM_{Host}}$ and $\rm{\tau_{Host}}$ at any given redshift $z_i$. 
Subsequently, for any mock event at a specific redshift $z_s$, we assign values of $\rm{SM_{Host,j}}$ and $\rm{\tau_{Host,j}}$  by randomly sampling from the interpolated distributions of $\rm{SM_{Host}}$ and $\rm{\tau_{Host}}$ at redshift $z_s$, using the inverse CDF method.

\subsubsection{Milky Way and its halo}

The contribution to DM and scattering by the electrons in the Milky Way can be estimated by using the NE2001 or YMW16 model. In practice, we calculate $\rm{DM_{MW,ISM}}$ and $\rm{\tau_{MW,ISM}}$ based on python package \texttt{Fruitbat}  \citep[][]{2019JOSS....4.1399B}\footnote{\href{https://fruitbat.readthedocs.io/en/latest/index.html}{https://fruitbat.readthedocs.io/en/latest/index.html}} , assuming the electron distribution in the Milky Way follows the NE2001 model. This computation requires the sky location of mock FRBs, that is, the right ascension and declination, or Galactic longitude and Galactic latitude. We generate the right ascension uniformly from $0^h$ to $24^h$ and determine the declination based on the probability density function illustrated in Figure 3 of \cite{2022ApJ...927...35C}.

The DM caused by the Milky Way's halo, denoted as $\rm{DM_{MW,Halo}}$, is assumed to follow the same distribution of $\rm{DM_{Host,Halo}}$ for disc galaxies in the TNG100 simulation with stellar mass ranging from $10^{10}\rm{M_{\odot}}$ to $10^{11}\rm{M_{\odot}}$ at redshift 0, i.e., follow a log-normal distribution,
\begin{equation}
    P(x; \mu, \sigma) = \frac{1}{x\sigma \sqrt{2\pi}} \rm{exp}\left( -\frac{(\rm{ln} \it(x)-\mu)^2}{2\sigma^2} \right),
\label{eqn:log-normal}
\end{equation}
with the fitting value of median of $e^{\mu}=21.162$ and variance of $\sigma=0.9787$ according to the calculation in Mo23. Note that these values are valid for the PSFR model, with slight variations for the PStar model (see Section 3.4 in Mo23 for more details). 

Based on the results in the previous subsection, we can estimate $\rm{\tau_{MW,Halo}}$ by the typical $\rm{\tau_{Host,Halo}}$ of MW-like galaxies in the TNG100 simulation. In our computation, we can separate $\rm{\tau_{Host}}$ into $\rm{\tau_{Host, ISM}}$ and $\rm{\tau_{Host, Halo}}$ in a method similar to $\rm{DM_{Host}}$ as described in in Mo23 (Section 3.4 therein). At 600 MHz, the 2.5th, median and 97.5th percentiles of $\rm{\tau_{Host,Halo}}$ for the PSFR population are $2.56\times 10^{-7}, \, 1.21\times 10^{-4}, \, 1.78\times 10^{-2}$ ms respectively. For the PStar population, these values are $2.10\times 10^{-8}, \, 4.68\times 10^{-4}, \, 1.01\times 10^{-2}$ ms, respectively. Consequently, $\rm{\tau_{MW,Halo}}$ appears to be negligible, given the electron distribution in the simulated MW-like galaxies in the TNG100 simulation. Therefore, $\rm{\tau_{MW,Halo}}$ is omitted in our model. This is consistent with \cite{2022ApJ...931...88C}, suggesting that the contribution to $\tau$ of FRB from the CGM of the Milky Way can be neglected. However, it is worth noting that the resolution of the TNG100 simulation may not adequately resolve clumps in the CGM of MW-like galaxies. As a result, $\rm{\tau_{MW,halo}}$ in our model may be underestimated in cases where the L.O.S. to certain FRBs passes through electron clumps in the Milky Way halo, although such occurrences are generally rare.

\subsubsection{The Intergalactic Medium}

The scattering caused by the IGM can be ignored. Following the convention procedure (e.g., \citealt[][]{2022ApJ...931...88C}), we sample $\rm{DM_{IGM}}$ for mock events at redshift $z$ using a log-normal function of the same form as Equation \ref{eqn:log-normal}, but with parameters $\sigma=\rm{\{ln[1+(\sigma_{DM_{IGM}} / \langle DM_{IGM} \rangle )^2]\}^2}, \, \mu=\rm{ln\langle DM_{IGM} \rangle - \sigma^2/2}$. In this work, we derive $\langle \rm{DM_{IGM}}(z) \rangle$ and $\rm{\sigma_{DM_{IGM}}}(z)$ as follows. 

The DM caused by the IGM as a function of redshift, denoted as the $\rm{DM_{IGM}}-z$ relation, and its variance have been extensively studied in theoretical and simulation work. However, there are notable discrepancies among various studies (e.g., \citealt{2019ApJ...886..135P, 2021ApJ...906...95Z}). An important factor contributing to these discrepancies is the uncertainty in the fraction of baryons that reside in the IGM, denoted as $f_{b,\rm{IGM}}$, which is not well constrained by observations. Theoretical works typically assume that $f_{b,\rm{IGM}}$ is around 0.80. For example, \cite{2018ApJ...867L..21Z} used $f_{b,\rm{IGM}}=0.83$. Furthermore, differences in the implemented stellar and AGN feedback models can lead to diverse results regarding the fraction of baryons contained by halos (\citealt{2023ARA&A..61..473C}), hence affecting $f_{b,\rm{IGM}}$ among different cosmological hydrodynamical simulations. Additionally, $f_{b,\rm{IGM}}$ is expected to evolve with redshift due to the growth of collapsed structures in the universe. Moreover, in some previous studies, electrons in foreground halos traversed by L.O.S. were also considered part of the IGM, which could moderately overestimate $\rm{DM_{IGM}}$.

\begin{figure}
    \includegraphics[width=1.0\columnwidth]{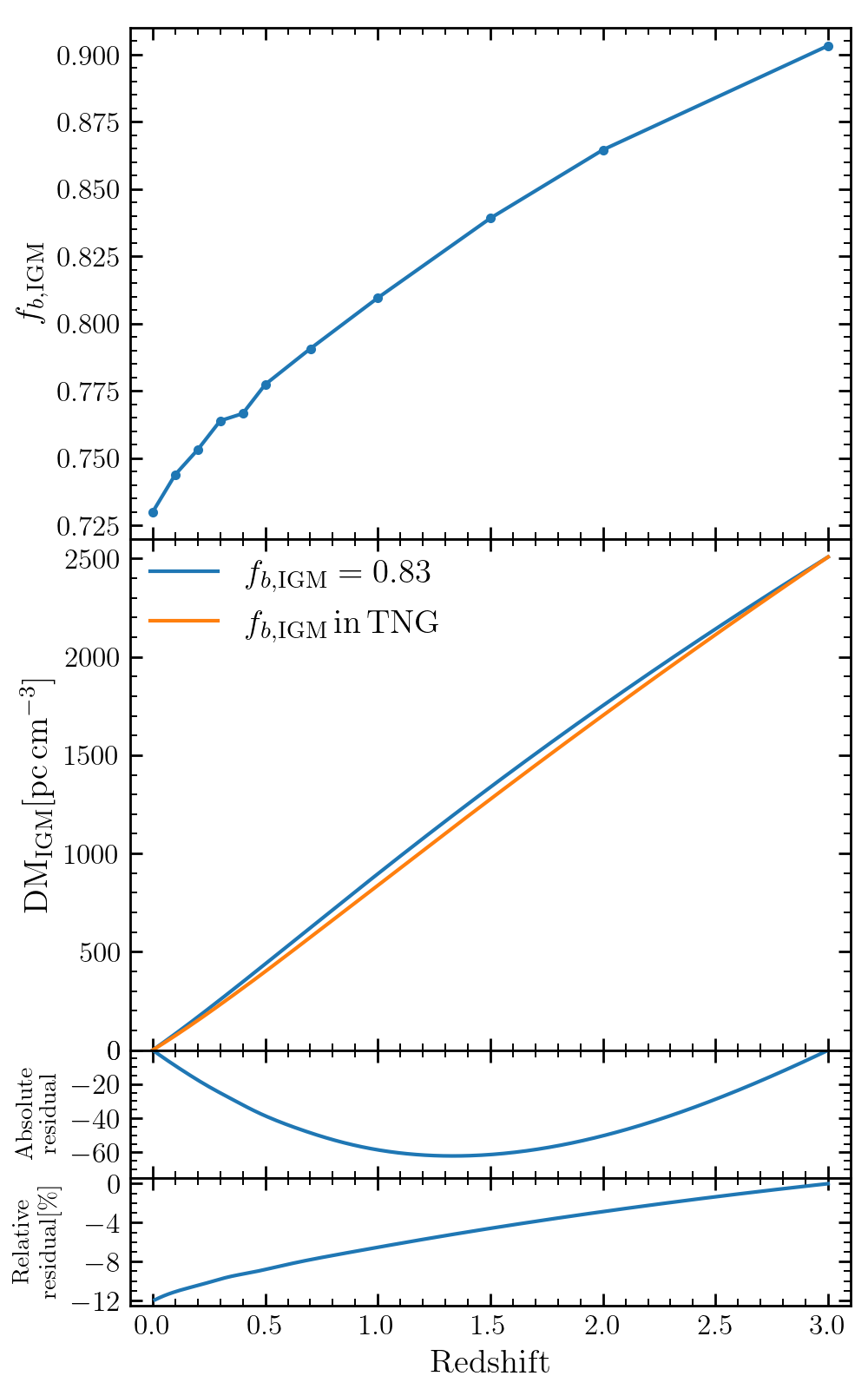}
    \caption{The top panel shows the evolution of baryon fraction in the IGM, $f_{b,\rm{IGM}}$, at redshifts z=0.0,0.1,0.2,0.3,0.4,0.5,0.7,1.0,1.5,2.0,3.0, which are calculated from the TNG100 simulation. The second panel shows $\rm{DM_{IGM}}$ as a function of redshift, in which the blue line is calculated by Equation \ref{eqn:DMigm} with $f_{b,\rm{IGM}}=0.83$, and the orange line is derived from the value in the top panel. The third panel presents the absolute residual of calculating $\rm{DM_{IGM}}$ by using the value of $f_{b,\rm{IGM}}$ in the TNG100 simulation, with respect to use $f_{b,\rm{IGM}}=0.83$. The fourth panel is the corresponding relative residual.}
    \label{fig:figm-DMigm-z}
\end{figure}

In this work, we define the IGM as baryons outside of dark matter halos. Baryons within collapsed halos are included in contributions from the hosts and intervening foreground halos, that is, $\rm{DM_{Fore}}$ and $\tau_{\rm{Fore}}$. We measure $f_{b,\rm{IGM}}$ by excluding baryons within the radius $R_{\rm{m200}}$ of halos in the TNG100 simulation. $R_{\rm{m200}}$ corresponds to the region where the density is 200 times the mean density of the universe. The values of $f_{b,\rm{IGM}}$ at redshifts 0.0, 0.1, 0.2, 0.3, 0.4, 0.5, 0.7, 1.0, 1.5, 2.0 and 3.0, extracted from the TNG100 simulation samples are shown in the upper panel of Figure \ref{fig:figm-DMigm-z}. To account for the evolution of $f_{b,\rm{IGM}}$ in the TNG100 simulation, we modify the Zhang18 model \citep[][]{2018ApJ...867L..21Z} to generate $\rm{DM_{IGM}(z)}$. Specifically, the mean value of $\rm{DM_{IGM}(z)}$ at a given redshift z is given by:
\begin{equation}
    \langle \rm{DM_{IGM}(z)} \rangle = \frac{3cH_0\Omega_{b,0} }{8\pi G m_p} \int^z_0\frac{\chi(z) f_{b,\rm{IGM}}(z) (1+z) dz}{[\Omega_m (1+z)^3+\Omega_\Lambda]^{1/2}} 
\label{eqn:DMigm}
\end{equation}
where $m_p$ is the proton mass, $\chi(z)=\frac{3}{4}y_1 \chi_{e,\rm{H}}(z)+\frac{1}{8}y_2 \chi_{e,\rm{He}}(z)$ is the number of free electrons per baryon in the universe, and $y_1, y_2 \sim 1$ denote the possible slight deviation from a hydrogen and helium mass abundance fraction 0.75, 0.25, respectively, and $\chi_{e,\rm{H}},\,\chi_{e,\rm{He}}$ represents the ionization fraction of hydrogen and helium respectively. At redshifts z<3, both hydrogen and helium are fully ionized, leading to $\chi(z)\sim 7/8$ \citep[][]{2018ApJ...867L..21Z}. The values of $f_{b,\rm{IGM}}(z)$ are interpolated from $f_{b,\rm{IGM}}(z_i)$ at different redshifts that were extracted from the TNG100 simulation. 

\begin{figure}
    \includegraphics[width=1.0\columnwidth]{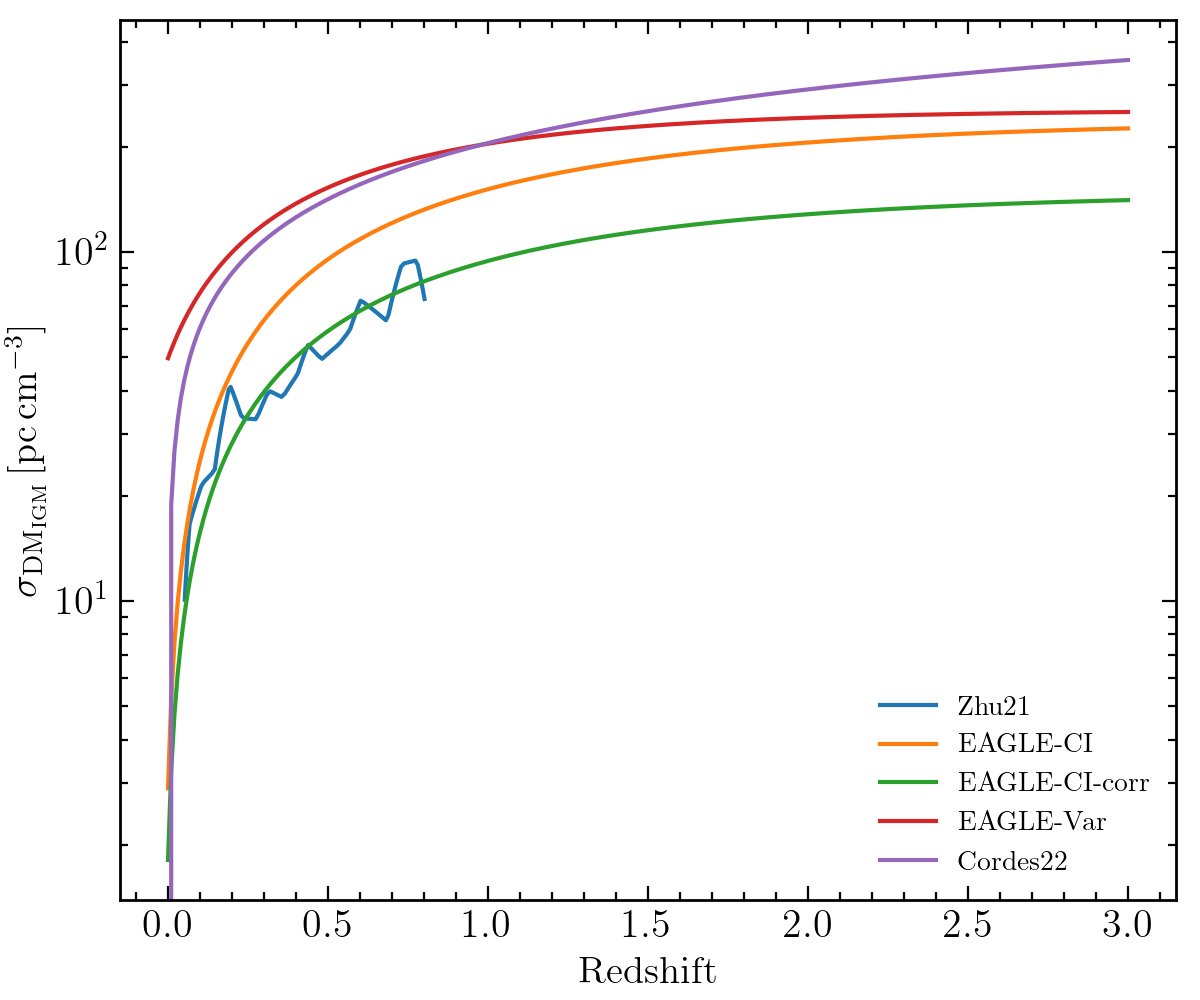}
    \caption{The blue, orange, red and purple lines are the evolution of $\sigma_{\rm{DM_{IGM}}}$ according to the results in \cite{2021ApJ...906...95Z}, best fitting results of EAGLE simulation given by the eqn. 12 and eqn. 11 of \cite{2021MNRAS.505.5356B}, and the simplified model used in \cite{2022ApJ...931...88C}, respectively. The green line is the formula of $\sigma_{\rm{DM_{IGM}}}$ we used in this work.}
    \label{fig:sigma DMigm z} 
\end{figure}

Meanwhile, due to the inhomogeneous distribution of the IGM, the value of $\rm{DM_{IGM}(z)}$ at a given redshift will fluctuate around $\langle \rm{DM_{IGM}(z)} \rangle$ along different L.O.S. \citep[][]{2014ApJ...780L..33M}, as estimated by cosmological hydrodynamic simulations (e.g. \citealt{2019ApJ...886..135P, 2021MNRAS.505.5356B, 2021ApJ...906...95Z, 2021ApJ...906...49Z}). However, \cite{2021MNRAS.505.5356B} and \cite{2021ApJ...906...49Z} did not exclude the contribution of the intervening foreground halos along the L.O.S. when estimating $\rm{DM_{IGM}}$ and its scatter. \cite{2021ApJ...906...95Z} separated the contribution from the diffuse IGM and the intervening foreground halos but only provided the results below redshift 0.8. We combine the results in \cite{2021ApJ...906...95Z} and \cite{2021MNRAS.505.5356B} with the following equation to describe the variance of $\rm{DM_{IGM}(z)}$ below z<3,
\begin{equation}
    \sigma_{\rm{DM_{IGM}}}(z) = 0.623\cdot (-234.3e^{1.0z}+237.2),
\label{eqn:sigma DMigm}
\end{equation}
where $-234.3e^{1.0z}+237.2$ is the fitting formula proposed in \cite{2021MNRAS.505.5356B}, and the factor 0.623 is introduced to exclude the effect of baryons in intervening halos based on the results in \cite{2021ApJ...906...95Z}. In comparison, the value given by Equation \ref{eqn:sigma DMigm} is lower than the simplified model used in \cite{2022ApJ...931...88C}, that is, $\sigma_{\rm{DM_{IGM}}}(z) = \sqrt{ 50\langle \rm{DM_{IGM}(z) \rangle}}$, as shown in Figure \ref{fig:sigma DMigm z}. 

\begin{figure}
    \includegraphics[width=1.0\columnwidth]{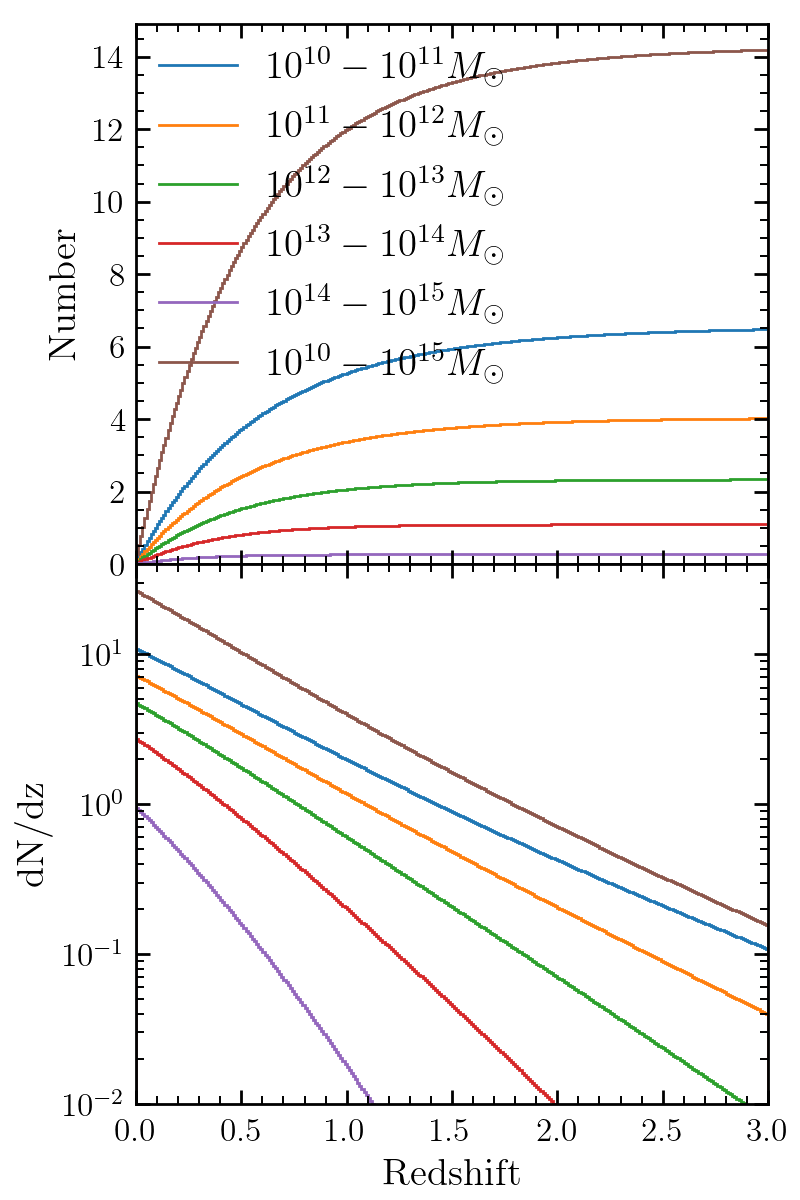}
    \caption{The upper and lower panels show the number of intervening extragalactic foreground halos and the differential number per unit redshift, as function of the redshift, respectively. The blue, orange, green, red, purple and brown lines indicate the results of halo mass with $10^{10}-10^{11},\, 10^{11}-10^{12},\, 10^{12}-10^{13},\, 10^{13}-10^{14},\, 10^{14}-10^{15},\, 10^{10}-10^{15}\, \rm{M_{\odot}}$, respectively.}
    \label{fig:num_halo_z}
\end{figure}

\subsubsection{extragalactic foreground halos and galaxies}

\begin{figure}
    \includegraphics[width=1.0\columnwidth]{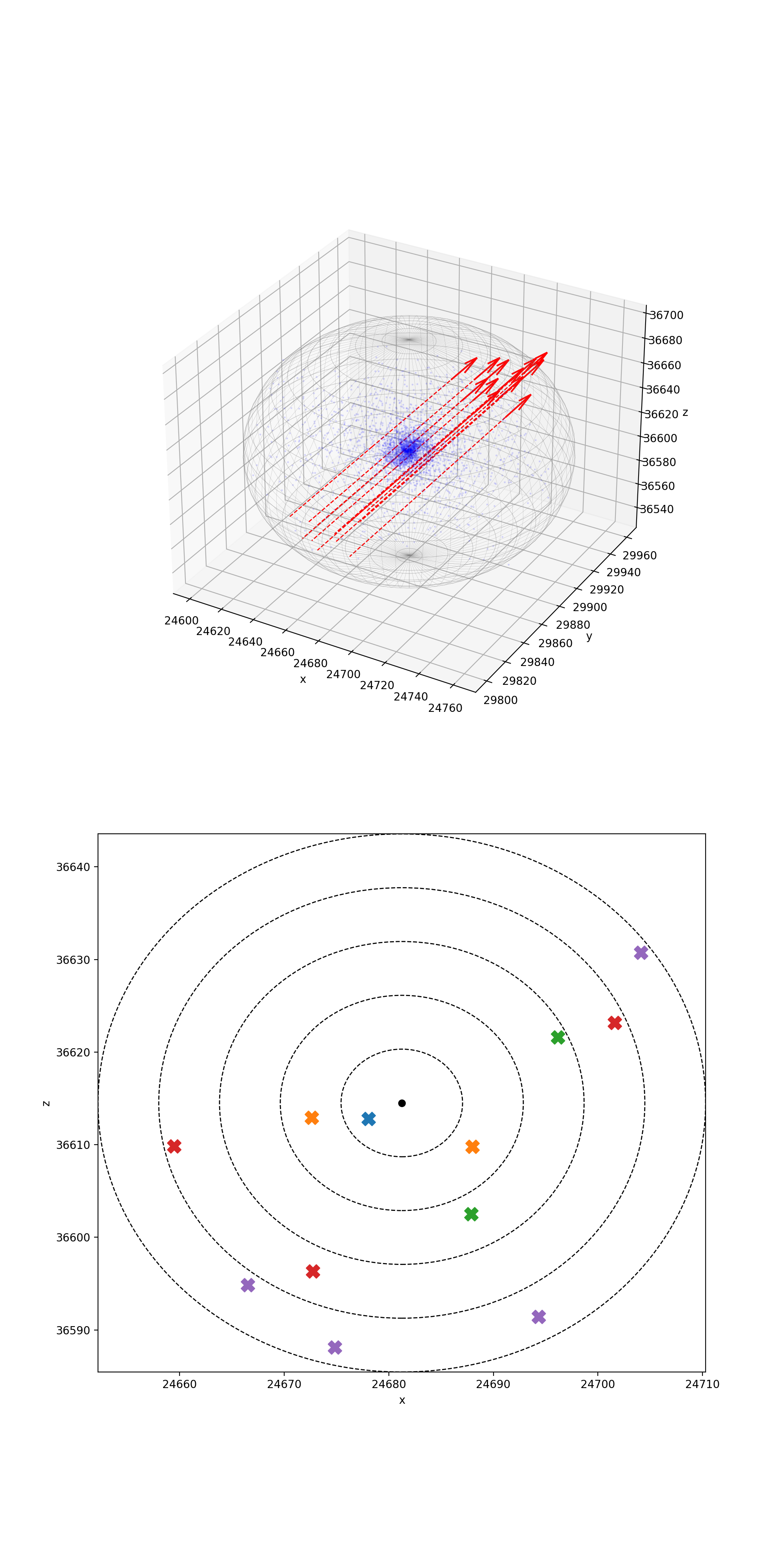}
    \caption{The upper panel is the 3D schematic diagram which illustrates the L.O.S. to FRBs that have passed through a foreground halo. The red dashed line with an arrow shows the L.O.S. and corresponding direction. The blue dots are gas particles in simulation, which indicates the spatial distribution of halo gas. The outer sphere is the halo boundary with radius $\rm{R_{m200}}$. The lower panel shows the projection of the L.O.S. to the plane that is perpendicular to them and across the halo center. Symbols 'x' with different colors represent the intersection points of different L.O.S.. The dotted circles mark radial bins.}
    \label{fig:passForeHalo example}
\end{figure}

For FRBs with $z > 0.2$, the contribution of extragalactic foreground halos and their galaxies to the DM and scattering cannot be ignored (e.g. \citealt{2013ApJ...776..125M, 2021ApJ...906...95Z}). \cite{2021ApJ...906...95Z} showed that the DM caused by the medium in foreground halos and galaxies is around 25$\%$ of that caused by diffuse IGM for a source at $z \sim 0.8$. \cite{2021ApJ...906...95Z} employed a direct method to estimate the contribution of foreground halos by stacking the simulation boxes at different redshift snapshots. They select a large number of sightlines passing through these boxes and identified the segments within halos for each sightline. We now adopt a method that significantly reduces the demand for computational resources. Firstly, one can estimate the mean number of intervening halos with different masses for a l.o.s ending at a specific redshift. For a sightline segment starting at redshift $z_i$ and ending at $z_f$, the number of halos with a mass $M_h$ that it will pass through can be given by \cite{padmanabhan2000theoretical},
\begin{equation}
    N_{\rm{obj}}=\int^{z_f}_{z_i} \frac{\pi r^2 n(z')d_H(z')}{1+z'}dz' ,
\label{eqn:number passing}
\end{equation}
where $r$ is the typical proper radius of halo with mass $M_h$, and here we use the virial radius of halos; 
$n(z)$ is the proper number density of halos, which can be calculated by the python package \texttt{hmf} \citep[][]{2013A&C.....3...23M, 2014ascl.soft12006M} \footnote{\href{https://hmf.readthedocs.io/en/latest/index.html}{https://hmf.readthedocs.io/en/latest/index.html}}. In \texttt{hmf}, the halo mass function is obtained according to the Extended Press-Schechter (EPS) formalism, which is consistent with the halo mass function of the TNG100 simulation; 
$d_{H}(z)=\frac{c}{H_0\sqrt{\Omega_\Lambda+\Omega_m (1+z)^3}}$ is the Hubble radius based on the standard cosmological model. 

Based on Equation \ref{eqn:number passing}, we derive the expected number of halos within the mass range $10^{10}-10^{15}\, \rm{M_\odot}$ that the light path starting from $z=0$ will pass through as a function of the source redshift, shown in the top panel of Figure \ref{fig:num_halo_z}. The signals emitted by an FRB source at $z\sim 1$ are expected to pass through around $7$ halos within the mass range $10^{10}-10^{15}\, \rm{M_\odot}$, consistent with previous studies \citep[e.g.,][]{2019MNRAS.483..971V}. The entire mass range is divided into 5 bins with an interval of $d\rm{log}(M_h)=1$, and the expected number of intervening halos in each mass bin is also displayed in Figure \ref{fig:num_halo_z}. Additionally, the bottom panel of Figure \ref{fig:num_halo_z} presents the differential number of intervening halos within different mass bins over redshift. Since the distribution of electrons in halos within the same mass bins also evolves with redshift, we divide the redshift range from the observer ($z=0$) to a source at $z_s$ (i.e., $0-z_s$) into many bins with an interval $\delta z=0.1$. In each redshift bin $z_{ibin}=[(i-1)*0.1, i*0.1), i=1, 2, 3, ...$, we can calculate the mean number of intervening halos within each halo mass bin $n(z_{ibin}, M_{jbin})$, where $M_{jbin}$ indicates the mass bin $10^{10+j-1}-10^{10+j} \rm{M_{\odot}}, j=1,2,3,...$. Furthermore, if the DM and scattering measure caused by the foreground halos in each redshift and mass bin ($z_{ibin}$, $M_{jbin}$), denoted as $\Delta \rm{DM_{Fore}}(i,j)$ and $\Delta\rm{SM_{Fore}}(i,j)$ respectively, are determined, it is straightforward to derive $\rm{DM_{Fore}}$ and $\rm{\tau_{Fore}}$ for a source at $z_s$. 

\begin{figure}
    \includegraphics[width=1.0\columnwidth]{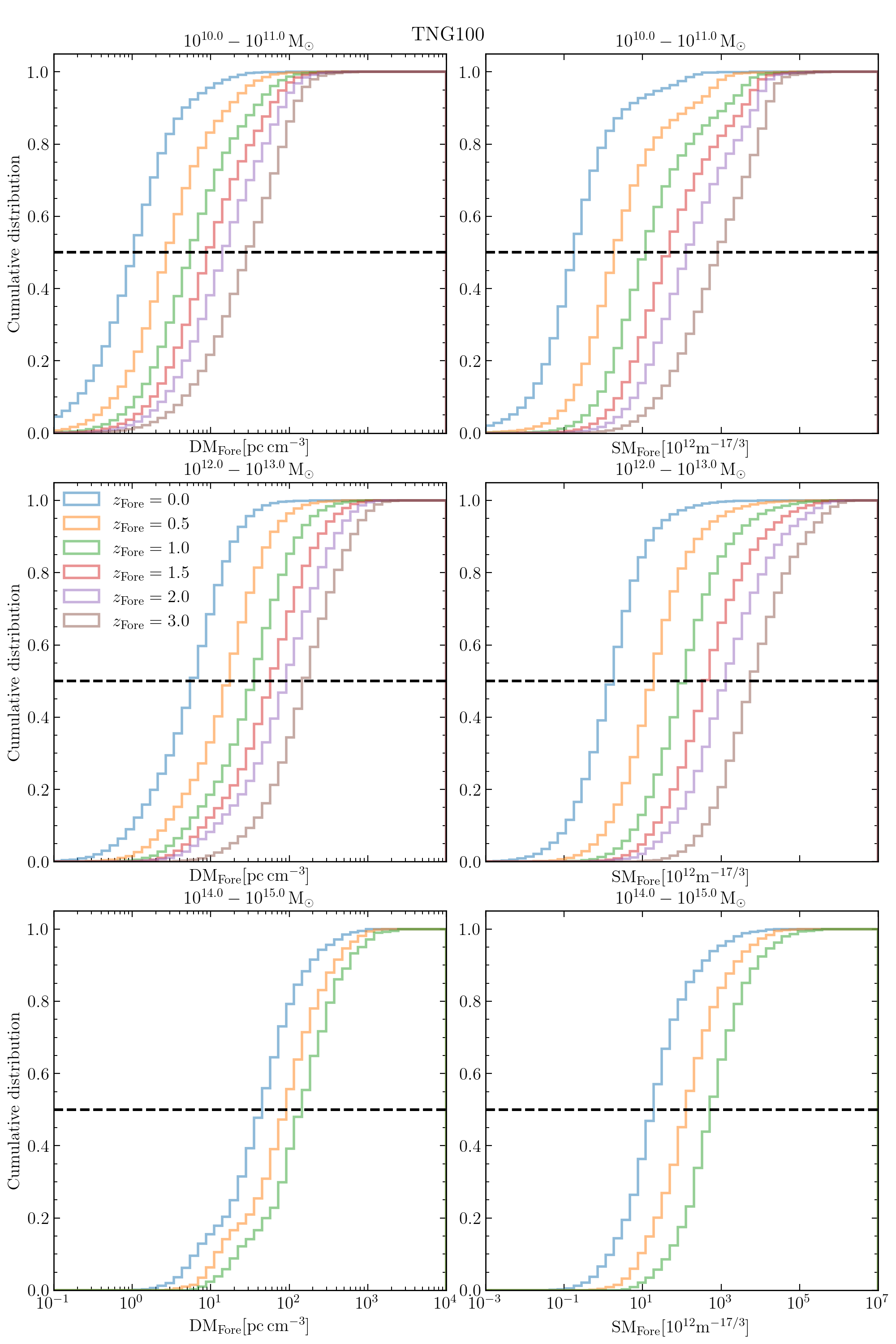}
    \caption{The cumulative distribution function of the dispersion measure and scattering measure caused by the foreground halos, i.e., $\rm{DM_{Fore}}$ (left) and $\rm{SM_{Fore}}$ (right), for halos with mass in the range $10^{10}-10^{11},\, 10^{12}-10^{13},\, 10^{14}-10^{15}\, \rm{M_{\odot}}$ (from top to bottom) respectively.}
    \label{fig:DMSM Fore massbin distribution}
\end{figure}

Actually, the expected number $n(z_{ibin}, M_{jbin})$ in a particular bin $(z_{ibin}, M_{jbin})$ is smaller than 1. Moreover, the DM and scattering measures along different light paths passing through the same halo can vary significantly. Therefore, our method to generate $\Delta \rm{DM_{Fore}}(i,j)$ and $\Delta\rm{SM_{Fore}}(i,j)$ is divided into two parts. The first part involves deriving the CDF of DM and SM along all possible sightlines passing through the foreground halos within the redshift and mass bin ($z_{ibin}$, $M_{jbin}$). These distributions are denoted as $\rm{CDF_{DM}(i,j)}$ and $\rm{CDF_{SM}(i,j)}$ respectively. In practice, we use interpolation to obtain $\rm{CDF_{DM}(i,j)}$ and $\rm{CDF_{SM}(i,j)}$ from the corresponding CDF for each mass bin at redshift $z=0.0, 0.5, 1.0, 1.5, 2.0, 3.0$.

In each of these six snapshots, we select the halos in the TNG100 simulation that fall into each mass bin $M_{jbin}$. For each selected halo, we randomly draw multiple sightlines that pass through it along the Y-axis of the simulation coordinate. In the two-dimensional projected plane, the probability that the sightline falls within a particular region within the halo is proportional to the area of the region. On average, we assume that an area of $\rm{(10kpc/h)^2}$ contains one sightline. The procedure for drawing lines is illustrated in Figure \ref{fig:passForeHalo example}. Our additional examination indicates that the projection direction, along the x, y, or z axis of the simulation coordinate, has a minor impact on the results. Next, we split each sightline into short segments with a length of $\Delta L=1\,\rm{kpc/h}$, and sum up the contribution to the DM and SM according to the electron distribution within the halo in every segment. Then, we collect the values of the DM and SM associated with all the sightlines drawn for halos in each mass bin at a specific redshift to build the corresponding CDF. Figure \ref{fig:DMSM Fore massbin distribution} shows the CDF for several mass bins at $z=0.0, 0.5, 1.0, 1.5, 2.0, 3.0$. 
From the CDFs at these redshifts, we interpolate $\rm{CDF_{DM}(i,j)}$ and $\rm{CDF_{SM}(i,j)}$. We have tested that the interpolated results (e.g. $z=0.1,0.2,0.7$) can agree well with the results derived directly using the method used above to calculate the CDFs at $z=0.0, 0.5, 1.0, 1.5, 2.0, 3.0$.  

The second part of generating $\Delta \rm{DM_{Fore}}(i,j)$ and $\Delta\rm{SM_{Fore}}(i,j)$ involves sampling from $\rm{CDF_{DM}(i,j)}$ and $\rm{CDF_{SM}(i,j)}$ with a probability equal to $n(z_{ibin}, M_{jbin})$. 
More specifically, for any particular bin $(z_{ibin}, M_{jbin})$ along the light path to any event, we draw two random numbers, $\eta$ and $\xi$, respectively, in the range (0,1). If $\eta$ is smaller than $n(z_{ibin}, M_{jbin})$, we deduce the DM and SM at which the cumulative probability of $\rm{CDF_{DM}(i,j)}$ and $\rm{CDF_{SM}(i,j)}$ is equal to $\xi$, and then assign them to be $\Delta \rm{DM_{Fore}}(i,j)$ and $\Delta\rm{SM_{Fore}}(i,j)$. That is, we use the inverse transform sampling method to generate $\Delta \rm{DM_{Fore}}(i,j)$ and $\Delta\rm{SM_{Fore}}(i,j)$.

For each mock source at $z_s$, we employ the procedures described above to produce all $\Delta \rm{DM_{Fore}}(i,j)$ and $\Delta\rm{SM_{Fore}}(i,j)$ along the sightline and then calculate the total $\rm{DM_{Fore}}$ and $\rm{\tau_{Fore}}$. For the latter, the effective distance $\rm{D_{eff}}$ in each redshift bin can be determined directly using the source redshift. 

\subsubsection{local environment}
\label{subsubsec:local env}
So far, the understanding of the local environment of FRB sources is rather poor. We assume that $\rm{DM_{Local}}$ follows a log-normal distribution, following the method used in \cite{2022ApJ...927...35C} and is based on the estimated local DM for Galactic pulses (\citealt{2003astro.ph..1598C}). We use the parameters $\mu=1.8,\, \sigma=0.8$ for the log-normal distribution of $\rm{DM_{Local}}$ for the PStar source population, and $\mu=2.8,\, \sigma=0.8$ for the PSFR source population, respectively. This assumption is made because of the expectation that star-forming environments tend to have higher gas and electron density. The 2.5th, median and 97.5th percentiles of $\rm{DM_{Local}}$ for the PStar population are $1.26, \, 6.05, \, 29.03 \, \rm{pc\,cm^{-3}}$, respectively, while for the PSFR population, they are $3.43, \, 16.44, \, 78.89 \, \rm{pc\,cm^{-3}}$ respectively. \cite{2002astro.ph..7156C} proposed that the local SM can be related to the $\rm{DM_{Local}}$ as,
\begin{equation}
\begin{split}
    \rm{SM_{Local}} = C_{SM} F n^2_e \Delta s \,\,\,\\
    = 10^{-5.55}F \rm{\frac{DM_{Local}}{(pc\,cm^{-3})}}{\frac{1}{\Delta s /(kpc)}} [kpc\,m^{-20/3}],
\end{split}
\label{eqn:sm local}
\end{equation}
where $F$ indicates the magnitude of the fluctuation in the electron density $n_e$, and $\Delta s$ denotes the proper size of clumps that contain electrons in the local environment. Usually, the local environment refers to the region very near the FRB source, i.e. within a few parsecs. Therefore, we assume that $\Delta s$ follows a uniform distribution from 0.001 to 0.01 kpc in this work. To estimate the scattering time, $D_{\rm{eff}}$ is required. \cite{2021ApJ...922..173C} estimated 
$D_{\rm{eff}}\sim 1\rm{kpc}$ for the screen of FRB190608, while \cite{2022ApJ...931...87O} suggested that $D_{\rm{eff}}$ could be under 0.1 kpc for FRB190520B. Hence, we assume that $D_{\rm{eff}}$ for the local screen follows a uniform distribution from 0.01 to 1 kpc. The fluctuation parameter $F$ can vary from $10^{-3}$ to 100 in different  media. For the ISM in host galaxies, \cite{2022ApJ...931...88C} assumed a uniform distribution of F within the range 0.01 to 10, or alternatively from 0.5 to 2. \cite{2022ApJ...934...71O} adopted a log-normal distribution for F, with a mean value between 0.003 and 1.0 for the ISM in various galaxies. The density fluctuation parameter of circumburst media is likely higher than the typical value of ISM, as suggested by previous studies of Parkes FRB (\citealt{2020MNRAS.498.4811H}), CHIME/FRB (\citealt{2022ApJ...927...35C}) and FRB 20190520B (\citealt{2022ApJ...931...87O}.) For simplicity, we assume that F follows a uniform distribution from 1 to $F_{\rm{max}}$ in our default scenario, where $F_{\rm{max}}$ is treated as a free parameter that will be determined by the Monte Carlo simulations detailed in Section \ref{sec:MCMC}.

Note that, due to the limited resolution of the TNG100 simulation, dense electron clumps with sizes ranging from a few to hundreds of parsecs cannot be resolved. These clumps may exist in the ISM and CGM of simulated galaxies. The contributions of unresolved nearby clumps to the scattering of mock FRBs can be accounted for by Equation \ref{eqn:sm local} and relatively small $D_{\rm{eff}}$. However, contributions from unresolved clumps with a large density fluctuation parameter in the ISM and CGM may not be adequately captured by the terms of $\tau_{\rm{Local}}$ and $\tau_{\rm{host}}$ in our default model. To address this limitation, we consider an alternative scenario in which $D_{\rm{eff}}$ of the local screen can range from 0.01 to 100 kpc when interpreting our results in the following sections. In this additional scenario, referred to as `Extended-local-scattering', we retain the term `local', which may encompass contributions from electron clumps located up to 100 kpc from the FRB source.

\section{Monte Carlo simulations}
\label{sec:MCMC}
Using the assumed population model and intrinsic properties of FRBs described in Section 2.3, we now generate approximately $10^8$ mock FRBs with different redshifts and energies. This process involves multiple free parameters, such as the fraction of sources tracing young progenitor $f_{\rm{PSFR}}$, the parameters in the FRB energy distribution $\gamma$ and $\rm{log_{10}(E_{*})}$, and the maximum fluctuation in the electron density of the local environment $F_{\rm{max}}$. From these mock FRBs, we select observable mock samples by applying the same fluence limit $>0.4 \rm{[Jy \, ms]}$ as \cite{2022ApJ...927...35C}, resulting in roughly $10^6$ selected mock FRBs. Subsequently, we compare the DM and $\tau$ distribution of the selected mock FRBs to those of the selection-corrected CHIME/FRB Catalog 1.
 
More specifically, we partition the DM ($\tau$ at 600 MHz) range from $100$ to $2300\, \rm{pc\,cm^{-3}}$ (0.1 ms to 10.0 ms) into 20 logarithmically spaced equal bins. Note that, the sources with $\tau>10.0$ ms are not included in our primary analysis due to significant uncertainties in the selection process of CHIME/FRB catalog. However, a brief discussion of the results, including such sources, is provided. We compare the distribution of observable mock sources in various DM ($\tau$) bins, $N_i, i=1,2,...,20$, with that of the selection-corrected CHIME/FRB Catalog 1, represented as $n_{i}$. These distributions are illustrated in Figure \ref{fig:compa mock obs}. The total number of observable mock sources has been re-scaled to match that of selection-corrected CHIME/FRB samples. To determine the optimal parameters for generating mock sources that can replicate the CHIME/FRB observations, we perform MCMC simulations by the python package \texttt{emcee} \citep[]{emcee}\footnote{\href{https://emcee.readthedocs.io/en/stable}{https://emcee.readthedocs.io/en/stable/}}. 

\begin{figure*}
    \centering
    \includegraphics[width=1.6\columnwidth]{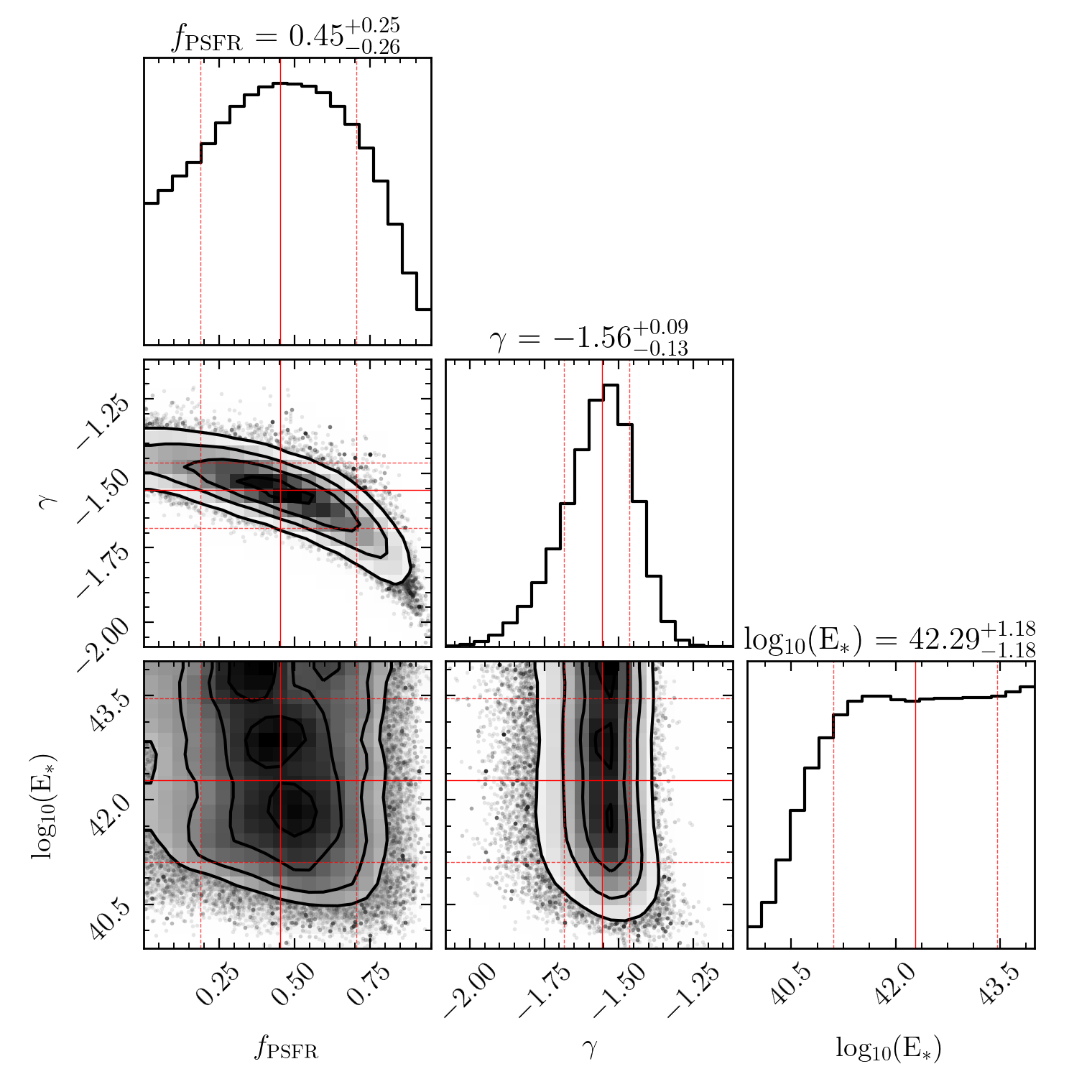}
    \caption{The posterior distribution of the MCMC results for the parameters $f_{\rm{PSFR}}, \, \gamma, \, \rm{log_{10}(E_*)}$ in DM-only model, thinned by a factor of 20 for the sake of clarity. The vertical and horizontal red lines denote the median(solid), 16th(dashed) and 84th(dashed) percentile of the distribution.}
    \label{fig:dm only mcmc res}
\end{figure*}

\begin{figure*}
    \centering
    \includegraphics[width=1.6\columnwidth]{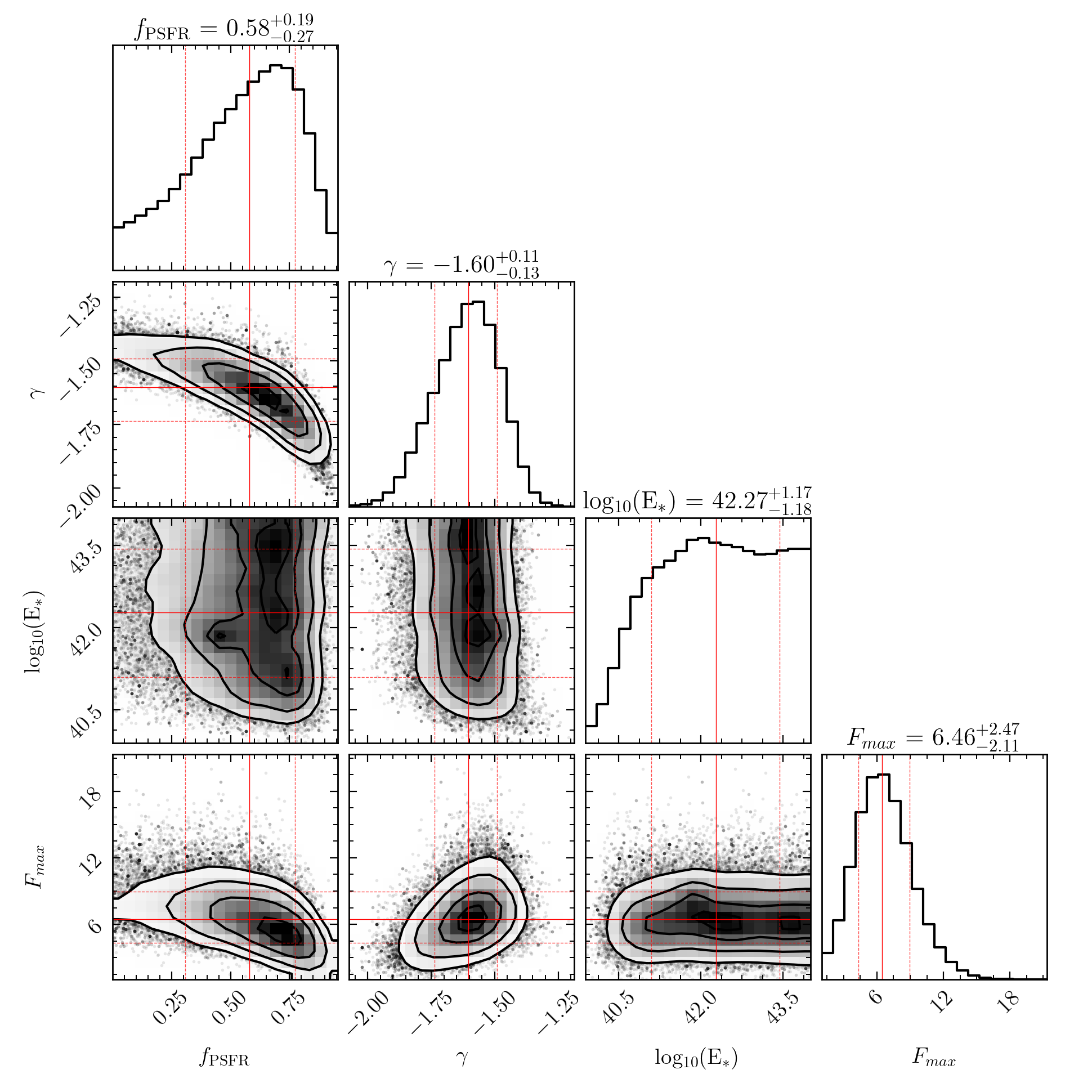}
    \caption{Similar to Figure \ref{fig:dm only mcmc res}, but for the parameters $f_{\rm{PSFR}}, \, \gamma, \, \rm{log_{10}(E_*)}, \, F_{\rm{max}}$ in the DM and $\tau$ combined  model. }
    \label{fig:mcmc res}
\end{figure*}


The log-likelihood function of MCMC reads as 
\begin{equation}
    \log \mathcal{L}=-0.5\cdot\chi^2=-0.5\cdot(\mathbf{N_i - n_i})^T Cov^{-1} (\mathbf{N_i - n_i}) ,
\label{eqn:likehood}
\end{equation}
where Cov is the covariance matrix.
All our parameters are sampled from uniform priors, as detailed in Table \ref{tab:dmtau mcmc res}. These priors are used to initialize the MCMC chains, which are then run with 50 walkers, 2000 burn-in steps, and a total of 20000 steps. According to \cite{emcee}, MCMC chains longer than approximately $50\tau_{f}$ typically ensure convergence, where $\tau_{f}$ represents the integrated auto-correlation time. Thus, our models might need an MCMC chain of 60,000 steps or more to converge. Alternatively, if the posterior distribution shapes in the first and second halves of the chain resemble each other, the chain can also be considered as converged \citep[][]{2018ApJS..236...11H}, which is the scenario for our chains. 

To explore the impact and potential benefits of combining DM and $\tau$, we performed MCMC simulations in two ways. One method only compares the DM distribution between observable mock events and the selection-corrected CHIME samples, while the other method will compare both the DM and $\tau$ distributions. 

\renewcommand{\arraystretch}{1.2} 
\begin{deluxetable*}{cccc}
	\caption{Table of results of the parameters fit in the DM and $\tau$ distribution. Best-fit results quote the median values of the posterior distributions, with the error bars indicating the 16th and 84th percentile of the posterior distributions.}
	\label{tab:dmtau mcmc res}
\tablehead{
\colhead{Parameter} & \colhead{Uniform log-prior range} & \colhead{Best-fit result(DM only) } & \colhead{Best-fit result (DM and $\tau$)}}
\startdata
            $f_{\rm{PSFR}}$ & [0.00, 1.00] & $0.45^{+0.25}_{-0.26}$ & $0.58^{+0.19}_{-0.27}$  \\
            $\gamma$ & [-3.00, -0.5] & $-1.56^{+0.09}_{-0.13}$ & $-1.60^{+0.11}_{-0.13}$  \\
            $\rm{log_{10}(E_{*})}$ & [37.00, 44.00] & $42.29^{+1.18}_{-1.18}$ & $42.27^{+1.17}_{-1.18}$  \\
            $F_{max}$ & [1.00, 500.00] & - & $6.46^{+2.47}_{-2.11}$  \\
\enddata
\end{deluxetable*}

\subsection{DM only}
First, we run MCMC with three free parameters ($f_{\rm{PSFR}}, \gamma, E_{*}$), to compare the observable mock sources with observations on the distribution of DM. The results are illustrated in Figure \ref{fig:dm only mcmc res}, created using the \texttt{corner} package \citep[][]{corner}. In this figure, the median of the posterior distributions is represented by solid lines, both horizontal and vertical, representing the optimal parameters. Additionally, the 16th and 84th percentiles of the posterior distributions are denoted by dashed lines, representing the lower and the upper $1\sigma$ error bars, respectively. These optimal parameters, along with their associated error bars, are listed in Table \ref{tab:dmtau mcmc res}. We find that with $f_{\rm{PSFR}}=0.45^{+0.25}_{-0.26}, \gamma=-1.56^{+0.09}_{-0.13}, \rm{log_{10}(E_{*})}=42.29^{+1.18}_{-1.18}$, the DM distribution of our observable mock FRBs is consistent with the selection-corrected CHIME/FRB observations. 

The parameter $\rm{log_{10}(E_*)}$ shows signs of non-convergence toward the end of the MCMC process, displaying a broad range of values from 42 to 44. The reason is that, when $\gamma$ is kept constant and $\rm{log_{10}(E_*)}$ exceeds 42, the distribution of FRBs as a function of energy changes slowly as $\rm{log_{10}(E_*)}$ increases.
Several previous works (e.g., \citealt{2022ApJ...927...35C}; \citealt{2022JCAP...01..040Q}; \citealt{2023ApJ...944..105S}) have also been able to generate mock events with a DM distribution consistent with the CHIME observations. It should be noted that the contributions from host galaxies and extragalactic foreground halos are usually estimated by theoretical models in the work of \cite{2022ApJ...927...35C}, while our work is based on the properties of galaxies and halos in the TNG100 simulation. 


The optimal parameters that we have determined are in partial agreement with certain previous studies. For example, \cite{2020MNRAS.494..665L} found a moderately higher index of $\gamma=-1.79^{+0.31}_{-0.35}$ and a substantially larger $\rm{log_{10}(E_{*})}=44.46^{+0.71}_{-0.38}$ for 46 known FRBs detected by 7 surveys that work at around 1GHz, such as the Parkes Magellanic Cloud Pulsar Survey, the Parkes Magellanic Cloud Pulsar Survey, Pulsar Arecibo L-band Feed Array, etc. Meanwhile, \cite{2023ApJ...944..105S} reported lower values for both $\gamma$ and $E_{*}$, with $\gamma=-1.3^{+0.7}_{-0.4}, \rm{log_{10}(E_{*})}=41.38^{+0.51}_{-0.50}$. The discrepancy between our findings and those of \cite{2023ApJ...944..105S} is mainly due to differences in the chosen values of the spectral index $\alpha$ and the frequency bandwidth $\Delta \nu$ in Equation \ref{eqn:fluence}. 
If we adopt the best-fit value of $\alpha=-1.39$ from \cite{2023ApJ...944..105S}, then our derived value of $\gamma$ would adjust to $\gamma=-1.34$, aligning more closely with the results of \cite{2023ApJ...944..105S}. However, it should be noted that there are still significant uncertainties about the spectral index of FRBs. \cite{2019ApJ...872L..19M} found the mean (median) spectral index is $\alpha=-1.5$ (-1.3) based on 23 FRBs detected by ASKAP. However, the values of $\alpha$ for individual FRBs can vary from -12 to +7.3. 

\begin{figure*}
    \centering
    \includegraphics[width=1.8\columnwidth]{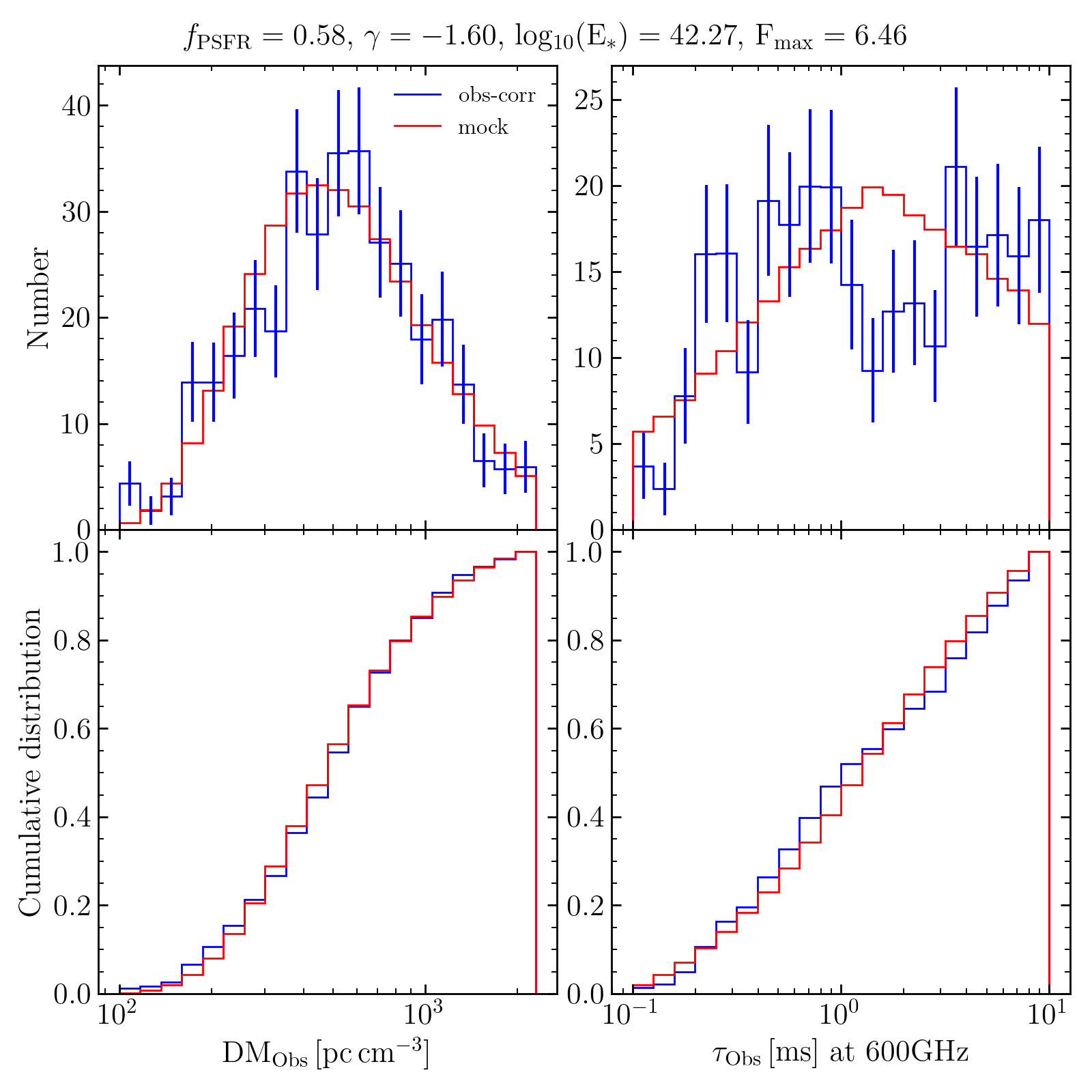}
    \caption{Comparison between the distribution of observed dispersion measure (left) and scattering time at 600 MHz (right) for the selection-corrected FRBs in CHIME/FRB catalog 1 (blue), and our mock FRBs, which is generated with parameter values given by the DM and $\tau$ combined MCMC simulation (red). The error bar in blue histogram shows the 68\% Poissonian confidence interval for each bin.}
    \label{fig:compa mock obs}
\end{figure*}

\subsection{DM and $\tau$ combined}
\label{subsec:dm tau mcmc res}
Furthermore, we utilize MCMC simulations to constrain the FRB population model parameter $f_{\rm{PSFR}}$, the FRB energy parameters $\gamma,\,\rm{log_{10}(E_{*})}$, and the local environment density fluctuation parameter of FRBs, $F_{\rm{max}}$, through a comparison of both the DM distribution and $\tau$ distribution of selection-corrected CHIME/FRB data and our observable mock events. The results are presented in Figure \ref{fig:mcmc res}. The best-fit parameters are estimated to be $f_{\rm{PSFR}}=0.58^{+0.19}_{-0.27},\, \gamma=-1.60^{+0.11}_{-0.13},\,\rm{log_{10}(E_{*})}=42.27^{+1.17}_{-1.18},\, F_{max}=6.46^{+2.47}_{-2.11} $, which are also listed in Table \ref{tab:dmtau mcmc res}. The DM and $\tau$ distribution of the observable mock events with these optimal parameters are shown in Figure \ref{fig:compa mock obs}. Visually, the observable mock events exhibit good agreement with the selection-corrected CHIME/FRB catalog, although there is discrepancy in certain regimes for the $\tau$ distribution. A quantitative evaluation of this comparison will be provided later in this section.

The values of the first three parameters are generally consistent with the results obtained from the DM-only analysis. However, the optimal value of the last parameter, $F_{\rm{max}}$, is relatively large, suggesting a significant impact of the local environment on the scattering within the current model. 
However, a value of $F_{max}=6.46^{+2.47}_{-2.11} $ may appear somewhat higher than the typical fluctuation parameter in the ISM of the MW (\citealt{2022ApJ...931...88C,2022ApJ...934...71O}), but it remains consistent with expectations found in the literature (e.g. \citealt{2020MNRAS.498.4811H,2022ApJ...927...35C}). On the other hand, as discussed in Section \ref{subsubsec:local env}, our default scenario for $\tau_{local}$ and $\tau_{host}$ may underestimate the scattering caused by unresolved small clumps in the ISM and CGM due to the limited resolution of the TNG simulation. To account for this, we could relax the ranges of $\rm{D_{eff}}$ for the local screen, expanding it from 0.01-1 kpc to 0.01-100 kpc. In this case, the predicted value of $F_{max}$ in the current model would decrease significantly to 0.17. In this `Extended-local-scattering' scenario, the local contribution could arise not only from nearby media within the local environment but also from more distant media in the ISM and CGM of host galaxies. We would like to remind the readers that the values of the four parameters we have derived are subject to some limitations. They were obtained by comparing observable mock sources with CHIME/FRB observations and are influenced by the reliability of our assumed properties of mock sources, as well as several observational factors such as the observed frequency band, band-averaged fluences, S/N, sample completeness, etc.

To evaluate the goodness of the best-fit parameters obtained from the MCMC simulation, we have conducted the Kolmogorov-Smirnov (KS) test and the Anderson-Darling (AD) test to compare the distribution of DM and $\tau$ between selection-corrected CHIME/FRB samples and our observable mock FRB samples. We utilize the Python package functions  \texttt{scipy.stats.ks\_2samp} and \texttt{scipy.stats.anderson\_ksamp} for these evaluations. The selection-corrected CHIME/FRB samples are generated using the inverse CDF method, with the corresponding CDF shown as the blue line in the bottom panels of Figure \ref{fig:compa mock obs}. To generate our observable mock FRB samples, we create approximately $2\times10^4$ mock samples based on the optimal parameters derived from MCMC. We then select a subsample of size equal to the selection-corrected CHIME/FRB samples using the bootstrap resampling method. This procedure is repeated 100 times, and the mean p-values derived from the KS and AD tests are summarized in Table \ref{tab:comparison of different models}.

The mean p-values for the KS test and the AD test in the DM distribution are 0.415 and 0.191, respectively, both exceeding the 0.05 threshold. This suggests that we cannot reject the null hypothesis that the two samples originate from the same distribution at the 95\% confidence level. For the $\tau$ distribution, the mean p-value of the KS test is $2\times10^{-4}$ when using the `two-sided’ hypothesis and 0.25 when using `greater/less' hypothesis in the \texttt{scipy.stats.ks\_2samp} function. Although the former result appears less consistent with the visual impression shown in Figure \ref{fig:compa mock obs}, the applicability of the `greater' hypothesis in this context may be somewhat questionable. Therefore, we will default to the results of the 'two-sided’ hypothesis in subsequent analyzes.

A mean p-value of $2\times10^{-4}$ for the $\tau$ distribution suggests that our mock FRB samples are marginally consistent with the selection-corrected CHIME/FRB samples. This value is slightly higher than that of the favored model in \cite{2022ApJ...927...35C}, which has a p-value of $10^{-4}$. However, this comparison is not entirely fair due to differences in the $\tau$ range: 0.1-10 ms in our default model versus 0.1-100 ms in \cite{2022ApJ...927...35C}. 
When we adopt the same $\tau$ range as \cite{2022ApJ...927...35C} (referred to as TNG100-$\tau$100ms model), two optimal parameters (i.e., $f_{\rm{PSFR}}$, $F_{\rm{max}}$) exhibit significant changes, $\gamma$ shows moderate change, while the remaining optimal parameter, $\rm{log_{10}(E_{*})}$, remains relatively unchanged. Additionally, the corresponding mean p-values for both the KS and AD tests, as well as the RMS error of localized FRB redshifts estimates (discussed in the next section), show minor variations. These results are summarized in Table \ref{tab:comparison of different models}. The distributions of DM and $\tau$ for selection-corrected CHIME/FRB samples and our observable mock FRB samples, derived from TNG100-$\tau$100ms model, are presented in the Figure \ref{fig:compa mock obs Lbin}.

For a $\tau$ range of $0.1-100$ ms, the mean p-values of the KS test on the $\tau$ distribution between CHIME/FRB samples and our mock FRB samples are around $4\times 10^{-4}$ for the `two-sided' hypothesis and 0.74 for the `greater/less' hypothesis. The former value is still higher than that of the favored model in \cite{2022ApJ...927...35C}, while the latter appears to be anomalously high. The AD test for the $\tau$ distribution between observations and our model yields a p-value of $10^{-3}$. Since the minimum output of p-value from \texttt{scipy.stats.anderson\_ksamp} is $10^{-3}$, the true p-value could potentially be lower than $10^{-3}$.

In short, it is a challenge to reproduce the distribution of $\tau$ of FRBs well in the CHIME / FRB catalog. However, it is essential to recognize that the observations of events with $\tau > 10$ ms (at 600MHz) and their corresponding selection function still have significant uncertainties \citep[][]{2021ApJS..257...59C}. Therefore, we argue that more robust observations are needed to improve constraints, particularly for events with $\tau > 10$ ms (at 600MHz). In addition, the KS and related tests have several important limitations: they are sensitive to sample size,  strongly influenced by outlying data, and more responsive to deviations near the center of the distribution than at the tails. Given the limited number of observed events and the considerable uncertainty in the selection function for highly scattered events, the KS test for the $\tau$ distribution may have very limited power in evaluating the null hypothesis, especially for $\tau > 10$ ms (at 600MHz). In the following context, we focus on the models that can broadly reproduce the joint distribution of DM and $\tau < 10$ ms.

\begin{figure*}
    \centering
    \includegraphics[width=1.8\columnwidth]{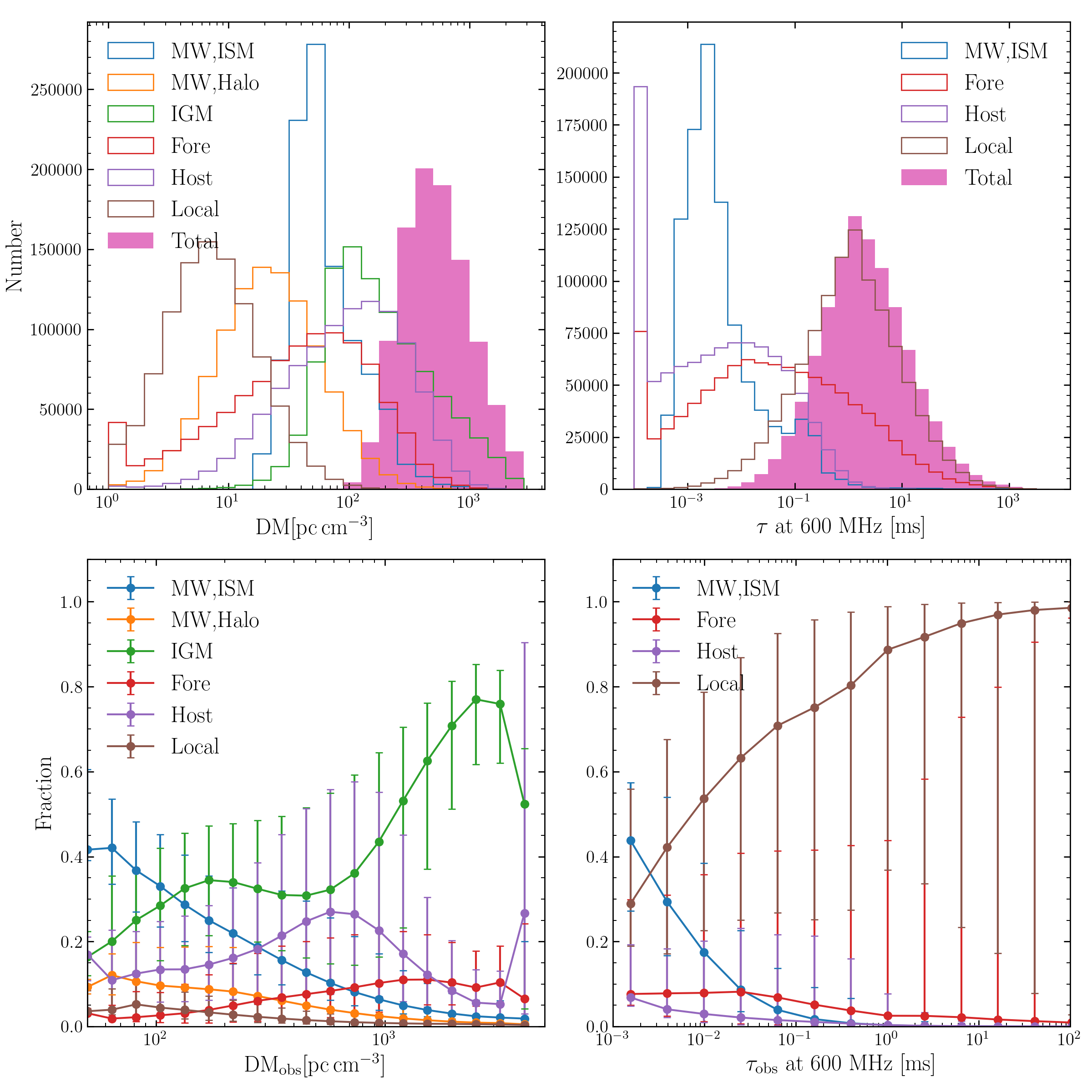}
    \caption{The upper row shows the distributions of dispersion measure (left) and scattering time at 600 MHz (right) of our $\sim10^6$ mock FRBs with fluence > 0.4 Jy ms, which are generated with parameter values given by the DM and $\tau$ combined MCMC simulation. The blue, orange, green, red, and purple histograms are the contribution of MW, IGM, extragalactic foreground halos and galaxies, host halos and galaxies, as well as the local environment, respectively. The brown full-filled histogram indicates the distribution of the total (i.e., observed) DM and scattering time. For the sake of clarity, $\tau<10^{-4}$ ms are set to $10^{-4}$ ms in the right panel . The lower row shows the relative contribution of each component to the total DM and $\tau$ in each bin, with the error bars enclosing the central 68\% of the samples. The meaning of different colored lines is the same as the upper panel.}
    \label{fig:multi component}
\end{figure*}

Based on the optimal parameters, we evaluate the relative contribution of multi-components to the DM and $\tau$ of FRBs in our models. To mitigate fluctuations due to the limited number of samples, we have generated around $10^6$ observed mock FRBs, and present their distribution of DM and $\tau$ in Figure \ref{fig:multi component}. The upper panel illustrates the number distribution of the contribution from each component, including the total DM and $\tau$. Meanwhile, the lower panel shows the relative contributions from various components as a function of the total DM and $\tau$. Comparing our results with \cite{2022ApJ...927...35C}, we find that the contributions from the Milky Way, host galaxy and halo, as well as the local environment, to DM are consistent with the `SFR-Spiral-short GRB-$\rm{DM_{Local}}>10\,pc\,cm^{-3}$' model in \cite{2022ApJ...927...35C}. However, our model indicates a more substantial contribution from the IGM to the DM of FRBs. In addition, the contribution from the extragalactic foreground halos to DM cannot be ignored. For $\tau$, significant contributions from the local environment are required to be consistent with CHIME observations by a high value of $F_{\rm{max}}$. Alternatively, as discussed in Section 2.4, this contribution could arise from unresolved electron clumps in the ISM and CGM of host galaxies. Our results on the primary contributor to $\tau$ are broadly aligned with those of \cite{2022ApJ...927...35C}. Meanwhile, for sources with $\tau (600\,\rm{MHz}) < 0.1$ ms, the diffuse medium in the Milky Way would be responsible for $5\%-45\%$ of $\tau$, a proportion that decreases as the total $\tau$ increases. On average, the electron clumps in foreground galaxies and halos are responsible for $5\%-15\%$ of $\tau$, for events with $0.01\,\rm{ms}<\tau< 3 \, \rm{ms}$. The ISM and CGM of foreground galaxies can be the dominant contributor to $\tau$ for a small fraction of the events.  

For certain events in the CHIME/FRB catalog, only upper limit values on $\tau$ are available. When comparing the distribution of scattering time, we have handled these upper limits by randomly selecting values from 0.001 to 1 times the upper limit as the `true' value. We have explored two other ways to deal with these upper-limit data. One approach considers the upper limits to be the true values, while the other excludes events with only upper limits of $\tau$. We find that the fluctuation parameters of the local environment, $F_{\rm{max}}$, increase further for these two scenarios. In addition, we examined the impact of the number of bins used when comparing the distribution of DM and $\tau$. We notice minor differences if the number of bins decreases from 20 to 15 or 10. 

\section{redshift estimator}
So far, reliable estimates of redshift are lacking for most FRBs. Based on our models of the contributions from various components to the DM and $\tau$ of FRBs, along with the optimal parameter values obtained from the MCMC simulations, we examine methods that utilize DM and $\tau$ to estimate the redshifts of FRBs.
Specifically, we explore the relationships among DM, $\tau$ and redshifts for approximately $10^6$ mock FRBs with a fluence > 0.4 Jy ms. These mock sources are selected with this fluence threshold from a raw sample numbered around $10^8$, generated using the model and procedures described in Section 2. These relationships can serve as redshift estimators when provided with the DM and $\tau$ of an FRB event. Furthermore, we apply these FRB redshift estimators to the currently localized FRB sources to assess their reliability. This assessment serves as a preliminary evaluation of whether these estimators are suitable for application to unlocalized FRBs. The main properties of the 71 localized FRBs are collected from the literature and presented in Table \ref{tab:localized FRB}. It is worth noting that the observed $\tau$, $\rm{\tau_{obs}}$, are scaled to 600 MHz from 
the referenced frequency in the literature using an index $x_\tau=4$ in $\tau(\nu) \propto \nu^{-x_{\tau}}$ (see relevant introduction in Section 2.4). 

\begin{figure}
    \centering
    \includegraphics[width=0.8\columnwidth]{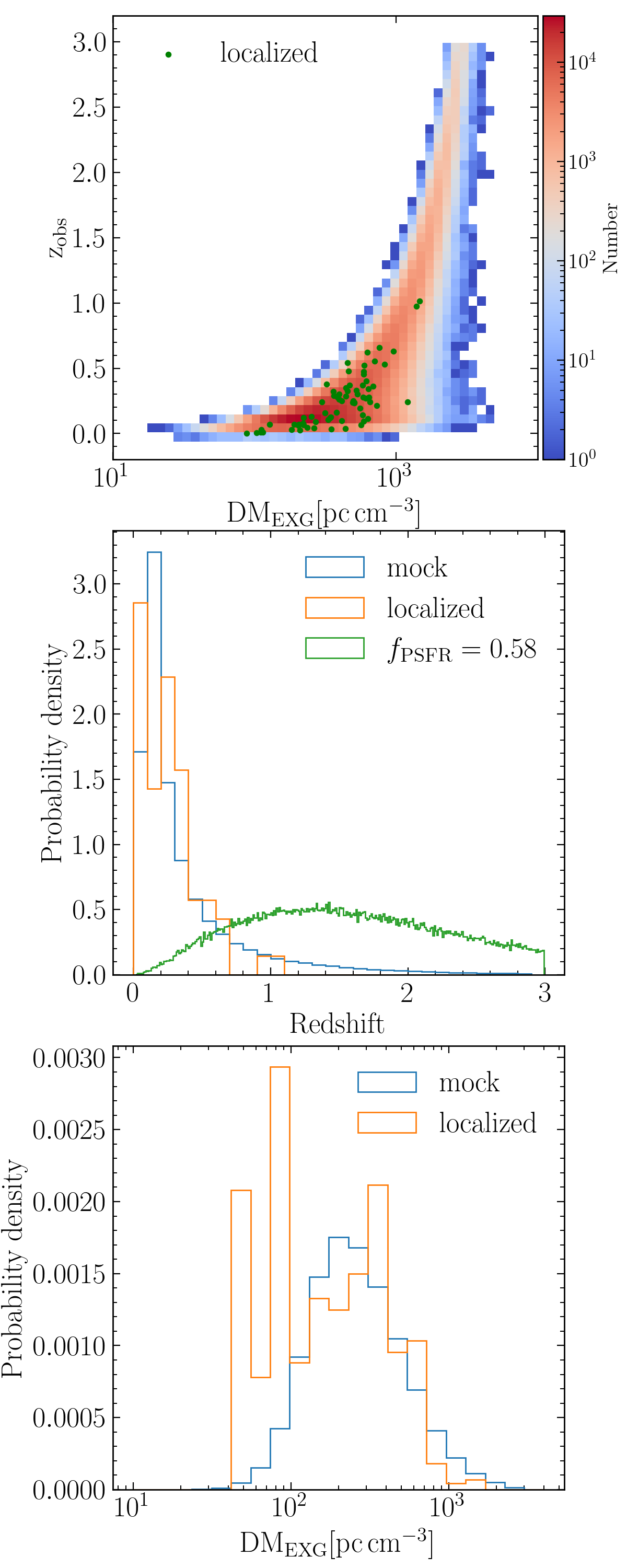}
    \caption{The top panel shows the distribution of the redshift and total dispersion measure of our $\sim 10^6$ mock FRBs with fluence above 0.4 Jy ms, superposed by 71 localized FRBs. The middle (bottom) panel shows the probability distribution of redshift (dispersion measure) of our mock FRBs with fluence above 0.4 Jy ms (blue) and localized FRBs (orange). The green histogram indicates the intrinsic redshift distribution of mock FRBs, i.e., the redshift distribution of mock FRBs in our model before applying a fluence limit > 0.4 Jy ms.}
    \label{fig:mock dm hist2D}
\end{figure}

Firstly, we explore the the extragalactic DM ($\rm{DM_{EXG}}$)-redshift distribution, as $\rm{DM_{EXG}}$ is often used independently to infer the redshift of FRBs. Here, $\rm{DM_{EXG}}$ is defined as $\rm{DM_{obs} - DM_{MW,ISM}}$, where $\rm{DM_{MW,ISM}}$ is estimated using the NE20001 model. We also have tested the definition of $\rm{DM_{EXG}}\equiv \rm{DM_{obs} - DM_{MW,ISM} - DM_{MW,Halo}}$, which leads to a mild difference in the RMS error of the redshift estimates by the DM-only or combined DM-$\tau$ method. Figure \ref{fig:mock dm hist2D} illustrates the frequency distribution of mock FRBs with fluence > 0.4 [Jy ms] in the $\rm{DM_{EXG}}$-redshift space. We find that 84\% of the mock FRBs with fluence > 0.4 Jy ms exhibit $\rm{DM_{EXG}}$ values ranging from $100-1000 \, \rm{pc\,cm^{-3}}$, and 83\% have redshifts between 0.0 and 0.6. These fractions are consistent with the corresponding fractions for the 71 localized FRBs, which are 83\%, 92\%, respectively. The lower two panels display the probability distribution of $\rm{DM_{EXG}}$ and the redshift for both mock and localized FRBs. Although we select observable mock FRBs based on the criteria of CHIME/FRB observations, the $\rm{DM_{EXG}}$ and redshift distributions for our mock FRBs generally match those of the localized FRBs detected by various telescopes. This consistency in the $\rm{DM_{EXG}}$-z distribution suggests that $\rm{DM_{EXG}}$ can serve as a reasonable redshift estimator. 

\begin{figure}
    \centering
    \includegraphics[width=1.\columnwidth]{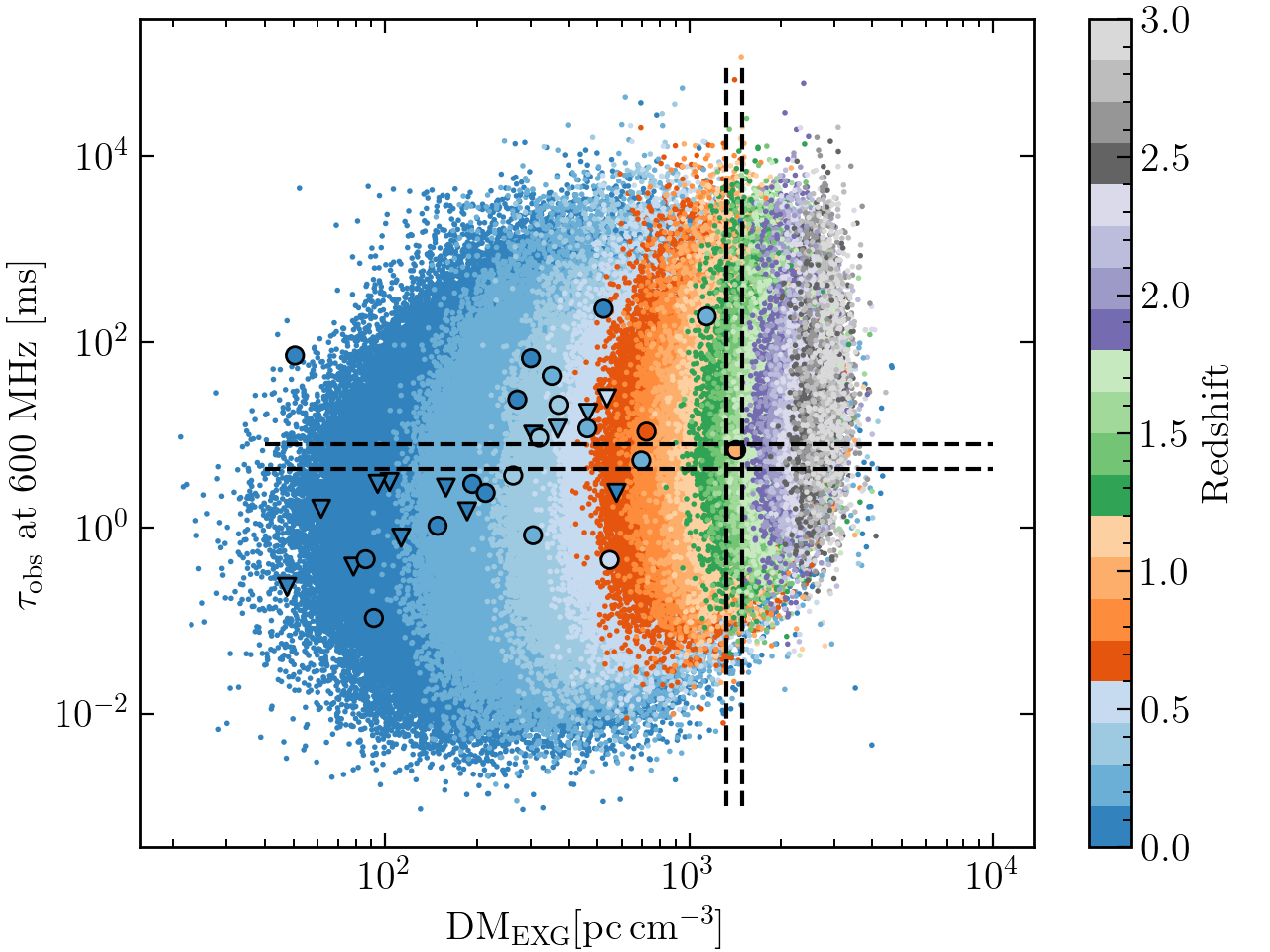}
    \caption{The distribution of the observed scattering time at 600 MHz and total dispersion measure of our $\sim 10^6$ mock FRBs with fluence > 0.4 Jy ms (dots), and 33 localized FRBs (solid circles and downward triangles represent sources with a scattering time of true value and upper limits, respectively). Redshift is indicated by the color. The vertical and horizontal dashed lines illustrate one of the bins of DM and $\tau$ that we employed to select mock FRBs to estimate the redshift of a FRB with given $\rm{DM_{obs}}$ and $\tau_{\rm{obs}}$. FRB 220610A is used as an example.}
    \label{fig:mock dm tau z}
\end{figure}

\begin{figure*}
    \centering
    \includegraphics[width=1.4\columnwidth]{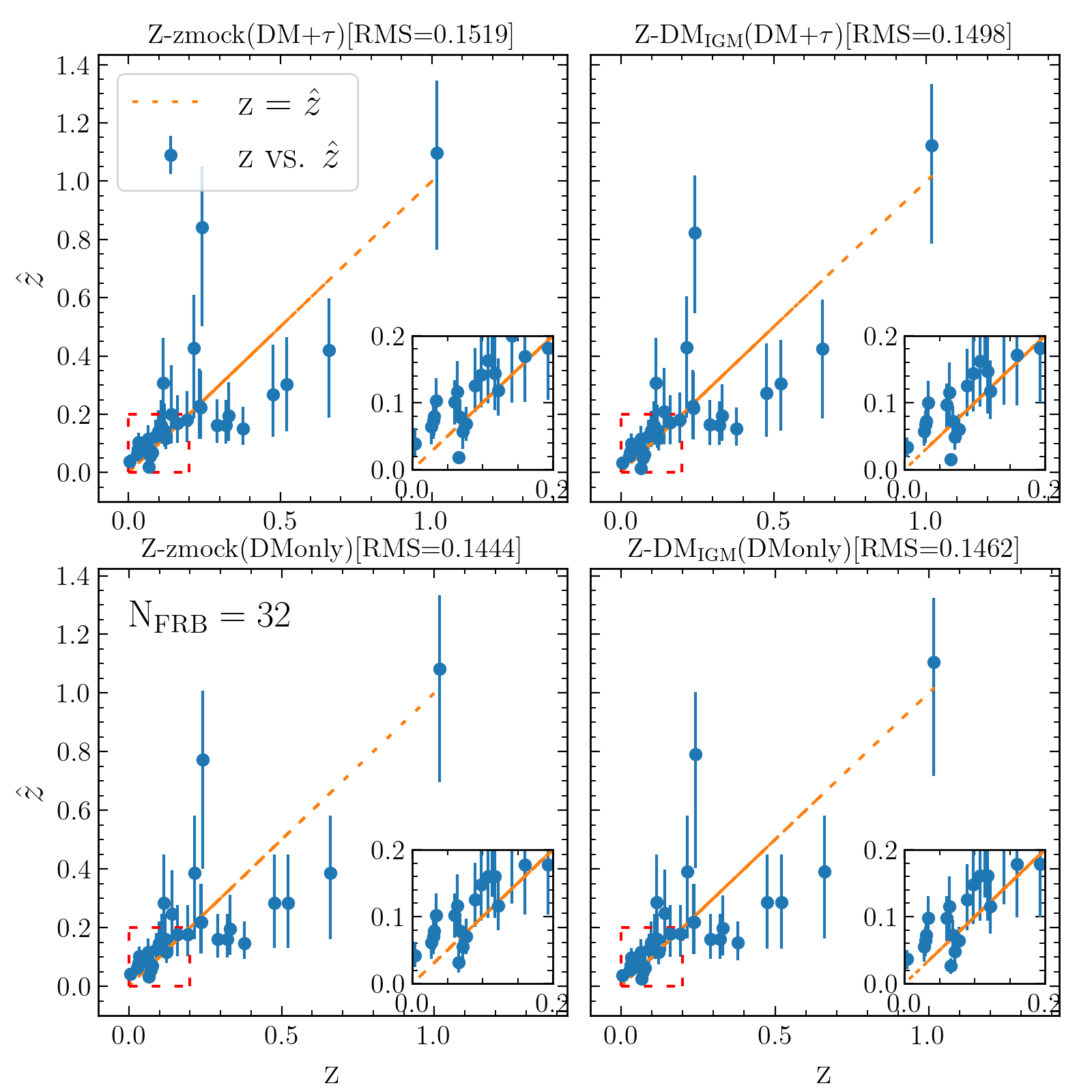}
    \caption{Redshift estimated by our procedures, $\hat{z}$ versus the true redshift, z, for 32 localized FRBs, of which both their DM and scattering time are available. The upper (lower) shows the results of the redshift estimator with combined DM and $\tau$ (DM-only) information. The left column indicates the method that estimates the redshift directly from the redshift of mock FRBs. The right column shows the results using an indirect method, through the $\rm{DM_{IGM}}$ of mock FRBs and Equation \ref{eqn:DMigm}. See Section 4 for more details. The insert in the bottom-right corner of each panel highlights the redshift range $0-0.2$.}
    \label{fig:FRB z estimator}
\end{figure*}

However, since we have not simulated the observed signal-to-noise ratio and central frequency, and the number of localized FRB events is limited, further examination is needed to confirm the agreement between our mock samples and actual observations, especially with an increased number of localized events in the future. Figure \ref{fig:mock dm hist2D} also displays the intrinsic redshift distribution of the raw sample of mock FRBs with $f_{\rm{PSFR}}=0.58$ , illustrating the effect of the observational selection criteria we employed, i.e., fluence > 0.4 Jy ms.

Furthermore, we investigate the distribution of mock events in the $\rm{DM_{EXG}}$-$\tau_{\rm{obs}}$ space, as shown in Figure \ref{fig:mock dm tau z}. Each dot represents a mock event and is color-coded according to its redshift. The 33 localized events with reported values of both DM and $\tau$ are also displayed as solid circles and downward triangles, color-coded by redshift. Visual inspection suggests that most of the localized events share a redshift comparable to neighboring mock events in the $\rm{DM_{EXG}}$-$\tau$ space, with a few exceptions.

Now, we introduce the specific procedures for constructing redshift estimators based on the $\rm{DM_{EXG}}$ and $\tau$ of mock events. We have developed four estimators, with the first two combining $\rm{DM_{EXG}}$ and $\tau$, while the remaining two rely solely on $\rm{DM_{EXG}}$. Let us start with the first two estimators. We divide the range of $\rm{DM_{EXG}}$ from 10 to 3000 $\rm{pc\,cm^{-3}}$ into 50 uniform bins in the logarithmic space and the range of $\tau$ from 0.01 to 1000 ms into 20 uniform bins in the logarithmic space. The intervals in both DM and $\tau$ are comparable to the typical observational uncertainties on $\rm{DM_{EXG}}$ and $\tau$. Subsequently, we group mock events that fall into each bin in the $\rm{DM_{EXG}}$-$\tau$ space as ($\rm{DM_{EXG,i}}$,$\tau_j$) where i=1,2,..50, j=1,2,...20. The redshift and $\rm{DM_{IGM}}$ distribution of mock events in each bin ($\rm{DM_{EXG,i}}$,$\tau_j$) are compiled to determine the median redshift and $\rm{DM_{IGM}}$, denoted as $z_{50}(\rm{DM_{EXG,i}},\tau_j)$,  $\rm{DM_{IGM,50}}(\rm{DM_{EXG,i}},\tau_j)$, respectively. 

The first estimator, named as `Z-zmock(DM+$\tau$)', directly assigns $z_{50}(\rm{DM_{EXG,i}},\tau_j)$ as the estimated redshift of an observed FRB event with a specified $\rm{DM_{EXG}}$ and $\tau$ that falls into the bin ($\rm{DM_{EXG,i}}$,$\tau_j$). The second estimator, named `Z-$\rm{DM_{IGM}}$(DM+$\tau$)', is an indirect method. That is, we use $\rm{DM_{IGM,50}}(\rm{DM_{EXG,i}},\tau_j)$ to inversely estimate redshift according to equation \ref{eqn:DMigm}, which is related to the free electron fraction $\chi(z)$, the baryon fraction in IGM, $f_{b,\rm{IGM}}$, and cosmological parameters. Figure \ref{fig:FRB z estimator} illustrates the results of these two estimators.  

For the other two estimators, we only use $\rm{DM_{EXG}}$ of mock FRBs to estimate redshift. We deduce the median redshift and $\rm{DM_{IGM}}$ of mock events falling into the bin ($\rm{DM_{EXG,i}}$), and denote them as $z_{50}(\rm{DM_{EXG,i}})$ and $\rm{DM_{IGM,50}}(\rm{DM_{EXG,i}})$, respectively. The third estimator, named `Z-zmock(DMonly)', assigns $z_{50}(\rm{DM_{EXG,i}})$ as the redshift of observed events falling into the bin ($\rm{DM_{EXG,i}}$). 
The fourth estimator is similar to the second estimator but uses $\rm{DM_{IGM,50}}(\rm{DM_{EXG,i}})$, and is named as `Z-$\rm{DM_{IGM}}$(DMonly).'

To evaluate the performance of these four redshift estimators, we apply them to all 32 localized FRBs with reported DM and $\tau$, excluding FRB200120E due to its observed redshift being less than zero. Figure \ref{fig:FRB z estimator} illustrates the results, comparing the predicted redshifts by these four estimators with the true redshifts of host galaxies reported in the literature. Overall, the redshifts predicted by all four estimators are generally consistent with the reported redshift of hosts, except for FRB190520B. This event has a large DM (1202 $\rm{pc\,cm^{-3}}$) but a relatively small redshift ($z=0.241$). The goodness of fit, indicated by the RMS residual of $\langle(\hat{z}-z)^2 \rangle^{1/2}$, is around $0.14-0.15$ for the four redshift estimators. If the event FRB190520B is further excluded, the RMS error for these estimators will decrease to around 0.11, which is approximately $20\%$ higher than the combined DM-$\tau$ method in \cite{2022ApJ...931...88C} (applied to 9 events). When our estimators are applied to those 25 events with secure host localizations, i.e., those with $\rm{P_{cc}}<0.05$ or $\rm{P_{host}}>0.95$, the RMS error is around $0.15-0.16$. If FRB190520B is excluded, the RMS error is $0.11$. 

There are notable errors in our estimator for several FRBs within the range of $0.3<z<0.7$, as shown in Figure \ref{fig:FRB z estimator}. This may be partly caused by the following reason. Our model has taken into account the contribution to DM from extragalactic foreground galaxies and their halos. For events where the L.O.S. passes through a smaller-than-expected amount of foreground medium, the redshift estimated by our model would be lower than the true redshift. For all of our estimators, the inclusion of $\tau$ has a minor impact on the accuracy, which differs from the results of \cite{2022ApJ...931...88C}. We will discuss this discrepancy in Section 5.3.

\begin{figure}
\centering
    \includegraphics[width=1.0\columnwidth]{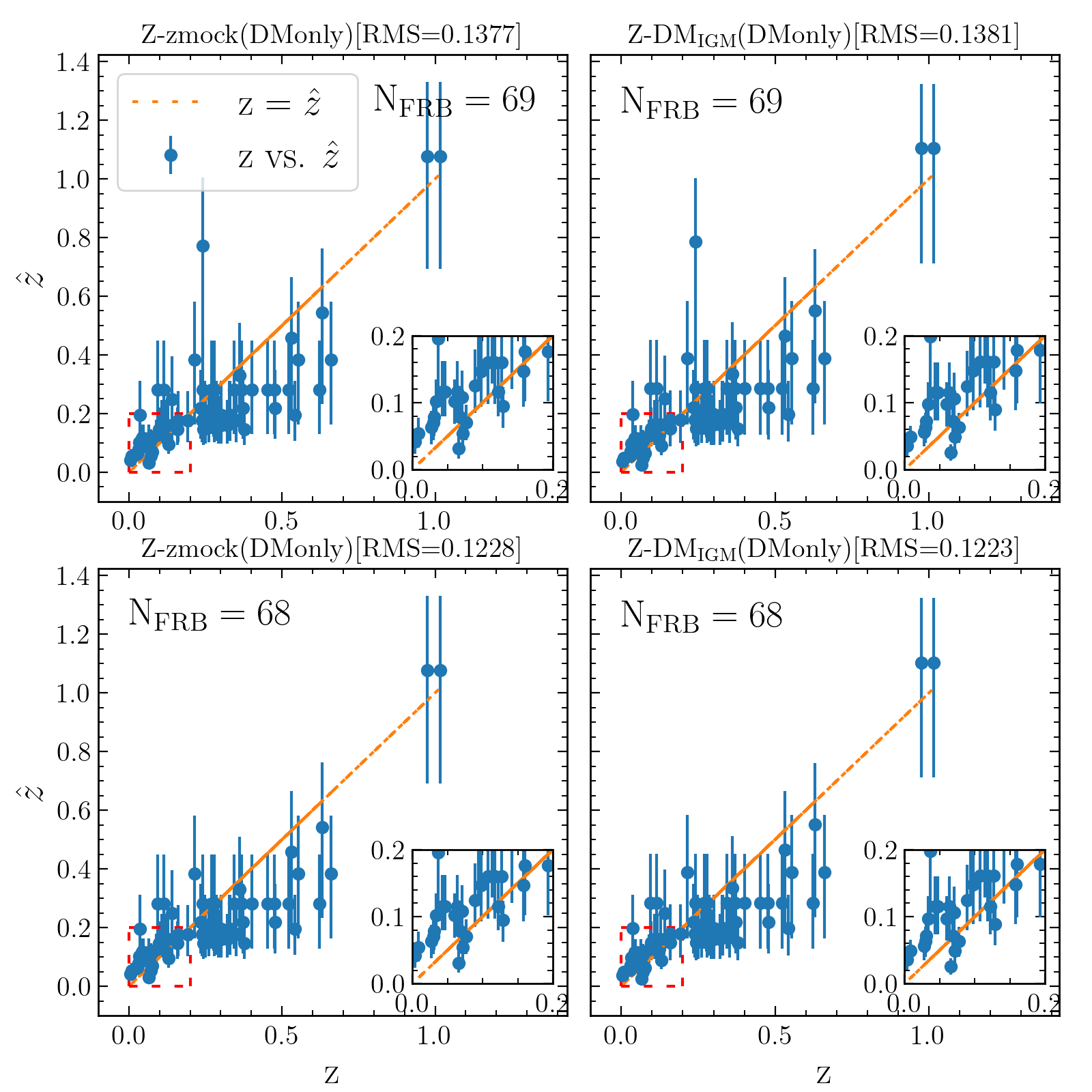}
    \caption{Similar to Figure \ref{fig:FRB z estimator}, but for 'Z-zmock(DMonly)' (left) and 'Z-$\rm{DM_{IGM}}$(DMonly)' (right) method. Upper: 69 FRB, excluding FRB200120E and FRB220319D. Lower: 68 FRB, excluding FRB200120E, FRB220319D and FRB190520B.}
    \label{fig:dmonly estimator}
\end{figure}

Since our DM-only and combined DM-$\tau$ redshift estimators show minor differences, we adopt the DM-only method to estimate the redshift of the 69 localized FRBs. FRB220319D is excluded because its $\rm{DM_{MW,ISM}}$ exceeds $\rm{DM_{obs}}$, and FRB200120E is excluded due to its observed negative redshift. The corresponding RMS values are around 0.14, which is slightly lower than those for the 32 FRBs with reported DM and $\tau$. Furthermore, when excluding the extreme event FRB190520B, the RMS error of our DM-only estimator for the remaining 68 events is around 0.12, which is comparable to that for the 31 events with both DM and $\tau$. These results are shown in Figure \ref{fig:dmonly estimator} and summarized in Table \ref{tab:comparison of different models}. 

\section{Discussion}

\begin{figure*}
    \centering
    \includegraphics[width=0.8\textwidth]{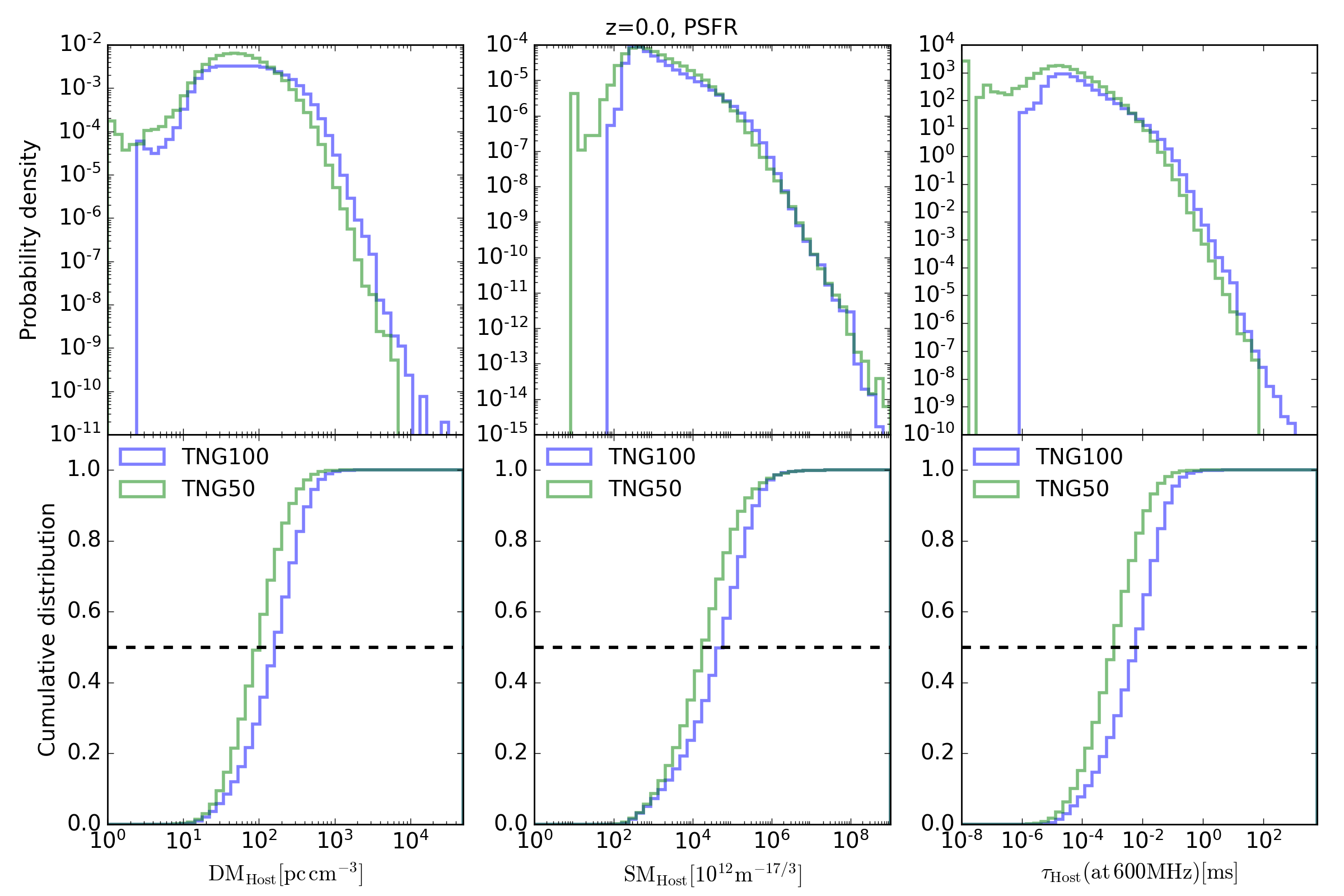}
    \caption{The probability (upper) and cumulative (lower) distribution of $\rm{DM_{Host}}$ (left), $\rm{SM_{Host}}$ (middle), $\rm{\tau_{Host}}$ (right)  based on TNG100 (blue) and TNG50 (green) simulation at z=0 for PSFR source population model.}
    \label{fig:compare dmtau of simu}
\end{figure*}

\subsection{Effect of simulation resolution}
Our results may be influenced by the limited resolution in the TNG100 simulation. To assess the potential impact of this resolution constraint, we perform the same analysis using the IllustrisTNG 50-1 (TNG50) simulation, which has a resolution approximately twice higher than that of TNG100. We find that there are moderate differences in the distribution of DM and the SM between the two simulations at redshifts below 0.5 The distributions of $\rm{DM_{Host}, \, SM_{Host}, \, \tau_{Host}}$ for the PSFR model at $z=0$ are shown in Figure \ref{fig:compare dmtau of simu}. The corresponding median values are $\rm{182\,pc\,cm^{-3},\, 6\times10^4\times10^{12}m^{-17/3},\, 0.007 }$ ms at 600 MHz for the TNG100 simulation, respectively, and $\rm{104\,pc\,cm^{-3},\, 2.3\times10^4\times10^{12}m^{-17/3},\, 0.001 }$ ms at 600 MHz for the TNG50 simulation, respectively.

The difference in the median value of the CDF of $\rm{DM_{Host}}$ ($\rm{\tau_{Host}}$) between the simulations of TNG50 and TNG100 of the PSFR model at low redshifts can moderately affect the best fit parameter of $\rm{f_{PSFR}}$ and $\gamma$. When using the same prior ranges as in Section 3, the best-fitting parameters are $f_{\rm{PSFR}}=0.33^{+0.25}_{-0.21}$, $\rm{\gamma}=-1.27^{+0.07}_{-0.09}$, $\rm{log_{10}(E_{*})}=41.93^{+1.38}_{-1.20}$ and $\rm{F_{max}}=12.30^{+3.98}_{-3.27}$, as listed in Table \ref{tab:comparison of different models}. However, there is a degeneracy between $f_{\rm{PSFR}}$ and $\gamma$. To extract the potential effect of resolution, we set a uniform log-prior range for $\gamma$ between -1.7 and -1.6 (labeled as TNG50-corr in Table \ref{tab:comparison of different models}) to match the best fit $\gamma$ from the TNG100 simulation. As a result, the best fit of $f_{\rm{PSFR}}$ for the TNG50-corr model becomes $0.85^{+0.04}_{-0.27}$, while $\rm{log_{10}(E_{*})}$ and $\rm{F_{max}}$ decrease slightly. This is in line with our expectations, since the median $\rm{DM_{Host}}$ and $\rm{\tau_{Host}}$ for the PSFR model of the TNG50 simulation are smaller than those of the TNG100 simulation, requiring a higher $f_{\rm{PSFR}}$ to compensate. 

Based on the best-fitting parameters of the TNG50-corr model, the results of the KS and AD tests for the DM and $\tau$ distribution, as well as the RMS error of the redshift estimates for 32 localized FRBs, are moderately worse compared to the results of TNG100, except for the RMS error of the redshift estimates for 9 FRBs used in \cite{2022ApJ...931...88C}. Our model, using optimal parameters based on the TNG50 simulation, yields an RMS error of approximately 0.1 for these 9 FRBs, which is comparable to the results of the combined DM-$\tau$ method in \cite{2022ApJ...931...88C}. However, as the number of FRBs increases, the accuracy of the redshift estimator based on the TNG50 simulation declines moderately. The corresponding results are also listed in Table \ref{tab:comparison of different models}.

\renewcommand{\arraystretch}{1.2} 
\begin{table*}
	\centering
	\caption{Table of parameter fit results for the combined DM and $\tau$ distribution, along with the corresponding KS and AD test results, and RMS error of the redshift estimator applied to localized FRB . For the RMS results, different FRB numbers (69,68,32,31,25,24,9) represent the following samples. 69 FRB indicates all the 71 events in Table \ref{tab:localized FRB}, but excluding FRB200120E and FRB220319D; 68 FRB indicates all the 71 events in Table \ref{tab:localized FRB}, but excluding FRB200120E, FRB220319D and FRB190520B; 32 FRB indicates all 33 events with both reported DM and $\tau$, but excluding FRB200120E; 31 FRB indicates all 33 events with both reported DM and $\tau$, but excluding FRB200120E and FRB190520B; 25 FRB indicates events with both reported DM and $\tau$, and also satisfy $\rm{P_{cc}}<0.05$ or $\rm{P_{host}}>0.95$; 24 FRB indicates events with both reported DM and $\tau$ (excluding FRB190520B), and also satisfy $\rm{P_{cc}}<0.05$ or $\rm{P_{host}}>0.95$; and 9 FRB samples corresponding to those used in \cite{2022ApJ...931...88C}. }
	\label{tab:comparison of different models}
	\begin{tabular}{cccccccc} 
		\hline
            \hline
		     & & & TNG100 & TNG100-$\tau$100ms & TNG50 & TNG50-corr & Ext-local  \\
		\hline
        & & $f_{\rm{PSFR}}$ & $0.58^{+0.16}_{-0.27}$ & $0.86^{+0.03}_{-0.04}$ & $0.33^{+0.25}_{-0.21}$ & $0.85^{+0.04}_{-0.27}$ & $0.45^{+0.22}_{-0.26}$  \\
            & & $\gamma$ & $-1.60^{+0.11}_{-0.13}$ & $-1.79^{+0.08}_{-0.07}$ & $-1.27^{+0.07}_{-0.09}$ & $-1.62^{+0.14}_{-0.04}$ & $-1.56^{+0.09}_{-0.11}$  \\
            & & $\rm{log_{10}(E_{*})}$ & $42.27^{+1.17}_{-1.18}$ & $42.44^{+1.06}_{-1.07}$ & $41.93^{+1.38}_{-1.20}$ & $41.93^{+1.29}_{-0.81}$ & $42.25^{+1.19}_{-1.17}$ \\
            & & $\rm{F_{max}}$ & $6.46^{+2.47}_{-2.11}$ & $30.67^{+6.66}_{-5.25}$ & $12.30^{+3.98}_{-3.27}$ & $5.81^{+3.98}_{-2.19}$ & 2 \\
            & & $\rm{D_{eff,max}}$ & 1 & 1 & 1 & 1 & $3.40^{+0.92}_{-0.77}$  \\
           \hline
            & \multirow{2}{*}{DM} & KS Test p-value & 0.415 & 0.456 & 0.415 & 0.215 & 0.441 \\
            & & AD Test p-value & 0.191 & 0.189 & 0.188 & 0.118 & 0.199 \\
            \hline
            & \multirow{2}{*}{$\tau$} & KS Test p-value & $2\times 10^{-4}$ & $4\times10^{-4}$ & $8\times 10^{-5}$ & $3\times10^{-5}$ & $6\times10^{-4}$  \\
            & & AD Test p-value & $10^{-3}$ & $10^{-3}$ & $10^{-3}$ & $10^{-3}$ & $10^{-3}$  \\
            \hline
            \multirow{12}{*}{\makecell{RMS error of \\redshift estimator\\ (Z$-$zmock)}} & 69 FRB & DMonly & 0.1377 & 0.1494 & 0.1710 & 0.1660 & 0.1352 \\
            & 68 FRB & DMonly & 0.1228 & 0.1371 & 0.1420 & 0.1355 & 0.1199 \\
            & \multirow{2}{*}{32 FRB} & DM+$\tau$ & 0.1519 & 0.1500 & 0.1979 & 0.1923 & 0.1516 \\
            & & DMonly & 0.1444 & 0.1490 & 0.1991 & 0.1933 & 0.1439 \\
            & \multirow{2}{*}{31 FRB} & DM+$\tau$ & 0.1104 & 0.1184 & 0.1402 & 0.1345 & 0.1084 \\
            &  & DMonly & 0.1112 & 0.1200 & 0.1415 & 0.1329 & 0.1097 \\
            & \multirow{2}{*}{25 FRB} & DM+$\tau$ & 0.1632 & 0.1572 & 0.2094 & 0.2026 & 0.1638 \\
            & & DMonly & 0.1523 & 0.1544 & 0.2113 & 0.2039 & 0.1525 \\
            & \multirow{2}{*}{24 FRB} & DM+$\tau$ & 0.1140 & 0.1181 & 0.1377 & 0.1258 & 0.1113 \\
            & & DMonly & 0.1111 & 0.1172 & 0.1395 & 0.1274 & 0.1106 \\
            & \multirow{2}{*}{9 FRB} & DM+$\tau$ & 0.1477 & 0.1726 & 0.0964 & 0.1011 & 0.1424 \\
            & & DMonly & 0.1537 & 0.1757 & 0.1007 & 0.1057 & 0.1488 \\
            \hline
            
	\end{tabular}
\end{table*}

However, it is important to note that the changes reported above are not solely a result of resolution differences. The sub-grid physics and associated parameters in TNG100 have been calibrated to match several key observations, including the galaxy stellar mass function at $z=0$, the cosmic star formation rate density, and the stellar mass-to-halo mass relation (\citealt{2018MNRAS.475..648P}). Other simulations within the TNG suite, such as TNG50, use the same galaxy formation model and related parameters as TNG100. As a result, the galaxy stellar mass function and star formation efficiency in TNG50 are higher than those in TNG100 and observations. This discrepancy leads to differences in the ionized gas fraction for halos of a given mass between the two simulations. Additionally, the smaller volume of TNG50 results in a lower number of more massive galaxies. Therefore, the changes in the best-fitted $\rm{f_{PSFR}}$ reported above arise from a combination of these factors and should be considered as an upper limit on the impact of increased resolution. We expect that a simulation with enhanced resolution, while still matching the key properties of galaxy populations, would lead to a moderate increase in $f_{\rm{PSFR}}$, with only minor changes to the other three parameters in our models.

\subsection{Extended-local-scattering scenario}
In Section 3, we adopted a uniform distribution for the effective distance of the local screen, $D_{\rm{eff}}$. In the default scenario, the range of $D_{\rm{eff}}$ is set to 0.01-1.0 kpc, which would require the circumburst media to be more turbulent than the ISM.  Alternatively, we show that if the range of $D_{\rm{eff}}$ is extended to 0.01-100.0 kpc, the optimal value of the density fluctuation parameter would decrease to 0.17, which is consistent with the value used for the ISM of disc galaxies in \citealt{2022ApJ...931...88C}. In this case, $\tau$ could be attributed to either the local medium or the ISM and CGM of the host galaxy. 

We further evaluate this Extended-local-scattering scenario by treating $D_{\rm{eff}}$ of the local environment as a free parameter in the MCMC simulations, while the fluctuation parameter F is assumed to follow a uniform distribution between 0.5 and 2, consistent with \cite{2022ApJ...931...88C}. This scenario is labeled as Ext-local in Table \ref{tab:comparison of different models}. The best-fit value for $D_{\rm{eff}}$ in this scenario is $3.40^{+0.92}_{-0.77}$ kpc, which is close to the typical distance from the ISM in the host galaxy to the FRB event. The other three parameters are similar to those in our default scenario, where the circumburst media within $\sim1$ kpc dominate $\tau$. The KS test of $\tau$ for this Ext-local scenario yields a p-value of $6\times 10^{-4}$, slightly higher than that for our default scenario. 

However, the redshift estimator based on this extended local scattering model achieves accuracy comparable to that of our default scenario. When FRB190520B is excluded (included), the RMS error is approximately 0.11 (0.15). Moreover, as shown in Table \ref{tab:comparison of different models}, including $\tau$ does not result in a reduction in the RMS error.

\subsection{Comparison with previous works}
 \cite{2022ApJ...927...35C} employed a largely theoretical framework incorporating contributions from multi-components to the DM and $\tau$, along with various models describing FRB source populations and host galaxies to simulate the DM and $\tau$ of FRBs. Through Monte Carlo simulations, they found that one of their nines source population models could marginally align with the CHIME/FRB catalog regarding the joint distribution of DM and scattering under the condition that the CGM of intervening foreground galaxies or the local environmental medium could substantially contribute to the scattering of certain FRBs. In contrast to their work, our study employs several distinct approaches and updates. First, we characterize the electron density in different types of host galaxies using state-of-the-art cosmological hydrodynamical simulations of galaxy formation, specifically the TNG100 simulation. Second, we put mock events in host galaxies according to the distribution of stellar mass or newly born stars that resolved in simulations. Third, we adjust the contributions from the IGM and intervening halos in our models to account for variations in the fraction of ionized baryons in both the IGM and the CGM of foreground halos as redshift evolves.

With these updates and alternative methods, our optimal model shows a slightly better agreement with the joint distribution of DM and $\tau$ of the CHIME/FRB catalog, as indicated by the KS test. However, our model still faces challenges in accurately reproducing the distribution of $\tau$ in the catalog. To improve the constraints and further refine our models, more robust observations of events with relatively large $\tau$ are needed. 

On the other hand, our DM-$\tau$ combined redshift estimator does not achieve substantial improvement compared to the DM alone estimator. In contrast, \cite{2022ApJ...931...88C} apply their DM-$\tau$ combined redshift estimator to 9 FRBs, and obtain a bias smaller by a factor of 4 to 10 than the DM-only estimator, along with a 40\% smaller RMS error. The estimated redshifts for the same nine FRB sources using both methods in our work and in \cite{2022ApJ...931...88C} are listed in Table \ref{tab:comparison with codes}. Additionally, to provide a more comprehensive comparison with the results of \cite{2022ApJ...931...88C}, the estimated values of $\rm{DM_h}=\rm{DM_{Host}+\rm{DM_{Local}}}$ are also included in Table \ref{tab:comparison with codes}. Several factors could potentially explain the discrepancy in the performance of our estimators and \cite{2022ApJ...931...88C}. 

Firstly, our models account for the scattering caused by the local media, ISM and CGM of the host galaxies in a statistical way and may be influenced by the limited resolution of the simulations. Future advanced simulations will be necessary to test the convergence and robustness of our models. In contrast, the method in \cite{2022ApJ...931...88C} involves a more detailed event-by-event analysis of $\tau$. Secondly, in our optimal model, for events with a considerable $\tau$, the dominant contribution to $\tau$ comes from the local media. However, the corresponding $\rm{DM_{Local}}$  is relatively small, typically on the order of $\sim 10\,\rm{ pc\, cm^{-3}}$. Consequently, $\rm{DM_{Local}}$ constitutes a very small fraction of the total DM for most FRB events, which generally have $\rm{DM}_{EXG}> 100\, \rm{pc\, cm^{-3}}$. Alternatively, the components $\rm{DM_{Local}}$ and $\rm{\tau_{Local}}$ in our model may be replaced by contributions from electron clumps in the ISM and CGM of the host, foreground, and the Milky Way. In either case, the DM component associated with the dominant portion of $\tau$ is relatively small. Therefore, the relative contribution and associated scatter from $\rm{DM_{Host}}$ to the total DM cannot be constrained, representing the largest uncertainty in the $\rm{DM}-z$ relation. However, sources with stronger scattering tend to have larger $\rm{DM_{Host}}$ in \cite{2022ApJ...931...88C}, potentially leading to a more precise redshift estimation for certain events by including the information of $\tau$. 

Nevertheless, our redshift estimators can generally provide reasonable estimates for relatively large samples and can be applied to events without $\tau$. For sample sizes greater than $20-30$, excluding extreme cases such as FRB 20190520B which has a high $\rm{DM_{obs}}=1202\, pc\,cm^{-3}$ but a low redshift ($\rm{z_{obs}=0.241}$), our estimators achieve an RMS error of $0.11-0.12$. This performance is approximately $20\%$ smaller and $20\%$ larger than the DM-only and DM–$\tau$ combined estimators in \cite{2022ApJ...931...88C}, respectively. Meanwhile, further verification is needed to determine whether the improvement provided by the $\rm{DM}-\tau$ combined method in \cite{2022ApJ...931...88C} will hold for large samples or not. Last but not least, the number of localized events is limited, and there are still some uncertainties on the galaxies associated to those localized FRBs, due to the limited precision of events and magnitude of galaxy catalog used. More localized events with high confidence levels in the future would give a more robust evaluation of these redshift estimators.

\subsection{Properties of host galaxies as constraint}

\begin{figure*}
\centering
    \includegraphics[width=2.\columnwidth]{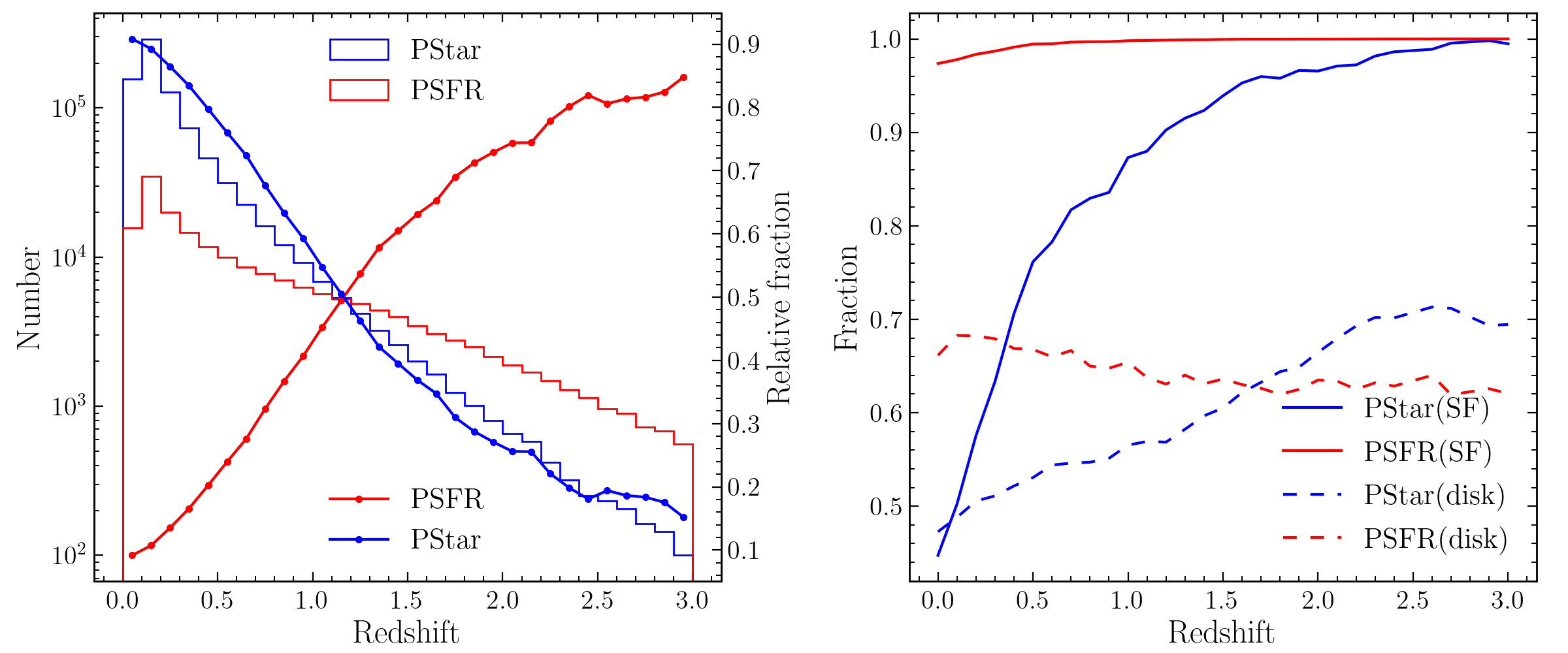}
    \caption{The left panel shows the redshift distribution of our $\sim 10^6$ mock FRB with fluence larger than 0.4 Jy ms, which are divided further into two populations. One is PSFR, i.e., tracing the cosmic star formation rate (red histogram), and the other is PStar, i.e., tracing the stellar mass (blue histogram). The red (blue) dotted line indicates the relative fraction, indicated by the right vertical axis, of mock FRBs that trace the PSFR (PStar) model in each redshift bin. In the right panel, the red solid (blue) line shows the fraction of mock PSFR (Pstar) FRBs that are hosted by star-forming galaxies as a function of redshift. The dashed line indicates the fraction of mock FRBs that are hosted by disk galaxies.}
    \label{fig:model fraction in each z bin}
\end{figure*}
Our exploration provides insights into the source population and intrinsic energy distribution of FRBs. As the number of observed events increases, it becomes crucial to verify the values of key parameters in our models of FRB source population and energy distribution, such as $f_{\rm{PSFR}}, \gamma$, and $\rm{log_{10}(E_{*})}$. In addition to examining the joint distribution of DM and $\tau$, another potential and straightforward approach for investigation is the proportion of FRBs in different galaxy types (star-forming, non-star-forming, or disk and non-disk galaxies), with a given value of $f_{\rm{PSFR}}$. To illustrate this, we attempt to predict the fractions of host galaxy types and compare them with the results of localized FRB events. For instance, \cite{2023arXiv231010018B} find that 18 localized FRBs are robustly associated with spiral- or late-type galaxies in the local Universe ($\rm{z_{host}\leq 0.1}$), including two that are disk-dominated galaxies.

The stellar mass function of galaxies in the TNG100 simulation has been tuned to agree with many observed properties of galaxies \citep[][]{2018MNRAS.475..648P}. We utilize the subhalo data, i.e., galaxies, of the TNG100 simulation to attribute the mock FRBs to different galaxy types. For simplicity, we assume that FRBs only originate from galaxies with a stellar mass larger than $10^8 M_{\odot}$. Currently, the host galaxy of FRB121102 has the lowest stellar mass at $1.3\pm0.4\times10^8 M_{\odot}$ \citep[][]{2017ApJ...843L...8B}, among known FRBs hosts. These galaxies are then categorized into star-forming and quiescent galaxies or disk and non-disk. There are multiple ways to determine whether a galaxy is star-forming or quiescent (e.g., \citealt{2011ApJ...730...61K, 2015ApJ...801L..29R, 2017MNRAS.466.1192M} ). 
Here, we adopt a conventional definition in which galaxies with a specific star formation rate $\rm{sSFR>10^{-11} \, yr^{-1}}$ are classified as star-forming \citep[][]{2011ApJ...730...61K}, while the rest are considered quiescent. Meanwhile, we identify disk and non-disk galaxies based on the method in \cite{2020ApJ...895...92Z}. That is, if more than $20\%$ of the stellar mass of one galaxy in TNG simulations behaves kinematically as a disk component and its flatness is less than 0.7, it will be considered as a disk galaxy.   
 
Next, we determine the fractions of stellar mass and star formation rate contributed by each type of galaxy at various redshifts using data from the TNG100 simulation. 
For instance, at redshift 0.1, approximately 50\% and 98\% of mock FRBs associated with PStar and PSFR populations, respectively, are allocated to star-forming galaxies. These fractions change slightly to 44\% and 97\%, respectively, at $z=0$, as illustrated in the right panel of Figure \ref{fig:model fraction in each z bin}. It is important to note that the definition of star-forming galaxies can significantly affect these fractions. For example, if we define star-forming galaxies as those with sSFR higher than $10^{-10}\,yr^{-1}$, these fractions at $z=0.1$ become 17\%, 68\%, respectively, as shown in the right panel of Figure \ref{fig:model fraction in each z bin of 1e-10}. Additionally, 49\% of the mock FRBs with old progenitor (PStar) and 68\% of the mock FRBs with young progenitor (PSFR) at $z=0.1$ are hosted by disk galaxies, respectively. In a simplified model, we extrapolate the fractions of host galaxies with different types at $z=0.1$ to represent the corresponding fractions in the redshift range $0<z<0.1$.
 
Finally, we can allocate mock FRBs from the two populations to different galaxy types across various redshift bins from $z=0$ to $z=3.0$. The number distribution of mock FRBs for both source populations, which trace the SFR and stellar mass, respectively, along with their relative fractions in each redshift bin, is illustrated in the left panel of Figure \ref{fig:model fraction in each z bin}. These distributions are derived under the condition of $f_{\rm{PSFR}}=0.58$, which is the overall fraction of FRBs tracing SFR in the redshift range of 0-3. In Figure \ref{fig:model fraction in each z bin}, we can see that 9\% and 91\% mock FRBs in redshift bin $0<z<0.1$ are attributed to young (PSFR) and old (PStar) progenitors. To determine the proportions of different types of host galaxies, we multiply the fractions of FRBs belonging to each population by the fractions of mock FRBs in different types of galaxies for each population. Then, we sum up these products. For instance, in the redshift bin $0<z<0.1$,  $(0.09\times0.98+0.91\times0.5)\times100\% = 54\%$ of mock FRBS are hosted by star-forming galaxies, whereas $51\%$ of FRB sources are hosted by disk galaxies in the same redshift bin. 

\begin{deluxetable*}{ccccccc}
\tablecaption{The comparison of the fraction of host galaxy type between our model prediction and localized FRBs. SF: star-forming.}
\label{tab:galaxy type predict}
\tablehead{
\colhead{}  & \multicolumn{3}{c}{Our model prediction } & \multicolumn{2}{c}{71 localized FRBs} & \colhead{54 secure localized FRBs} 
}
\startdata
        Redshift range & $0-0.1$ & $0-1.1$ & $0-3$ & $0-0.1$ & $0-1.1$ & $0-0.7$ \\
        Fraction of SF galaxy ($\rm{sSFR>10^{-11}yr^{-1}}$) & 54\% & 68\% & 70\% & 86\% & 83\% & 81\% \\
        Fraction of SF galaxy ($\rm{sSFR>10^{-10}yr^{-1}}$) & 22\% & 34\% & 38\% & 38\% & 44\% & 44\% \\
        Fraction of disk galaxy & 51\% & 54\% & 54\% & $\geq86$\% & $\geq32$\% & $\geq30$\%  \\
\enddata
\end{deluxetable*}

This prediction appears to contradict the observations reported in \cite{2023arXiv231010018B}, where 18 localized FRBs are found to be robustly associated with disk galaxies in the local Universe ($\rm{z_{obs}\leq 0.1}$). To date, there are 21 localized FRBs within this redshift range, as listed in Table \ref{tab:localized FRB}. Of these, aside from the 18 events identified by \cite{2023arXiv231010018B}, the host galaxy morphologies of the remaining 3 FRBs remain unclear. Therefore, at least $86\%$ (18/21) of the currently observed FRB sources in the range $0<z<0.1$ are hosted by disk galaxies. Similarly, $86\%$ (18/21) of localized FRBs in this range are hosted by star-forming galaxies ($\rm{sSFR>10^{-11}yr^{-1}}$), which exceeds our predicted value of $54\%$. This discrepancy may be partially attributed to the limited sample size of observed events. In the following discussion, we explore the implications of expanding the sample of FRBs across a broader redshift range.

Using the method described above, we can assess the overall fractions of host galaxies with different types in our model within the redshift range 0-1.1, considering that the maximum redshift of localized FRBs is 1.016 (FRB220610A). Within the redshift range $0<z<1.1$, 68\% of the mock FRBs in our model reside in star-forming galaxies, while $54\%$ are situated in disk galaxies. Extending this analysis to the redshift range of 0-3, these ratios will be $70\%$ and $54\%$, respectively. To date, the total number of localized FRBs is 71. The properties of these localized events and their host galaxies are listed in Table \ref{tab:localized FRB}. By utilizing the reported SFR and stellar mass of identified FRB host galaxies, we calculate sSFR, thereby determining whether these galaxies are star-forming galaxies or not. According to the criterion of $\rm{sSFR>10^{-11}\,yr^{-1}}$, we find that $83\%$ (59/71) of the known host galaxies can be classified as star-forming galaxies, which is generally aligns with our prediction. This percentage decreases to $44\%$ (31/71) for $\rm{sSFR>10^{-10}\,yr^{-1}}$. Meanwhile, for all of the localized FRBs reported so far, the fraction of FRBs located in spiral galaxies is $32\%$ (23/71), which generally agrees with our prediction for the ratio of FRBs located in disc galaxies. For 54 secure hosts of localized FRBs ($\rm{P_{cc}}<0.05$ or $\rm{P_{host}}>0.95$), the corresponding fraction of star-forming galaxy ($\rm{sSFR>10^{-11}\,yr^{-1}}$), star-forming galaxy ($\rm{sSFR>10^{-10}\,yr^{-1}}$) and disk galaxy are $81\%$, $44\%$, $\geq30\%$, respectively. These comparisons of the host galaxy type between our model prediction and localized FRBs are summarized in Table \ref{tab:galaxy type predict}. 

The fraction of localized events may represent a lower limit due to the absence of morphological information for the host galaxies of some FRBs. Additionally, the sample is incomplete, particularly for high-redshift events, and some host galaxies may be misidentified. As a result, the constraints imposed by current observations on the types of host galaxies remain relatively weak. To further validate our model, more localized events with detailed host galaxy morphology information are required.

On the other hand, while the morphologies of galaxies in TNG100 align well with observations, the simulation contains a relatively higher number of compact, concentrated galaxies and fewer merging systems (\citealt{2019MNRAS.483.4140R}), with these accounting for less than $5\%$ of the total galaxies. Additionally, the fraction of quenched galaxies as a function of stellar mass in TNG100 generally agrees with observations, but the overall star-forming and quenched fractions of all galaxies exhibit a potential discrepancy of around $10\%$ compared to observations (\citealt{2021MNRAS.506.4760D}). These limitation of TNG100 simulation would also lead to the discrepancy between our model prediction and observations.

With more localized FRB events coming in the near future and enhanced simulations of galaxy formation, we anticipate obtaining a more robust comparison, which will subsequently enhance our constraints on the parameters in our models. This includes refining the fractions of FRBs that trace different progenitors. As our models gain tighter constraints, they will offer deeper insights into the properties of FRBs and their physical origins.

\begin{longrotatetable}
\begin{deluxetable*}{cccccccccccc}
\tablecaption{The properties of 71 localised FRBs and their host galaxies.}
\label{tab:localized FRB} 
\tablehead{
\colhead{FRB} & \colhead{redshift} & \colhead{$\rm{DM_{obs}}$ }  & \colhead{$\rm{DM_{MW,ISM}}^{\rm{a}}$ } & \colhead{$\rm{\tau_{obs}}^{\rm{b}}$}   & \colhead{$\rm{\tau_{MW,ISM}}^{\rm{c}}$}  & \colhead{fluence } & \colhead{$\rm{p_{cc}}$/$\rm{P_{host}}$$^{\rm{d}}$} & \colhead{host type}  & \colhead{host SFR$^{\rm{e}}$ }   & \colhead{host $\rm{M_{*}}$ }  & \colhead{Ref.} \\
\colhead{} & \colhead{} & \colhead{$[\rm{pc\,cm^{-3}}]$}  & \colhead{$[\rm{pc\,cm^{-3}}]$} & \colhead{$[\rm{ms}]$}   & \colhead{$[\rm{ms}]$}  & \colhead{ $[\rm{Jy\, ms}]$} & \colhead{} & \colhead{}  & \colhead{ $[\rm{M_{\odot}\,yr^{-1}}]$}   & \colhead{$[\rm{10^{9}M_{\odot}}]$}  & \colhead{} 
}

\startdata
    \textbf{121102A} & 0.19273 & 557 & 188/287 & <11.6 & 0.2/13 & 0.02-7 & 0.0003/0.99 & irregular dwarf(SF) & 0.13 & 0.13 & 1-4   \\ 
    171020A & 0.00867 & 114.1 & 37/25 & - & - & 200 & -/1 & spiral & 0.13 & 0.9 & 5,6 \\
    \textbf{180301A} & 0.3304 & 522 & 151/253 & 21 & 0.07/7.6 & 1.3-4.9 & -/0.999 & (SF) & 1.93 & 2.3 & 7,8  \\ 
    \textbf{180814A} & 0.06835 & 190.9 & 88/108 & <3.1 & 0.008/0.2 & 2.6 & -/- & early-type spiral & <0.32 & 60.26 & 12,20  \\ 
    \textbf{180916B} & 0.0337 & 347.8 & 199/325 & 1.05 & 0.17/22 & 0.2-2.5 & 0.01/- & spiral & 0.06 & 2.15 & 9-12   \\ 
    180924B & 0.3214 & 361.42 & 40/28 & 9.28 & 0.001/0.001 & 16 & 0.0018/- & early-type spiral & 0.88 & 13.2 & 10,13  \\ 
    \textbf{181030A} & 0.0039 & 101.9 & 40/32 & <1.6 & 0.001/0.002 & 4.5-7.3 & 0.0025/- & spiral(SF) & 0.35 & 5.8 & 11,14  \\ 
    181112A & 0.4755 & 589.27 & 42/29 & 0.45 & 0.001/0.001 & 26 & 0.0257/- & -(SF) & 0.6 & 2.6 & 15,16   \\ 
    181220A & 0.0275 & 208.7 & 123/119 & 0.457 & 0.02/0.33 & 3.0 & 0.03/- & spiral & 2.9  & 7.24 & 12,17 \\ 
    181223C & 0.0302 & 111.6 & 20/19 & 0.107 & 0.0004/0.0004 & 2.84 & 0.04/- & spiral & 0.15  & 1.95 & 12,17 \\ 
    190102C & 0.2913 & 363.6 & 57/43 & 0.83 & 0.003/0.006 & 14 & 0.005/- & (SF) & 1.5 & 3.16 & 18,19,23   \\
    \textbf{190110C} & 0.12244 & 221.6 & 36/29 & <1.5 & 0.001/0.001 & 1.4 & 0.102/0.779 & irregular & 0.54 & 25 & 29,30   \\ 
    \textbf{190303A}$^{\rm{f}}$ & 0.064 & 223.2 & 30/22 & 2.96 & 0.0007/0.0005 & 2.54 & -/- & spiral & 6.92 & 56.23 & 12,20  \\ 
    190418A & 0.0312 & 182.8 & 70/86 & <0.78 & 0.004/0.087 & 2.2 & 0.08/- & spiral & 0.8  & 18.62 & 12,17 \\
    190425A & 0.0715 & 127.8 & 49/39 & <0.38 & 0.003/0.004 & 31.6 & 0.0012/- & spiral & 1.6  & 18.2 & 12,17 \\
    \textbf{190520B} & 0.241 & 1202 & 60/50 & 188 & 0.004/0.01 & 0.03-0.33 & 0.008/- & dwarf & 0.41 & 0.6 & 21  \\ 
    190523A & 0.66 & 760.8 & 37/30 & 10.8 & 0.001/0.002 & 280 & 0.0733/- & - & <1.3 & 117.5 & 22   \\ 
    190608B & 0.1178 & 338.7 & 37/27 & 66.76 & 0.001/0.001 & 26 & 0.0016/- & spiral(SF) & 1.2 & 25.1 & 18,19,23   \\ 
    190611B & 0.378 & 321.4 & 58/44 & 3.64 & 0.003/0.006 & 10 & 0.0169/- & - & 0.27 & 0.8 & 10,19,23   \\
    190614D$^{\rm{g}}$ & 0.63 & 959.2 & 88/109 & - & - & 0.63 & 0.0708/- & early-type & $\sim0$ & 3.98 & 24  \\
    \textbf{190711A} & 0.522 & 593.1 & 56/43 & <24.7 & 0.003/0.006 & 34 & 0.0106/- & regular(SF) & 0.42 & 0.81 & 10,19,25  \\
    190714A & 0.2365 & 504 & 38/31 & <17.2 & 0.001/0.002 & 8 & 0.005/- & possibly spiral & 0.65 & 14.9 & 10,25   \\ 
    191001A & 0.234 & 506.92 & 44/31 & 11.74 & 0.002/0.002 & 143 & 0.0031/- & spiral & 8.06 & 46.4 & 10,26   \\
    \textbf{191106C} & 0.10775 & 331.6 & 25/21 & <10 & 0.0006/0.0004 & 1.48 & 0.063/0.815 & possibly elliptical & 4.75 & 45 & 29,30   \\ 
    191228A & 0.2432 & 297.9 & 33/20 & - & - & 40 & 22.9 & - & 0.5 & 5.4 & 7   \\
    \textbf{200120E} & -0.0001$^{\rm{h}}$ & 87.82 & 41/33 & <0.23 & 0.001/0.002 & 0.09-2.25 & 0.0006/- & spiral/GC$^{\rm{i}}$ & 0.4-0.8 & 72 & 27,28 \\ 
    \textbf{200223B} & 0.0602 & 201.8 & 46/37 & <2.7 & 0.002/0.003 & 1.06 & 0.01/0.899 & spiral & 0.59 & 56 & 29,30   \\ 
    200430A & 0.16 & 380.1 & 27/26 & 43.2 & 0.001/0.001 & 35 & 0.0051/- & - & 0.2 & 1.3  & 10,31 \\
    200906A & 0.3688 & 577.8 & 36/38 & - & - & 59 & -/1 & - & 0.48 & 13.3 & 7   \\
    \textbf{201124A} & 0.098 & 412 & 140/197 & 24 & 0.05/2.67 & 0.005-187 & $6.9\times 10^{-5}$/0.997 & spiral(SF) & 2.43 & 19 & 31-33   \\
    210117A & 0.214 & 730 & 34/23. & 5.28 & 0.001/0.001 & 36 & -/0.9984 & dwarf & 0.014 & 0.36 & 34   \\
    210320C & 0.2796 & 384.8 & 39/30 & - & - & - & -/0.999 & - & 3.51 & 23.44 & 35,36   \\ 
    210405I & 0.066 & 566.43 & 516/349 & 71.4 & 0.9/29 & 120.8 & 0.0025/- & spiral & >0.3 & 177.8 & 37 \\ 
    210410D & 0.1415 & 578.78 & 56/42 & 226.9 & 0.003/0.005 & 35.4 & -/0.9957 & - & 0.03 & 2.88 & 38   \\ 
    210807D & 0.1293 & 251.9 & 121/94 & - & - & 113 & -/0.957 & - & 0.63 & 93.3 & 35,36   \\
    211127I & 0.0469 & 234.83 & 42/31 & - & - & 31 & -/0.998 & spiral & 0.45 & 3 & 35,39   \\
    211203C & 0.3437 & 636.2 & 64/48 & - & - & 28 & -/1 & - & 15.91 & 5.75 & 35,36   \\
    211212A & 0.0707 & 206 & 39/27 & - & - & - & -/0.998 & - & 0.73 & 19.05 & 35,36   \\ 
    220105A & 0.2784 & 583 & 22/21 & - & - & - & -/1 & - & 0.42 & 10.23 & 36   \\
    220207C & 0.04304 & 262.38 & 76/83 & - & - & 16.2 & -/0.97 & disk-dominated & 2.14 & 7.94 & 40 \\
    220307B & 0.24812 & 499.27 & 128/187 & - & - & 3.2 & -/0.98 & - & 3.52 & 537 & 40  \\
    220310F & 0.47796 & 462.24 & 46/40 & - & - & 26.2 & -/0.99 & - & 0.15 & 56.23 & 40 \\
    220319D & 0.01123 & 110.98 & 140/211 & - & - & 8-11.2 & -/0.99 & spiral(SF) & 1.8 & 8.51 & 40,41 \\ 
    220418A & 0.622 & 623.25 & 37/30 & - & - & 4.2 & -/0.97 & - & 0.37 & 67.6 & 40 \\
    220506D & 0.30039 & 396.97 & 85/98 & - & - & 13.2 & -/0.98 & - & 7.01 & 28.2 & 40 \\
    220509G & 0.0894 & 269.53 & 56/52 & 2.37 & 0.003/0.012 & 5.8 & -/0.97 & spiral$^{\rm{j}}$ & 0.08  & 134.9 & 17,40,42,43 \\ 
    220610A & 1.016 & 1458.1 & 31/14 & 6.84 & 0.0007/0.0001 & 45 & -/0.87 & (SF) & 1.7 & 5 & 44,45  \\
    220825A & 0.2414 & 651.24 & 78/87 & - & - & 5.8 & -/0.99 & - & 1.34 & 138 & 40 \\
    \textbf{220912A} & 0.0771 & 219.46 & 125/122 & <2.96 & 0.02/0.37 & 0.02-431 & -/0.95 & disk-dominated & >0.1 & 10 & 46-48  \\ 
    220914A & 0.1139 & 631.28 & 55/51 & <2.37 & 0.003/0.011 & 2.6 & -/0.97 & spiral(SF) & 1.45 & 9.77 & 40,42,43 \\
    220920A & 0.15824 & 341.99 & 40/33 & - & - & 3.9 & -/0.98 & - & 0.39 & 70.8 & 40 \\
    221012A & 0.28467 & 441.08 & 54/51 & - & - & 5.1 & -/1 & - & 0.49 & 199.5 & 40 \\ 
    220204A & 0.4012 & 612.6 & 51/46 & - & - & - & -/0.99 & - & 4.31 & 5.01 & 49 \\
    220208A & 0.351 & 440.73 & 102/129 & - & - & - & -/0.56 & - & 0.68 & 12.02 & 49 \\
    220330D & 0.3714 & 467.79 & 39/30 & - & - & - & -/0.63 & - & 2.01 & 31.62 & 49 \\
    220726A & 0.3619 & 686.23 & 90/111 & - & - & - & -/0.99 & - & 0.71 & 15.14 & 49 \\
    221027A & 0.5422 & 452.72 & 47/41 & - & - & - & -/0.62 & - & 0.56 & 2.95 & 49 \\
    221029A & 0.975 & 1391.7 & 44/36 & - & - & - & -/0.92 & - & 5.21 & 38.9 & 49 \\
    221101B & 0.2395 & 491.55 & 131/192 & - & - & - & -/0.99 & - & 12.27 & 162.2 & 49 \\
    221113A & 0.2505 & 411.03 & 92/115 & - & - & - & -/0.99 & - & 0.24 & 3.02 & 49 \\
    221116A & 0.2764 & 643.45 & 132/196 & - & - & - & -/0.94 & - & 22.58 & 102.3 & 49 \\
    221219A & 0.553 & 706.7 & 44/39 & - & - & - & -/0.99 & - & 1.78 & 16.22 & 49 \\
    230124 & 0.0939 & 590.57 & 39/32 & - & - & - & -/0.99 & - & 0.75 & 2.88 & 49 \\
    230216A & 0.531 & 828.29 & 38/27 & - & - & - & -/0.42 & - & 20.81 & 6.61 & 49 \\
    230307A & 0.2706 & 608.85 & 38/29 & - & - & - & -/0.97 & - & 0.46 & 57.54 & 49 \\
    230501A & 0.3015 & 532.47 & 126/180 & - & - & - & -/0.99 & - & 4.1 & 19.5 & 49 \\
    230626A & 0.327 & 452.72 & 39/33 & - & - & - & -/0.99 & - & 0.98 & 27.54 & 49 \\
    230628A & 0.127 & 344.95 & 39/31 & - & - & - & -/0.95 & - & 0.14 & 1.95 & 49 \\
    230712A & 0.4525 & 587.57 & 39/31 & - & - & - & -/0.99 & - & 29.91 & 134.9 & 49 \\
    231120A & 0.0368 & 437.7 & 44/36 & - & - & - & -/0.99 & - & 0.4 & 25.12 & 49 \\
    231123B & 0.2621 & 396.86 & 40/34 & - & - & - & -/0.9 & - & 4.85 & 109.65 & 49 \\
\enddata
\tablecomments{Boldface represents repeating FRBs; SF: star-forming. (a) Calculated using Ne2001 model (left) and YMW16 model (right). (b) Scaled at 600 MHz based on $\tau \propto \nu^{-4}$. (c) Estimated using NE2001/YMW16 model, also scaled at 600 MHz. (d) $\rm{P_{cc}}$ (left) is the probability of chance coincidence between FRBs and their hosts; $\rm{P_{host}}$ (right) is the probability of the Probabilistic Association of Transients to their Hosts (PATH) methodology \citep{2021ApJ...911...95A} .
(e) Estimated using stellar population synthesis code or spectroscopy. For more details, see Ref. (f) It resides in a merging pair of spiral galaxies. Here, we adopt the data of a galaxy with a larger stellar mass. (g) There are two possible sources, and here we adopt the larger stellar mass one. (h) Located in M81 with distance $\sim 3.6$ Mpc. (i) Ref.28 found it's located in a globular cluster (GC) in M81.  (j) Ref.42 and 43 think it's located in an early-type or elliptical galaxy. \\
References:
(1) \cite{2017ApJ...843L...8B}. (2) \cite{2017ApJ...834L...7T}. (3) \cite{2014ApJ...790..101S}  (4) \cite{2021Natur.598..267L} (5) \cite{2018Natur.562..386S} (6) \cite{2018ApJ...867L..10M} (7) \cite{2022AJ....163...69B} (8) \cite{2019MNRAS.486.3636P} (9) \cite{2020Natur.577..190M} (10) \cite{2020ApJ...903..152H} (11) \cite{2019ApJ...885L..24C} (12) \cite{2021ApJS..257...59C} (13) \cite{2019Sci...365..565B} (14) \cite{2021ApJ...919L..24B} (15) \cite{2019Sci...366..231P} (16) \cite{2020ApJ...891L..38C} (17) \cite{2023arXiv231010018B} (18) \cite{2020ApJ...895L..37B} (19) \cite{2020Natur.581..391M} (20) \cite{2023ApJ...950..134M} (21) \cite{2022Natur.606..873N} (22) \cite{2019Natur.572..352R} (23) \cite{2020MNRAS.497.3335D} (24) \cite{2020ApJ...899..161L} (25) \cite{2020MNRAS.497.1382Q} (26) \cite{2020ApJ...901L..20B} (27) \cite{2021ApJ...910L..18B} (28) \cite{2022Natur.602..585K} (29) \cite{2023arXiv230402638I} (30) \cite{2023ApJ...947...83C} (31) \cite{2022ApJ...931...88C} (32) \cite{2021ApJ...919L..23F} (33) \cite{2022Natur.609..685X} (34) \cite{2023ApJ...948...67B} (35) \cite{2022MNRAS.516.4862J} (36) \cite{2023ApJ...954...80G} (37) \cite{2024MNRAS.527.3659D} (38) \cite{2023MNRAS.524.2064C} (39) \cite{2023ApJ...949...25G} (40) \cite{2023arXiv230703344L} (41) \cite{2023arXiv230101000R} (42) \cite{2023ApJ...949L..26C} (43) \cite{2023ApJ...950..175S} (44) \cite{2023Sci...382..294R} (45) \cite{2023arXiv231110815G} (46) \cite{2023ApJ...955..142Z} (47) \cite{2023ApJ...949L...3R} (48) \cite{2024MNRAS.52710425S} (49)\cite{2024Natur.635...61S} } 
\end{deluxetable*}
\end{longrotatetable}

\section{Conclusion}
In this work, we use a simplified model of the FRB source population, and a comprehensive assessment of the contributions from various components to the DM and scattering of the FRBs along the line of sights, to replicate the joint distribution of DM and $\tau$ of FRBs observed in the CHIME/FRB catalog. Our estimation of contributions from the Milky Way, local environment, and IGM is grounded in updated theoretical models incorporating recent progress. Additionally, we have determined the contributions from the Milky Way's halo, foreground galaxies, and halos, as well as host galaxies and halos, based on the properties of galaxies in the state-of-the-art cosmological hydrodynamical simulations of TNG100 and TNG50. We produce mock FRB sources between redshift $0-3$, taking into account two possible source populations and their intrinsic energy distribution. These mock samples are then used to compare with the CHIME/FRB catalog on the number distribution of DM and $\tau$, taking into account the observation selection effect. Through MCMC simulations, we match the CHIME/FRB observations and determine the optimal parameters in our models.

Based on our framework, we have developed redshift estimators for FRBs using DM-only information and DM combined with $\tau$ information.
We evaluated these estimators using data from around 70 localized FRB events. We argue that the properties of host galaxies could serve as a means to validate the optimal parameters in our model, which is generally consistent with currently localized FRB events. We summarize our findings as follows. 

\begin{itemize}
    \item If around $30\%-80\%$ of the FRBs sources are associated with the young progenitors, and their intrinsic energy distribution follows a Schechter function characterized by $\gamma=-1.60^{+0.11}_{-0.13},\,\rm{log_{10}(E_{*})}=42.27^{+1.17}_{-1.18},\,$, our model can well reproduce the observed distribution of DM and broadly matches the distribution of $\tau$ in the CHIME/FRB catalog, while accounting for selection effects. More events with robust measurements of $\tau$ are needed to to further refine the constraints and improve our model. 
    \item In our model with optimized parameters, the primary contribution to DM comes from the IGM, consistent with previous findings in the literature. Foreground galaxies and their halos make a non-negligible contribution to the DM, particularly for events with $\rm{DM_{obs}}>100\, pc\,cm^{-3}$ and $z > 0.3$. Regarding $\tau$, the dominant contribution arises from the `local' environment, due to electron clumps with highly turbulent and dense ionized clumps, typically on the scale of parsecs. Alternatively, electron clumps with moderate density fluctuations within the host's ISM and halo can dominate $\tau$. Meanwhile, electron clumps in the MW's halo or in the CGM of intervening foreground galaxies, could also account for the observed $\tau$ in some FRBs. Overall, the diffuse medium in foreground galaxies and halos are responsible for $5\%-15\%$ of the scattering, specifically for events with $0.01\,\rm{ms}<\tau< 3 \, \rm{ms}$. 
    \item Our model, which accounts for the contributions of multiple components to the DM and scattering of FRBs based on TNG100 simulation, is combined with optimized parameters for the source population fraction and intrinsic energy distribution. Using this framework, we find that both estimation methods - one utilizing DM alone and the other incorporating DM along with $\tau$ - yield reasonable redshift estimates for 31 localized FRBs with both measured DM and $\tau$,  as well as for 68 of 71 localized FRBS. If the extreme event FRB190520B is excluded, our redshift estimators achieve an RMS error of approximately $0.11-0.12$. However, incorporating $\tau$ does not significantly enhance the accuracy.
    
    \item Assuming that 30\%-80\% of FRBs within the redshift range of 0-3 originate from young progenitors, we anticipate that 54\% of FRBs at $z<0.1$ will be hosted by star-forming galaxies, and the corresponding fraction is 68\% for events within the redshift range of 0-1.1. Similarly, the fraction for disk galaxies is 51\% and 54\%, respectively. The fractions predicted by our model for the redshift range 0–1.1 generally align with the observed values from currently localized FRBs. However, caution is warranted when making this comparison, as the results may be influenced by sample biases arising from incompleteness and the lack of morphological classification for most high-redshift FRBs, and the limitation of TNG100 simulation.
\end{itemize}

A significantly larger sample of localized FRBs with high-confidence information in the future will provide more robust observational constraints. This will enable the validation of our models and associated parameters, enhancing the accuracy and reliability of our estimates for the contributions of various components to the dispersion measure (DM) and scattering of FRBs, their intrinsic properties. Consequently, this will enhance the precision of our redshift estimators. Additionally, future galaxy formation simulations with higher resolution and improved sub-grid physics will further refine our models. Moreover, our study could address fundamental questions in galaxy formation and evolution, such as the fraction of baryonic matter within halos and the roles of different feedback mechanisms therein (\citealt{2023ARA&A..61..473C}). 

\section{Acknowledgments}
 The authors thank the anonymous referee for her/his very useful comments and suggestions that improved the manuscript. This work is supported by the National Natural Science Foundation of China (NFSC) through grant 11733010. W.S.Z. is supported by NSFC grant 12173102. Analysis carried in this work was completed on the HPC facility of the School of Physics and Astronomy, Sun Yat-Sen University.

%

\vspace{5mm}






\appendix

\section{scattering measure and pulse broadening time}
\label{app:sm and tau}

We assume that the CGM and ISM of the intervening galaxies and halos, as well as the host halos and galaxies can be described by the Kolmogorov turbulence model  between the inner scale, $l_0$, and outer scale, $L_0$. The electron density power spectrum of the inhomogeneous medium  is assumed to follow a power law \citep[][]{2013ApJ...776..125M, 2021ApJ...906...95Z}, that is,
\begin{equation}
    P(k) = C^2_N k^{-\beta} e^{-(kl_0)^2}, \, k>L_0^{-1},
    \label{eqn:ne power spectrum}
\end{equation}
where $C^2_N$ is the amplitude of the turbulence; k is wavenumber; $\beta=11/3$ for Kolmogorov turbulence. Here, we adopt values of $l_0=10^6\, \rm{m}, \, L_0=5\,\rm{pc}$ for the medium in the foreground and host galaxy and halos, as the 'Model-A' in \cite{2021ApJ...906...95Z}.
For index $\beta > 3$ ,$C^2_N$ can be written as 
\begin{equation}
    C^2_N \approx \frac{\beta-3}{2(2\pi)^{4-\beta}} \langle \delta n^2_e(z) \rangle L_0^{3-\beta},
    \label{eqn:turbulence amplitude}
\end{equation}
where $\langle \delta n^2_e(z) \rangle$ is root-mean-square electron density. In our calculation, we adopt the assumption of $\langle \delta n^2_e(z) \rangle \approx n^2_e(z)$, following \cite{2013ApJ...776..125M}.
 For a FRB source at $z_S$, the temporal broadening time caused by a medium with a size range from $z$ to $z+\Delta z$ is given by \citep[][]{2013ApJ...776..125M, 2021ApJ...906...95Z}:
\begin{equation}
\begin{split}
    \Delta \tau=3.32\times 10^{-4}(1+z)^{-1}\left(\frac{\lambda_0}{30\rm{cm}} \right)^4 \left(\frac{D_{\rm{eff}}}{1\rm{Gpc}} \right)   \\   \times \left(\frac{\rm{\Delta SM_{eff}}}{10^{12}\rm{m}^{-17/3}} \right) \left(\frac{l_0}{1\rm{AU}} \right)^{-1/3} \, [\rm{ms}], \, r_{\rm{diff}}<l_0,
    \label{eqn:delta tau for r small}
\end{split}
\end{equation}

\begin{equation}
\begin{split}
    \Delta \tau=9.50\times 10^{-4}(1+z)^{-1} \left(\frac{\lambda_0}{30\rm{cm}} \right)^{22/5} \left(\frac{D_{\rm{eff}}}{1\rm{Gpc}} \right)  \\  \times \left(\frac{\rm{\Delta SM_{eff}}}{10^{12}\rm{m}^{-17/3}} \right)^{6/5}  \, [\rm{ms}], \, r_{\rm{diff}}>l_0,
    \label{eqn:delta tau for r large}
\end{split}
\end{equation}
where $\lambda_0$ is the observed wavelength, $\lambda_0=30\, \rm{cm}$ is equivalent to observed at 1GHz; $D_{\rm{eff}}$ is effective angular distance to which is equal $D_{LS}D_L/D_S$, where $D_{LS}, \, D_L, \, D_S$ are the angular diameter distances from the scattering layer (media) to the source, from the scattering layer to the observer, and from the source to the observer, respectively; $\rm{\Delta SM_{eff}} = \int \frac{C^2_N(\it{l})}{(1+z')^2} \it{dl}  = \int^{z+\Delta z}_z \frac{C^2_N(z') d_H(z')}{(1+z')^3} \it{dz'}$ 
is the effective measure(SM) in the observer's frame, and the definition of SM in the rest frame (e.g. host galaxy) is $\rm{SM}=\int C^2_N \it{dl}$; $r_{\rm{diff}}$ is the diffractive length scale, which is adopted from \citep[][]{2013ApJ...776..125M, 2021ApJ...906...95Z}:

\begin{equation}
\begin{split}
    r_{\rm{diff}}=2.7\times10^{10} \left(\frac{\lambda_0}{30\,\rm{cm}} \right)^{-1} \left(\frac{\rm{SM_{eff}}}{10^{12}\rm{m}^{-17/3}} \right)^{-1/2}  \\  \times \left(\frac{l_0}{1\rm{AU}} \right)^{1/6} \, [\rm{m}], \, r_{\rm{diff}}<l_0,
    \label{eqn:rdiff small}
\end{split}
\end{equation}

\begin{equation}
\begin{split}
    r_{\rm{diff}}=1.6\times10^{10}\left(\frac{\lambda_0}{30\,\rm{cm}}\right)^{-6/5} \left(\frac{\rm{SM_{eff}}}{10^{12}\rm{m}^{-17/3}}\right)^{-3/5} \, [\rm{m}],  \\   r_{\rm{diff}}>l_0,
    \label{eqn:rdiff large}
\end{split}
\end{equation}

\section{Miscellaneous}

\begin{figure}
    \centering
    \includegraphics[width=0.8\columnwidth]{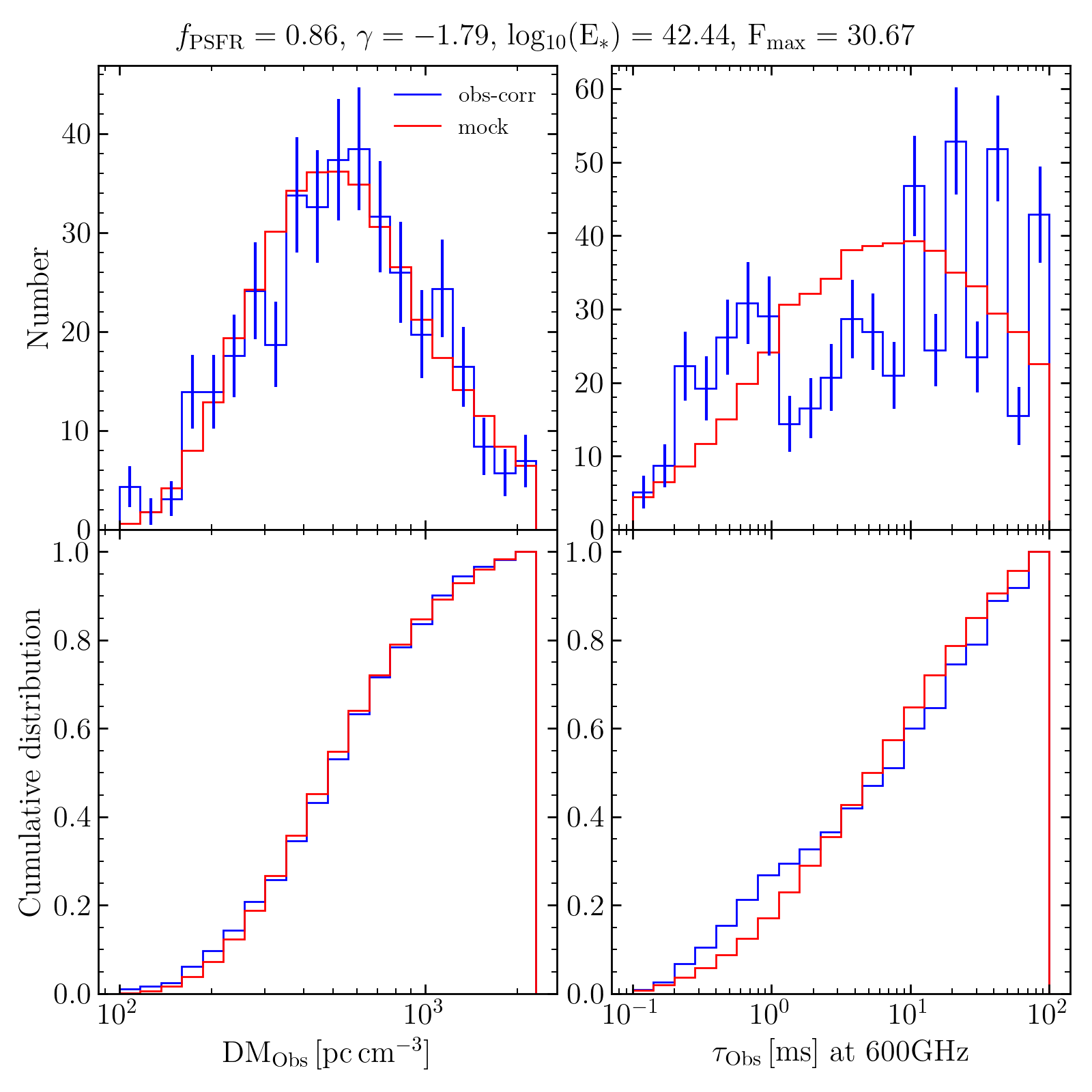}
    \caption{Similar to Figure \ref{fig:compa mock obs}, but for a larger $\tau$ range of 0.1-100 ms at 600 MHz. The title denotes the corresponding optimal parameters from MCMC simulation with 20000 steps. }
    \label{fig:compa mock obs Lbin}
\end{figure}

Figure \ref{fig:compa mock obs Lbin} indicates the comparison of the joint DM and $\tau$ distribution of FRBs between our mock results and the selection-corrected observations from CHIME/FRB, where $\tau$ at 600 MHz ranges from 0.1 ms to 100 ms (i.e., TNG100-$\tau$100ms model in Table \ref{tab:comparison of different models}). Our mock results are generated using the optimal parameters of $f_{\rm{PSFR}}=0.86,\,\gamma=-1.79,\,\rm{log_{10}(E_*)}=42.44$ and $F_{\rm{max}}=30.67$. derived from the MCMC simulation with 20000 steps. 

Table \ref{tab:comparison of different models} lists all our parameter fit results for different models, along with the corresponding KS and AD test results, and root mean square (RMS) of the localized FRB redshift estimation. 

Table \ref{tab:comparison with codes} provides a comparison between the results of our redshift estimator and that in \cite{2022ApJ...931...88C} for the same nine FRBs.

\begin{figure}
    \includegraphics[width=1.\columnwidth]{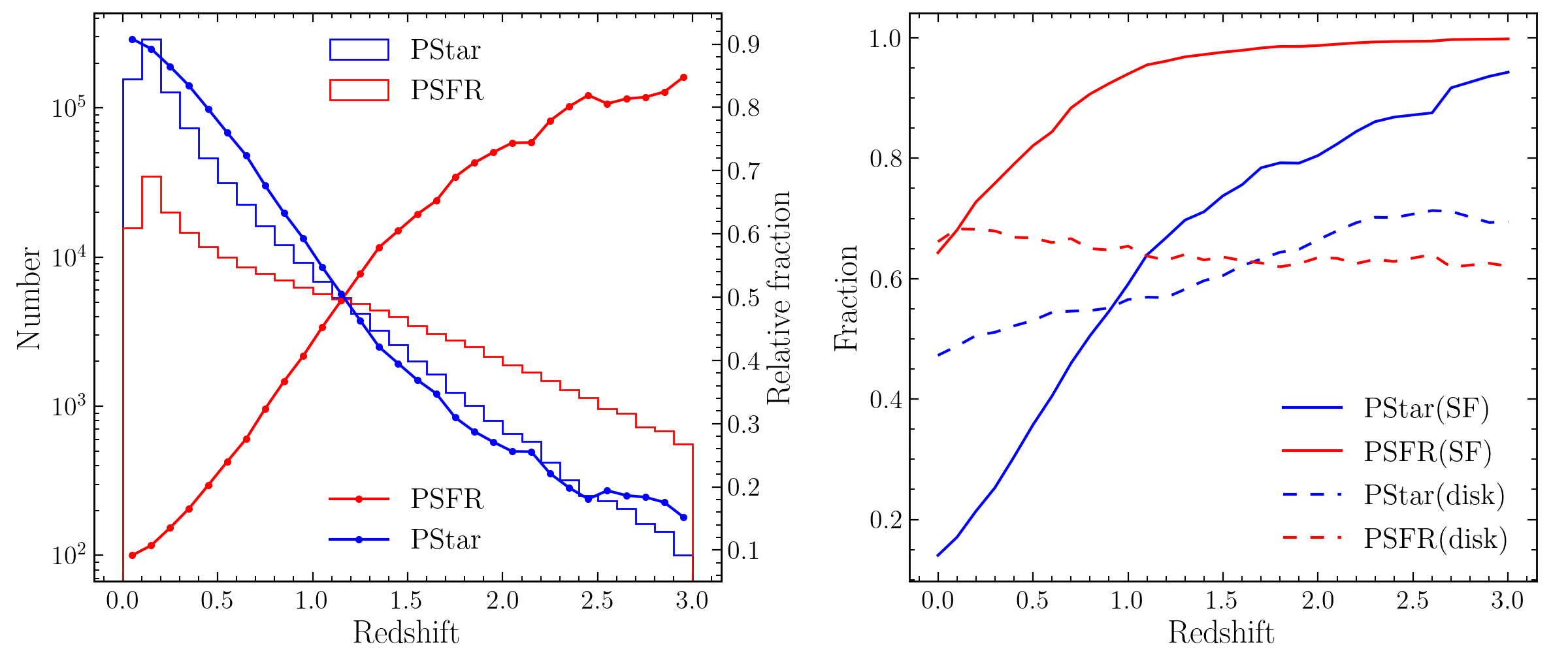}
    \caption{Similar to Figure \ref{fig:model fraction in each z bin},  but using the following sSFR criterion $>10^{-10} yr^{-1}$. }
    \label{fig:model fraction in each z bin of 1e-10}
\end{figure}

Figure \ref{fig:model fraction in each z bin of 1e-10} shows the results regarding the proportion of mock FRBs in different galaxy types, if a higher threshold of $\rm{sSFR>10^{-10}yr^{-1}}$ is adotped when a galaxy is considered to be star forming or not.

\renewcommand{\arraystretch}{1.2} 
\begin{table}
	\centering
	\caption{The comparison of the same nine FRB redshift estimation between \cite{2022ApJ...931...88C} and our results.}
	\label{tab:comparison with codes}
	\begin{tabular}{ccc|ccc|ccc} 
		\hline
		     & & & \multicolumn{3}{c|}{ \cite{2022ApJ...931...88C}}   & \multicolumn{3}{c}{Our results}   \\
            FRB & $\rm{z_{obs}}$ & $\rm{\tau_{obs}}$[ms] at 600MHz & $\rm{z_{est}}$(DMonly) & $\rm{z_{est}}$(DM+$\tau$) & $\rm{DM_{h}}$[$\rm{pc\,cm^{-3}}$] & $\rm{z_{est}}$(DMonly) & $\rm{z_{est}}$(DM+$\tau$) & $\rm{DM_{h}}$[$\rm{pc\,cm^{-3}}$] \\ 
		\hline
            180924B & 0.321 & 9.3 & 0.319 & 0.216 & 99 & 0.161 & 0.162 & 125 \\
            181112A & 0.475 & 0.5 & 0.571 & 0.561 & 206 & 0.283 & 0.271 & 258\\
            190102C & 0.291 & 0.8 & 0.302 & 0.292 & 100 & 0.161 & 0.162 & 112 \\
            190523A & 0.660 & 10.8 & 0.762 & 0.623 & 261 & 0.384 & 0.422 & 288 \\
            190608B & 0.118 & 66.8 & 0.297 & 0.111 & 190 & 0.161 & 0.149 & 147 \\
            190611B & 0.378 & 3.6 & 0.253 & 0.207 & 58 & 0.147 & 0.151 & 100 \\
            191001A & 0.234 & 11.7 & 0.477 & 0.356 & 287 & 0.218 & 0.226 & 194 \\
            200430A & 0.160 & 43.2 & 0.356 & 0.205 & 217 & 0.177 & 0.169 & 163 \\
            201124A & 0.098 & 24.0 & 0.305 & 0.165 & 172 & 0.147 & 0.144 & 111 \\
            \hline
            \multicolumn{3}{c|}{Root mean square (RMS) error}  & 0.152 & 0.09 &  & 0.154 & 0.147 &  \\
            \hline
	\end{tabular}
\end{table}


\bibliography{scatter_dm}{}

\begin{thebibliography}{}
\expandafter\ifx\csname natexlab\endcsname\relax\def\natexlab#1{#1}\fi
\providecommand{\url}[1]{\href{#1}{#1}}
\providecommand{\dodoi}[1]{doi:~\href{http://doi.org/#1}{\nolinkurl{#1}}}
\providecommand{\doeprint}[1]{\href{http://ascl.net/#1}{\nolinkurl{http://ascl.net/#1}}}
\providecommand{\doarXiv}[1]{\href{https://arxiv.org/abs/#1}{\nolinkurl{https://arxiv.org/abs/#1}}}

\bibitem[{{Aggarwal} {et~al.}(2021){Aggarwal}, {Budav{\'a}ri}, {Deller},
  {Eftekhari}, {James}, {Prochaska}, \& {Tendulkar}}]{2021ApJ...911...95A}
{Aggarwal}, K., {Budav{\'a}ri}, T., {Deller}, A.~T., {et~al.} 2021, \apj, 911,
  95, \dodoi{10.3847/1538-4357/abe8d2}

\bibitem[{{Bannister} {et~al.}(2019){Bannister}, {Deller}, {Phillips},
  {Macquart}, {Prochaska}, {Tejos}, {Ryder}, {Sadler}, {Shannon}, {Simha},
  {Day}, {McQuinn}, {North-Hickey}, {Bhandari}, {Arcus}, {Bennert}, {Burchett},
  {Bouwhuis}, {Dodson}, {Ekers}, {Farah}, {Flynn}, {James}, {Kerr}, {Lenc},
  {Mahony}, {O'Meara}, {Os{\l}owski}, {Qiu}, {Treu}, {U}, {Bateman}, {Bock},
  {Bolton}, {Brown}, {Bunton}, {Chippendale}, {Cooray}, {Cornwell}, {Gupta},
  {Hayman}, {Kesteven}, {Koribalski}, {MacLeod}, {McClure-Griffiths},
  {Neuhold}, {Norris}, {Pilawa}, {Qiao}, {Reynolds}, {Roxby}, {Shimwell},
  {Voronkov}, \& {Wilson}}]{2019Sci...365..565B}
{Bannister}, K.~W., {Deller}, A.~T., {Phillips}, C., {et~al.} 2019, Science,
  365, 565, \dodoi{10.1126/science.aaw5903}

\bibitem[{{Baptista} {et~al.}(2023){Baptista}, {Prochaska}, {Mannings},
  {James}, {Shannon}, {Ryder}, {Deller}, {Scott}, {Glowacki}, \&
  {Tejos}}]{2023arXiv230507022B}
{Baptista}, J., {Prochaska}, J.~X., {Mannings}, A.~G., {et~al.} 2023, arXiv
  e-prints, arXiv:2305.07022, \dodoi{10.48550/arXiv.2305.07022}

\bibitem[{{Bassa} {et~al.}(2017){Bassa}, {Tendulkar}, {Adams}, {Maddox},
  {Bogdanov}, {Bower}, {Burke-Spolaor}, {Butler}, {Chatterjee}, {Cordes},
  {Hessels}, {Kaspi}, {Law}, {Marcote}, {Paragi}, {Ransom}, {Scholz},
  {Spitler}, \& {van Langevelde}}]{2017ApJ...843L...8B}
{Bassa}, C.~G., {Tendulkar}, S.~P., {Adams}, E.~A.~K., {et~al.} 2017, \apjl,
  843, L8, \dodoi{10.3847/2041-8213/aa7a0c}

\bibitem[{{Batten}(2019)}]{2019JOSS....4.1399B}
{Batten}, A. 2019, The Journal of Open Source Software, 4, 1399,
  \dodoi{10.21105/joss.01399}

\bibitem[{{Batten} {et~al.}(2021){Batten}, {Duffy}, {Wijers}, {Gupta}, {Flynn},
  {Schaye}, \& {Ryan-Weber}}]{2021MNRAS.505.5356B}
{Batten}, A.~J., {Duffy}, A.~R., {Wijers}, N.~A., {et~al.} 2021, \mnras, 505,
  5356, \dodoi{10.1093/mnras/stab1528}

\bibitem[{{Bennett} \& {Sijacki}(2020)}]{2020MNRAS.499..597B}
{Bennett}, J.~S., \& {Sijacki}, D. 2020, \mnras, 499, 597,
  \dodoi{10.1093/mnras/staa2835}

\bibitem[{{Bhandari} {et~al.}(2020{\natexlab{a}}){Bhandari}, {Sadler},
  {Prochaska}, {Simha}, {Ryder}, {Marnoch}, {Bannister}, {Macquart}, {Flynn},
  {Shannon}, {Tejos}, {Corro-Guerra}, {Day}, {Deller}, {Ekers}, {Lopez},
  {Mahony}, {Nu{\~n}ez}, \& {Phillips}}]{2020ApJ...895L..37B}
{Bhandari}, S., {Sadler}, E.~M., {Prochaska}, J.~X., {et~al.}
  2020{\natexlab{a}}, \apjl, 895, L37, \dodoi{10.3847/2041-8213/ab672e}

\bibitem[{{Bhandari} {et~al.}(2020{\natexlab{b}}){Bhandari}, {Bannister},
  {Lenc}, {Cho}, {Ekers}, {Day}, {Deller}, {Flynn}, {James}, {Macquart},
  {Mahony}, {Marnoch}, {Moss}, {Phillips}, {Prochaska}, {Qiu}, {Ryder},
  {Shannon}, {Tejos}, \& {Wong}}]{2020ApJ...901L..20B}
{Bhandari}, S., {Bannister}, K.~W., {Lenc}, E., {et~al.} 2020{\natexlab{b}},
  \apjl, 901, L20, \dodoi{10.3847/2041-8213/abb462}

\bibitem[{{Bhandari} {et~al.}(2022){Bhandari}, {Heintz}, {Aggarwal}, {Marnoch},
  {Day}, {Sydnor}, {Burke-Spolaor}, {Law}, {Xavier Prochaska}, {Tejos},
  {Bannister}, {Butler}, {Deller}, {Ekers}, {Flynn}, {Fong}, {James}, {Lazio},
  {Luo}, {Mahony}, {Ryder}, {Sadler}, {Shannon}, {Han}, {Lee}, \&
  {Zhang}}]{2022AJ....163...69B}
{Bhandari}, S., {Heintz}, K.~E., {Aggarwal}, K., {et~al.} 2022, \aj, 163, 69,
  \dodoi{10.3847/1538-3881/ac3aec}

\bibitem[{{Bhandari} {et~al.}(2023){Bhandari}, {Gordon}, {Scott}, {Marnoch},
  {Sridhar}, {Kumar}, {James}, {Qiu}, {Bannister}, {T. Deller}, {Eftekhari},
  {Fong}, {Glowacki}, {Prochaska}, {Ryder}, {Shannon}, \&
  {Simha}}]{2023ApJ...948...67B}
{Bhandari}, S., {Gordon}, A.~C., {Scott}, D.~R., {et~al.} 2023, \apj, 948, 67,
  \dodoi{10.3847/1538-4357/acc178}

\bibitem[{{Bhardwaj} {et~al.}(2021{\natexlab{a}}){Bhardwaj}, {Gaensler},
  {Kaspi}, {Landecker}, {Mckinven}, {Michilli}, {Pleunis}, {Tendulkar},
  {Andersen}, {Boyle}, {Cassanelli}, {Chawla}, {Cook}, {Dobbs}, {Fonseca},
  {Kaczmarek}, {Leung}, {Masui}, {Mnchmeyer}, {Ng}, {Rafiei-Ravandi}, {Scholz},
  {Shin}, {Smith}, {Stairs}, \& {Zwaniga}}]{2021ApJ...910L..18B}
{Bhardwaj}, M., {Gaensler}, B.~M., {Kaspi}, V.~M., {et~al.} 2021{\natexlab{a}},
  \apjl, 910, L18, \dodoi{10.3847/2041-8213/abeaa6}

\bibitem[{{Bhardwaj} {et~al.}(2021{\natexlab{b}}){Bhardwaj}, {Kirichenko},
  {Michilli}, {Mayya}, {Kaspi}, {Gaensler}, {Rahman}, {Tendulkar}, {Fonseca},
  {Josephy}, {Leung}, {Merryfield}, {Petroff}, {Pleunis}, {Sanghavi}, {Scholz},
  {Shin}, {Smith}, \& {Stairs}}]{2021ApJ...919L..24B}
{Bhardwaj}, M., {Kirichenko}, A.~Y., {Michilli}, D., {et~al.}
  2021{\natexlab{b}}, \apjl, 919, L24, \dodoi{10.3847/2041-8213/ac223b}

\bibitem[{{Bhardwaj} {et~al.}(2023){Bhardwaj}, {Michilli}, {Kirichenko},
  {Modilim}, {Shin}, {Kaspi}, {Andersen}, {Cassanelli}, {Brar}, {Chatterjee},
  {Cook}, {Dong}, {Fonseca}, {Gaensler}, {Ibik}, {Kaczmarek}, {Lanman},
  {Leung}, {Masui}, {Pandhi}, {Pearlman}, {Pleunis}, {Prochaska},
  {Rafiei-Ravandi}, {Sand}, {Scholz}, \& {Smith}}]{2023arXiv231010018B}
{Bhardwaj}, M., {Michilli}, D., {Kirichenko}, A.~Y., {et~al.} 2023, arXiv
  e-prints, arXiv:2310.10018, \dodoi{10.48550/arXiv.2310.10018}

\bibitem[{{Bochenek} {et~al.}(2020){Bochenek}, {Ravi}, {Belov}, {Hallinan},
  {Kocz}, {Kulkarni}, \& {McKenna}}]{2020Natur.587...59B}
{Bochenek}, C.~D., {Ravi}, V., {Belov}, K.~V., {et~al.} 2020, \nat, 587, 59,
  \dodoi{10.1038/s41586-020-2872-x}

\bibitem[{{Caleb} {et~al.}(2023){Caleb}, {Driessen}, {Gordon}, {Tejos},
  {Bernales}, {Qiu}, {Chibueze}, {Stappers}, {Rajwade}, {Cavallaro}, {Wang},
  {Kumar}, {Majid}, {Wharton}, {Naudet}, {Bezuidenhout}, {Jankowski},
  {Malenta}, {Morello}, {Sanidas}, {Surnis}, {Barr}, {Chen}, {Kramer}, {Fong},
  {Kilpatrick}, {Prochaska}, {Simha}, {Venter}, {Heywood}, {Kundu}, \&
  {Schussler}}]{2023MNRAS.524.2064C}
{Caleb}, M., {Driessen}, L.~N., {Gordon}, A.~C., {et~al.} 2023, \mnras, 524,
  2064, \dodoi{10.1093/mnras/stad1839}

\bibitem[{{Chawla} {et~al.}(2022){Chawla}, {Kaspi}, {Ransom}, {Bhardwaj},
  {Boyle}, {Breitman}, {Cassanelli}, {Cubranic}, {Dong}, {Fonseca}, {Gaensler},
  {Giri}, {Josephy}, {Kaczmarek}, {Leung}, {Masui}, {Mena-Parra}, {Merryfield},
  {Michilli}, {M{\"u}nchmeyer}, {Ng}, {Patel}, {Pearlman}, {Petroff},
  {Pleunis}, {Rahman}, {Sanghavi}, {Shin}, {Smith}, {Stairs}, \&
  {Tendulkar}}]{2022ApJ...927...35C}
{Chawla}, P., {Kaspi}, V.~M., {Ransom}, S.~M., {et~al.} 2022, \apj, 927, 35,
  \dodoi{10.3847/1538-4357/ac49e1}

\bibitem[{{CHIME/FRB Collaboration} {et~al.}(2018){CHIME/FRB Collaboration},
  {Amiri}, {Bandura}, {Berger}, {Bhardwaj}, {Boyce}, {Boyle}, {Brar},
  {Burhanpurkar}, {Chawla}, {Chowdhury}, {Cliche}, {Cranmer}, {Cubranic},
  {Deng}, {Denman}, {Dobbs}, {Fandino}, {Fonseca}, {Gaensler}, {Giri},
  {Gilbert}, {Good}, {Guliani}, {Halpern}, {Hinshaw}, {H{\"o}fer}, {Josephy},
  {Kaspi}, {Landecker}, {Lang}, {Liao}, {Masui}, {Mena-Parra}, {Naidu},
  {Newburgh}, {Ng}, {Patel}, {Pen}, {Pinsonneault-Marotte}, {Pleunis}, {Rafiei
  Ravandi}, {Ransom}, {Renard}, {Scholz}, {Sigurdson}, {Siegel}, {Smith},
  {Stairs}, {Tendulkar}, {Vanderlinde}, \& {Wiebe}}]{2018ApJ...863...48C}
{CHIME/FRB Collaboration}, {Amiri}, M., {Bandura}, K., {et~al.} 2018, \apj,
  863, 48, \dodoi{10.3847/1538-4357/aad188}

\bibitem[{{CHIME/FRB Collaboration} {et~al.}(2019){CHIME/FRB Collaboration},
  {Andersen}, {Bandura}, {Bhardwaj}, {Boubel}, {Boyce}, {Boyle}, {Brar},
  {Cassanelli}, {Chawla}, {Cubranic}, {Deng}, {Dobbs}, {Fandino}, {Fonseca},
  {Gaensler}, {Gilbert}, {Giri}, {Good}, {Halpern}, {Hill}, {Hinshaw},
  {H{\"o}fer}, {Josephy}, {Kaspi}, {Kothes}, {Landecker}, {Lang}, {Li}, {Lin},
  {Masui}, {Mena-Parra}, {Merryfield}, {Mckinven}, {Michilli}, {Milutinovic},
  {Naidu}, {Newburgh}, {Ng}, {Patel}, {Pen}, {Pinsonneault-Marotte}, {Pleunis},
  {Rafiei-Ravandi}, {Rahman}, {Ransom}, {Renard}, {Scholz}, {Siegel}, {Singh},
  {Smith}, {Stairs}, {Tendulkar}, {Tretyakov}, {Vanderlinde}, {Yadav}, \&
  {Zwaniga}}]{2019ApJ...885L..24C}
{CHIME/FRB Collaboration}, {Andersen}, B.~C., {Bandura}, K., {et~al.} 2019,
  \apjl, 885, L24, \dodoi{10.3847/2041-8213/ab4a80}

\bibitem[{{CHIME/FRB Collaboration} {et~al.}(2020){CHIME/FRB Collaboration},
  {Andersen}, {Bandura}, {Bhardwaj}, {Bij}, {Boyce}, {Boyle}, {Brar},
  {Cassanelli}, {Chawla}, {Chen}, {Cliche}, {Cook}, {Cubranic}, {Curtin},
  {Denman}, {Dobbs}, {Dong}, {Fandino}, {Fonseca}, {Gaensler}, {Giri}, {Good},
  {Halpern}, {Hill}, {Hinshaw}, {H{\"o}fer}, {Josephy}, {Kania}, {Kaspi},
  {Landecker}, {Leung}, {Li}, {Lin}, {Masui}, {McKinven}, {Mena-Parra},
  {Merryfield}, {Meyers}, {Michilli}, {Milutinovic}, {Mirhosseini},
  {M{\"u}nchmeyer}, {Naidu}, {Newburgh}, {Ng}, {Patel}, {Pen},
  {Pinsonneault-Marotte}, {Pleunis}, {Quine}, {Rafiei-Ravandi}, {Rahman},
  {Ransom}, {Renard}, {Sanghavi}, {Scholz}, {Shaw}, {Shin}, {Siegel}, {Singh},
  {Smegal}, {Smith}, {Stairs}, {Tan}, {Tendulkar}, {Tretyakov}, {Vanderlinde},
  {Wang}, {Wulf}, \& {Zwaniga}}]{2020Natur.587...54C}
{CHIME/FRB Collaboration}, {Andersen}, B.~C., {Bandura}, K.~M., {et~al.} 2020,
  \nat, 587, 54, \dodoi{10.1038/s41586-020-2863-y}

\bibitem[{{CHIME/FRB Collaboration} {et~al.}(2021){CHIME/FRB Collaboration},
  {Amiri}, {Andersen}, {Bandura}, {Berger}, {Bhardwaj}, {Boyce}, {Boyle},
  {Brar}, {Breitman}, {Cassanelli}, {Chawla}, {Chen}, {Cliche}, {Cook},
  {Cubranic}, {Curtin}, {Deng}, {Dobbs}, {Dong}, {Eadie}, {Fandino}, {Fonseca},
  {Gaensler}, {Giri}, {Good}, {Halpern}, {Hill}, {Hinshaw}, {Josephy},
  {Kaczmarek}, {Kader}, {Kania}, {Kaspi}, {Landecker}, {Lang}, {Leung}, {Li},
  {Lin}, {Masui}, {McKinven}, {Mena-Parra}, {Merryfield}, {Meyers}, {Michilli},
  {Milutinovic}, {Mirhosseini}, {M{\"u}nchmeyer}, {Naidu}, {Newburgh}, {Ng},
  {Patel}, {Pen}, {Petroff}, {Pinsonneault-Marotte}, {Pleunis},
  {Rafiei-Ravandi}, {Rahman}, {Ransom}, {Renard}, {Sanghavi}, {Scholz}, {Shaw},
  {Shin}, {Siegel}, {Sikora}, {Singh}, {Smith}, {Stairs}, {Tan}, {Tendulkar},
  {Vanderlinde}, {Wang}, {Wulf}, \& {Zwaniga}}]{2021ApJS..257...59C}
{CHIME/FRB Collaboration}, {Amiri}, M., {Andersen}, B.~C., {et~al.} 2021,
  \apjs, 257, 59, \dodoi{10.3847/1538-4365/ac33ab}

\bibitem[{{Chime/Frb Collaboration} {et~al.}(2023){Chime/Frb Collaboration},
  {Andersen}, {Bandura}, {Bhardwaj}, {Boyle}, {Brar}, {Cassanelli},
  {Chatterjee}, {Chawla}, {Cook}, {Curtin}, {Dobbs}, {Dong}, {Faber},
  {Fandino}, {Fonseca}, {Gaensler}, {Giri}, {Herrera-Martin}, {Hill}, {Ibik},
  {Josephy}, {Kaczmarek}, {Kader}, {Kaspi}, {Landecker}, {Lanman}, {Lazda},
  {Leung}, {Lin}, {Masui}, {McKinven}, {Mena-Parra}, {Meyers}, {Michilli},
  {Ng}, {Pandhi}, {Pearlman}, {Pen}, {Petroff}, {Pleunis}, {Rafiei-Ravandi},
  {Rahman}, {Ransom}, {Renard}, {Sand}, {Sanghavi}, {Scholz}, {Shah}, {Shin},
  {Siegel}, {Smith}, {Stairs}, {Su}, {Tendulkar}, {Vanderlinde}, {Wang},
  {Wulf}, \& {Zwaniga}}]{2023ApJ...947...83C}
{Chime/Frb Collaboration}, {Andersen}, B.~C., {Bandura}, K., {et~al.} 2023,
  \apj, 947, 83, \dodoi{10.3847/1538-4357/acc6c1}

\bibitem[{{Chittidi} {et~al.}(2021){Chittidi}, {Simha}, {Mannings},
  {Prochaska}, {Ryder}, {Rafelski}, {Neeleman}, {Macquart}, {Tejos},
  {Jorgenson}, {Day}, {Marnoch}, {Bhandari}, {Deller}, {Qiu}, {Bannister},
  {Shannon}, \& {Heintz}}]{2021ApJ...922..173C}
{Chittidi}, J.~S., {Simha}, S., {Mannings}, A., {et~al.} 2021, \apj, 922, 173,
  \dodoi{10.3847/1538-4357/ac2818}

\bibitem[{{Cho} {et~al.}(2020){Cho}, {Macquart}, {Shannon}, {Deller},
  {Morrison}, {Ekers}, {Bannister}, {Farah}, {Qiu}, {Sammons}, {Bailes},
  {Bhandari}, {Day}, {James}, {Phillips}, {Prochaska}, \&
  {Tuthill}}]{2020ApJ...891L..38C}
{Cho}, H., {Macquart}, J.-P., {Shannon}, R.~M., {et~al.} 2020, \apjl, 891, L38,
  \dodoi{10.3847/2041-8213/ab7824}

\bibitem[{{Connor} {et~al.}(2023){Connor}, {Ravi}, {Catha}, {Chen}, {Faber},
  {Lamb}, {Hallinan}, {Harnach}, {Hellbourg}, {Hobbs}, {Hodge}, {Hodges},
  {Law}, {Rasmussen}, {Sayers}, {Sharma}, {Sherman}, {Shi}, {Simard},
  {Somalwar}, {Squillace}, {Weinreb}, {Woody}, {Yadlapalli}, \& {Deep Synoptic
  Array Team}}]{2023ApJ...949L..26C}
{Connor}, L., {Ravi}, V., {Catha}, M., {et~al.} 2023, \apjl, 949, L26,
  \dodoi{10.3847/2041-8213/acd3ea}

\bibitem[{{Cook} {et~al.}(2023){Cook}, {Bhardwaj}, {Gaensler}, {Scholz},
  {Eadie}, {Hill}, {Kaspi}, {Masui}, {Curtin}, {Dong}, {Fonseca},
  {Herrera-Martin}, {Kaczmarek}, {Lanman}, {Lazda}, {Leung}, {Meyers},
  {Michilli}, {Pandhi}, {Pearlman}, {Pleunis}, {Ransom}, {Rahman}, {Sand},
  {Shin}, {Smith}, {Stairs}, \& {Stenning}}]{2023ApJ...946...58C}
{Cook}, A.~M., {Bhardwaj}, M., {Gaensler}, B.~M., {et~al.} 2023, \apj, 946, 58,
  \dodoi{10.3847/1538-4357/acbbd0}

\bibitem[{{Cordes} \& {Chatterjee}(2019)}]{2019ARA&A..57..417C}
{Cordes}, J.~M., \& {Chatterjee}, S. 2019, \araa, 57, 417,
  \dodoi{10.1146/annurev-astro-091918-104501}

\bibitem[{{Cordes} \& {Lazio}(2002)}]{2002astro.ph..7156C}
{Cordes}, J.~M., \& {Lazio}, T.~J.~W. 2002, arXiv e-prints, astro.
\newblock \doarXiv{astro-ph/0207156}

\bibitem[{{Cordes} \& {Lazio}(2003)}]{2003astro.ph..1598C}
---. 2003, arXiv e-prints, astro, \dodoi{10.48550/arXiv.astro-ph/0301598}

\bibitem[{{Cordes} {et~al.}(2022){Cordes}, {Ocker}, \&
  {Chatterjee}}]{2022ApJ...931...88C}
{Cordes}, J.~M., {Ocker}, S.~K., \& {Chatterjee}, S. 2022, \apj, 931, 88,
  \dodoi{10.3847/1538-4357/ac6873}

\bibitem[{{Crain} \& {van de Voort}(2023)}]{2023ARA&A..61..473C}
{Crain}, R.~A., \& {van de Voort}, F. 2023, \araa, 61, 473,
  \dodoi{10.1146/annurev-astro-041923-043618}

\bibitem[{{Day} {et~al.}(2020){Day}, {Deller}, {Shannon}, {Qiu(邱昊)},
  {Bannister}, {Bhandari}, {Ekers}, {Flynn}, {James}, {Macquart}, {Mahony},
  {Phillips}, \& {Xavier Prochaska}}]{2020MNRAS.497.3335D}
{Day}, C.~K., {Deller}, A.~T., {Shannon}, R.~M., {et~al.} 2020, \mnras, 497,
  3335, \dodoi{10.1093/mnras/staa2138}

\bibitem[{{Deng} \& {Zhang}(2014)}]{2014ApJ...783L..35D}
{Deng}, W., \& {Zhang}, B. 2014, \apjl, 783, L35,
  \dodoi{10.1088/2041-8205/783/2/L35}

\bibitem[{{Dolag} {et~al.}(2015){Dolag}, {Gaensler}, {Beck}, \&
  {Beck}}]{2015MNRAS.451.4277D}
{Dolag}, K., {Gaensler}, B.~M., {Beck}, A.~M., \& {Beck}, M.~C. 2015, \mnras,
  451, 4277, \dodoi{10.1093/mnras/stv1190}

\bibitem[{{Donnari} {et~al.}(2021){Donnari}, {Pillepich}, {Nelson},
  {Marinacci}, {Vogelsberger}, \& {Hernquist}}]{2021MNRAS.506.4760D}
{Donnari}, M., {Pillepich}, A., {Nelson}, D., {et~al.} 2021, \mnras, 506, 4760,
  \dodoi{10.1093/mnras/stab1950}

\bibitem[{{Driessen} {et~al.}(2024){Driessen}, {Barr}, {Buckley}, {Caleb},
  {Chen}, {Chen}, {Gromadzki}, {Jankowski}, {Kraan-Korteweg}, {Palmerio},
  {Rajwade}, {Tremou}, {Kramer}, {Stappers}, {Vergani}, {Woudt},
  {Bezuidenhout}, {Malenta}, {Morello}, {Sanidas}, {Surnis}, \&
  {Fender}}]{2024MNRAS.527.3659D}
{Driessen}, L.~N., {Barr}, E.~D., {Buckley}, D.~A.~H., {et~al.} 2024, \mnras,
  527, 3659, \dodoi{10.1093/mnras/stad3329}

\bibitem[{{Faucher-Gigu{\`e}re} \& {Oh}(2023)}]{2023ARA&A..61..131F}
{Faucher-Gigu{\`e}re}, C.-A., \& {Oh}, S.~P. 2023, \araa, 61, 131,
  \dodoi{10.1146/annurev-astro-052920-125203}

\bibitem[{{Fong} {et~al.}(2021){Fong}, {Dong}, {Leja}, {Bhandari}, {Day},
  {Deller}, {Kumar}, {Prochaska}, {Scott}, {Bannister}, {Eftekhari}, {Gordon},
  {Heintz}, {James}, {Kilpatrick}, {Mahony}, {Rouco Escorial}, {Ryder},
  {Shannon}, \& {Tejos}}]{2021ApJ...919L..23F}
{Fong}, W.-f., {Dong}, Y., {Leja}, J., {et~al.} 2021, \apjl, 919, L23,
  \dodoi{10.3847/2041-8213/ac242b}

\bibitem[{Foreman-Mackey(2016)}]{corner}
Foreman-Mackey, D. 2016, The Journal of Open Source Software, 1, 24,
  \dodoi{10.21105/joss.00024}

\bibitem[{{Foreman-Mackey} {et~al.}(2013){Foreman-Mackey}, {Hogg}, {Lang}, \&
  {Goodman}}]{emcee}
{Foreman-Mackey}, D., {Hogg}, D.~W., {Lang}, D., \& {Goodman}, J. 2013, PASP,
  125, 306, \dodoi{10.1086/670067}

\bibitem[{{Glowacki} {et~al.}(2023){Glowacki}, {Lee-Waddell}, {Deller}, {Deg},
  {Gordon}, {Grundy}, {Marnoch}, {Shen}, {Ryder}, {Shannon}, {Wong},
  {D{\'e}nes}, {Koribalski}, {Murugeshan}, {Rhee}, {Westmeier}, {Bhandari},
  {Bosma}, {Holwerda}, \& {Prochaska}}]{2023ApJ...949...25G}
{Glowacki}, M., {Lee-Waddell}, K., {Deller}, A.~T., {et~al.} 2023, \apj, 949,
  25, \dodoi{10.3847/1538-4357/acc1e3}

\bibitem[{{Gordon} {et~al.}(2023{\natexlab{a}}){Gordon}, {Fong}, {Kilpatrick},
  {Eftekhari}, {Leja}, {Prochaska}, {Nugent}, {Bhandari}, {Blanchard}, {Caleb},
  {Day}, {Deller}, {Dong}, {Glowacki}, {Gourdji}, {Mannings}, {Mahoney},
  {Marnoch}, {Miller}, {Paterson}, {Rastinejad}, {Ryder}, {Sadler}, {Scott},
  {Sears}, {Shannon}, {Simha}, {Stappers}, \& {Tejos}}]{2023ApJ...954...80G}
{Gordon}, A.~C., {Fong}, W.-f., {Kilpatrick}, C.~D., {et~al.}
  2023{\natexlab{a}}, \apj, 954, 80, \dodoi{10.3847/1538-4357/ace5aa}

\bibitem[{{Gordon} {et~al.}(2023{\natexlab{b}}){Gordon}, {Fong}, {Simha},
  {Dong}, {Kilpatrick}, {Deller}, {Ryder}, {Eftekhari}, {Glowacki}, {Marnoch},
  {Muller}, {Nugent}, {Palmese}, {Prochaska}, {Rafelski}, {Shannon}, \&
  {Tejos}}]{2023arXiv231110815G}
{Gordon}, A.~C., {Fong}, W.-f., {Simha}, S., {et~al.} 2023{\natexlab{b}}, arXiv
  e-prints, arXiv:2311.10815, \dodoi{10.48550/arXiv.2311.10815}

\bibitem[{{Hackstein} {et~al.}(2020){Hackstein}, {Br{\"u}ggen}, {Vazza}, \&
  {Rodrigues}}]{2020MNRAS.498.4811H}
{Hackstein}, S., {Br{\"u}ggen}, M., {Vazza}, F., \& {Rodrigues}, L.~F.~S. 2020,
  \mnras, 498, 4811, \dodoi{10.1093/mnras/staa2572}

\bibitem[{{Hagstotz} {et~al.}(2022){Hagstotz}, {Reischke}, \&
  {Lilow}}]{2022MNRAS.511..662H}
{Hagstotz}, S., {Reischke}, R., \& {Lilow}, R. 2022, \mnras, 511, 662,
  \dodoi{10.1093/mnras/stac077}

\bibitem[{{Hashimoto} {et~al.}(2022){Hashimoto}, {Goto}, {Chen}, {Ho}, {Hsiao},
  {Wong}, {On}, {Kim}, {Kilerci-Eser}, {Huang}, {Santos}, \&
  {Yamasaki}}]{2022MNRAS.511.1961H}
{Hashimoto}, T., {Goto}, T., {Chen}, B.~H., {et~al.} 2022, \mnras, 511, 1961,
  \dodoi{10.1093/mnras/stac065}

\bibitem[{{Heintz} {et~al.}(2020){Heintz}, {Prochaska}, {Simha}, {Platts},
  {Fong}, {Tejos}, {Ryder}, {Aggerwal}, {Bhandari}, {Day}, {Deller},
  {Kilpatrick}, {Law}, {Macquart}, {Mannings}, {Marnoch}, {Sadler}, \&
  {Shannon}}]{2020ApJ...903..152H}
{Heintz}, K.~E., {Prochaska}, J.~X., {Simha}, S., {et~al.} 2020, \apj, 903,
  152, \dodoi{10.3847/1538-4357/abb6fb}

\bibitem[{{Hogg} \& {Foreman-Mackey}(2018)}]{2018ApJS..236...11H}
{Hogg}, D.~W., \& {Foreman-Mackey}, D. 2018, \apjs, 236, 11,
  \dodoi{10.3847/1538-4365/aab76e}

\bibitem[{{Ibik} {et~al.}(2023){Ibik}, {Drout}, {Gaensler}, {Scholz},
  {Michilli}, {Bhardwaj}, {Kaspi}, {Pleunis}, {Cassanelli}, {Cook}, {Dong},
  {Leung}, {Masui}, {Kaczmarek}, {Lu}, {Pearlman}, {Rafiei-Ravandi}, {Sand},
  {Shin}, {Smith}, \& {Stairs}}]{2023arXiv230402638I}
{Ibik}, A.~L., {Drout}, M.~R., {Gaensler}, B.~M., {et~al.} 2023, arXiv
  e-prints, arXiv:2304.02638, \dodoi{10.48550/arXiv.2304.02638}

\bibitem[{{Ioka}(2003)}]{2003ApJ...598L..79I}
{Ioka}, K. 2003, \apjl, 598, L79, \dodoi{10.1086/380598}

\bibitem[{{James} {et~al.}(2022{\natexlab{a}}){James}, {Prochaska}, {Macquart},
  {North-Hickey}, {Bannister}, \& {Dunning}}]{2022MNRAS.509.4775J}
{James}, C.~W., {Prochaska}, J.~X., {Macquart}, J.~P., {et~al.}
  2022{\natexlab{a}}, \mnras, 509, 4775, \dodoi{10.1093/mnras/stab3051}

\bibitem[{{James} {et~al.}(2022{\natexlab{b}}){James}, {Prochaska}, {Macquart},
  {North-Hickey}, {Bannister}, \& {Dunning}}]{2022MNRAS.510L..18J}
---. 2022{\natexlab{b}}, \mnras, 510, L18, \dodoi{10.1093/mnrasl/slab117}

\bibitem[{{James} {et~al.}(2022{\natexlab{c}}){James}, {Ghosh}, {Prochaska},
  {Bannister}, {Bhandari}, {Day}, {Deller}, {Glowacki}, {Gordon}, {Heintz},
  {Marnoch}, {Ryder}, {Scott}, {Shannon}, \& {Tejos}}]{2022MNRAS.516.4862J}
{James}, C.~W., {Ghosh}, E.~M., {Prochaska}, J.~X., {et~al.}
  2022{\natexlab{c}}, \mnras, 516, 4862, \dodoi{10.1093/mnras/stac2524}

\bibitem[{{Jaroszynski}(2019)}]{2019MNRAS.484.1637J}
{Jaroszynski}, M. 2019, \mnras, 484, 1637, \dodoi{10.1093/mnras/sty3529}

\bibitem[{{Jaroszy{\'n}ski}(2020)}]{2020AcA....70...87J}
{Jaroszy{\'n}ski}, M. 2020, \actaa, 70, 87, \dodoi{10.32023/0001-5237/70.2.1}

\bibitem[{{Karim} {et~al.}(2011){Karim}, {Schinnerer},
  {Mart{\'\i}nez-Sansigre}, {Sargent}, {van der Wel}, {Rix}, {Ilbert},
  {Smol{\v{c}}i{\'c}}, {Carilli}, {Pannella}, {Koekemoer}, {Bell}, \&
  {Salvato}}]{2011ApJ...730...61K}
{Karim}, A., {Schinnerer}, E., {Mart{\'\i}nez-Sansigre}, A., {et~al.} 2011,
  \apj, 730, 61, \dodoi{10.1088/0004-637X/730/2/61}

\bibitem[{{Kirsten} {et~al.}(2022){Kirsten}, {Marcote}, {Nimmo}, {Hessels},
  {Bhardwaj}, {Tendulkar}, {Keimpema}, {Yang}, {Snelders}, {Scholz},
  {Pearlman}, {Law}, {Peters}, {Giroletti}, {Paragi}, {Bassa}, {Hewitt},
  {Bach}, {Bezrukovs}, {Burgay}, {Buttaccio}, {Conway}, {Corongiu}, {Feiler},
  {Forss{\'e}n}, {Gawro{\'n}ski}, {Karuppusamy}, {Kharinov}, {Lindqvist},
  {Maccaferri}, {Melnikov}, {Ould-Boukattine}, {Possenti}, {Surcis}, {Wang},
  {Yuan}, {Aggarwal}, {Anna-Thomas}, {Bower}, {Blaauw}, {Burke-Spolaor},
  {Cassanelli}, {Clarke}, {Fonseca}, {Gaensler}, {Gopinath}, {Kaspi}, {Kassim},
  {Lazio}, {Leung}, {Li}, {Lin}, {Masui}, {Mckinven}, {Michilli}, {Mikhailov},
  {Ng}, {Orbidans}, {Pen}, {Petroff}, {Rahman}, {Ransom}, {Shin}, {Smith},
  {Stairs}, \& {Vlemmings}}]{2022Natur.602..585K}
{Kirsten}, F., {Marcote}, B., {Nimmo}, K., {et~al.} 2022, \nat, 602, 585,
  \dodoi{10.1038/s41586-021-04354-w}

\bibitem[{{Law} {et~al.}(2020){Law}, {Butler}, {Prochaska}, {Zackay},
  {Burke-Spolaor}, {Mannings}, {Tejos}, {Josephy}, {Andersen}, {Chawla},
  {Heintz}, {Aggarwal}, {Bower}, {Demorest}, {Kilpatrick}, {Lazio}, {Linford},
  {Mckinven}, {Tendulkar}, \& {Simha}}]{2020ApJ...899..161L}
{Law}, C.~J., {Butler}, B.~J., {Prochaska}, J.~X., {et~al.} 2020, \apj, 899,
  161, \dodoi{10.3847/1538-4357/aba4ac}

\bibitem[{{Law} {et~al.}(2023){Law}, {Sharma}, {Ravi}, {Chen}, {Catha},
  {Connor}, {Faber}, {Hallinan}, {Harnach}, {Hellbourg}, {Hobbs}, {Hodge},
  {Hodges}, {Lamb}, {Rasmussen}, {Sherman}, {Shi}, {Simard}, {Squillace},
  {Weinreb}, {Woody}, \& {Yadlapalli}}]{2023arXiv230703344L}
{Law}, C.~J., {Sharma}, K., {Ravi}, V., {et~al.} 2023, arXiv e-prints,
  arXiv:2307.03344, \dodoi{10.48550/arXiv.2307.03344}

\bibitem[{{Lee} {et~al.}(2023){Lee}, {Khrykin}, {Simha}, {Ata}, {Huang},
  {Prochaska}, {Tejos}, {Cooke}, {Nagamine}, \& {Zhang}}]{2023arXiv230605403L}
{Lee}, K.-G., {Khrykin}, I.~S., {Simha}, S., {et~al.} 2023, arXiv e-prints,
  arXiv:2306.05403, \dodoi{10.48550/arXiv.2306.05403}

\bibitem[{{Li} {et~al.}(2021){Li}, {Wang}, {Zhu}, {Zhang}, {Zhang}, {Duan},
  {Zhang}, {Feng}, {Tang}, {Chatterjee}, {Cordes}, {Cruces}, {Dai}, {Gajjar},
  {Hobbs}, {Jin}, {Kramer}, {Lorimer}, {Miao}, {Niu}, {Niu}, {Pan}, {Qian},
  {Spitler}, {Werthimer}, {Zhang}, {Wang}, {Xie}, {Yue}, {Zhang}, {Zhi}, \&
  {Zhu}}]{2021Natur.598..267L}
{Li}, D., {Wang}, P., {Zhu}, W.~W., {et~al.} 2021, \nat, 598, 267,
  \dodoi{10.1038/s41586-021-03878-5}

\bibitem[{{Lorimer} {et~al.}(2007){Lorimer}, {Bailes}, {McLaughlin},
  {Narkevic}, \& {Crawford}}]{2007Sci...318..777L}
{Lorimer}, D.~R., {Bailes}, M., {McLaughlin}, M.~A., {Narkevic}, D.~J., \&
  {Crawford}, F. 2007, Science, 318, 777, \dodoi{10.1126/science.1147532}

\bibitem[{{Luo} {et~al.}(2018){Luo}, {Lee}, {Lorimer}, \&
  {Zhang}}]{2018MNRAS.481.2320L}
{Luo}, R., {Lee}, K., {Lorimer}, D.~R., \& {Zhang}, B. 2018, \mnras, 481, 2320,
  \dodoi{10.1093/mnras/sty2364}

\bibitem[{{Luo} {et~al.}(2020){Luo}, {Men}, {Lee}, {Wang}, {Lorimer}, \&
  {Zhang}}]{2020MNRAS.494..665L}
{Luo}, R., {Men}, Y., {Lee}, K., {et~al.} 2020, \mnras, 494, 665,
  \dodoi{10.1093/mnras/staa704}

\bibitem[{{Macquart} \& {Ekers}(2018)}]{2018MNRAS.480.4211M}
{Macquart}, J.~P., \& {Ekers}, R. 2018, \mnras, 480, 4211,
  \dodoi{10.1093/mnras/sty2083}

\bibitem[{{Macquart} \& {Koay}(2013)}]{2013ApJ...776..125M}
{Macquart}, J.-P., \& {Koay}, J.~Y. 2013, \apj, 776, 125,
  \dodoi{10.1088/0004-637X/776/2/125}

\bibitem[{{Macquart} {et~al.}(2019){Macquart}, {Shannon}, {Bannister}, {James},
  {Ekers}, \& {Bunton}}]{2019ApJ...872L..19M}
{Macquart}, J.~P., {Shannon}, R.~M., {Bannister}, K.~W., {et~al.} 2019, \apjl,
  872, L19, \dodoi{10.3847/2041-8213/ab03d6}

\bibitem[{{Macquart} {et~al.}(2020){Macquart}, {Prochaska}, {McQuinn},
  {Bannister}, {Bhandari}, {Day}, {Deller}, {Ekers}, {James}, {Marnoch},
  {Os{\l}owski}, {Phillips}, {Ryder}, {Scott}, {Shannon}, \&
  {Tejos}}]{2020Natur.581..391M}
{Macquart}, J.~P., {Prochaska}, J.~X., {McQuinn}, M., {et~al.} 2020, \nat, 581,
  391, \dodoi{10.1038/s41586-020-2300-2}

\bibitem[{{Madau} \& {Dickinson}(2014)}]{2014ARA&A..52..415M}
{Madau}, P., \& {Dickinson}, M. 2014, \araa, 52, 415,
  \dodoi{10.1146/annurev-astro-081811-125615}

\bibitem[{{Mahony} {et~al.}(2018){Mahony}, {Ekers}, {Macquart}, {Sadler},
  {Bannister}, {Bhandari}, {Flynn}, {Koribalski}, {Prochaska}, {Ryder},
  {Shannon}, {Tejos}, {Whiting}, \& {Wong}}]{2018ApJ...867L..10M}
{Mahony}, E.~K., {Ekers}, R.~D., {Macquart}, J.-P., {et~al.} 2018, \apjl, 867,
  L10, \dodoi{10.3847/2041-8213/aae7cb}

\bibitem[{{Maragkoudakis} {et~al.}(2017){Maragkoudakis}, {Zezas}, {Ashby}, \&
  {Willner}}]{2017MNRAS.466.1192M}
{Maragkoudakis}, A., {Zezas}, A., {Ashby}, M.~L.~N., \& {Willner}, S.~P. 2017,
  \mnras, 466, 1192, \dodoi{10.1093/mnras/stw3180}

\bibitem[{{Marcote} {et~al.}(2020){Marcote}, {Nimmo}, {Hessels}, {Tendulkar},
  {Bassa}, {Paragi}, {Keimpema}, {Bhardwaj}, {Karuppusamy}, {Kaspi}, {Law},
  {Michilli}, {Aggarwal}, {Andersen}, {Archibald}, {Bandura}, {Bower}, {Boyle},
  {Brar}, {Burke-Spolaor}, {Butler}, {Cassanelli}, {Chawla}, {Demorest},
  {Dobbs}, {Fonseca}, {Giri}, {Good}, {Gourdji}, {Josephy}, {Kirichenko},
  {Kirsten}, {Landecker}, {Lang}, {Lazio}, {Li}, {Lin}, {Linford}, {Masui},
  {Mena-Parra}, {Naidu}, {Ng}, {Patel}, {Pen}, {Pleunis}, {Rafiei-Ravandi},
  {Rahman}, {Renard}, {Scholz}, {Siegel}, {Smith}, {Stairs}, {Vanderlinde}, \&
  {Zwaniga}}]{2020Natur.577..190M}
{Marcote}, B., {Nimmo}, K., {Hessels}, J.~W.~T., {et~al.} 2020, \nat, 577, 190,
  \dodoi{10.1038/s41586-019-1866-z}

\bibitem[{{Margalit} {et~al.}(2019){Margalit}, {Berger}, \&
  {Metzger}}]{2019ApJ...886..110M}
{Margalit}, B., {Berger}, E., \& {Metzger}, B.~D. 2019, \apj, 886, 110,
  \dodoi{10.3847/1538-4357/ab4c31}

\bibitem[{{McGregor} \& {Lorimer}(2023)}]{2023arXiv230911522M}
{McGregor}, K., \& {Lorimer}, D.~R. 2023, arXiv e-prints, arXiv:2309.11522,
  \dodoi{10.48550/arXiv.2309.11522}

\bibitem[{{McQuinn}(2014)}]{2014ApJ...780L..33M}
{McQuinn}, M. 2014, \apjl, 780, L33, \dodoi{10.1088/2041-8205/780/2/L33}

\bibitem[{{Michilli} {et~al.}(2023){Michilli}, {Bhardwaj}, {Brar}, {Gaensler},
  {Kaspi}, {Kirichenko}, {Masui}, {Mckinven}, {Ng}, {Patel}, {Sand}, {Scholz},
  {Shin}, {Siegel}, {Stairs}, {Cassanelli}, {Cook}, {Dobbs}, {Dong}, {Fonseca},
  {Ibik}, {Kaczmarek}, {Leung}, {Pearlman}, {Petroff}, {Pleunis},
  {Rafiei-Ravandi}, {Sanghavi}, {Shaw}, \& {Tendulkar}}]{2023ApJ...950..134M}
{Michilli}, D., {Bhardwaj}, M., {Brar}, C., {et~al.} 2023, \apj, 950, 134,
  \dodoi{10.3847/1538-4357/accf89}

\bibitem[{{Mo} {et~al.}(2023){Mo}, {Zhu}, {Wang}, {Tang}, \&
  {Feng}}]{2023MNRAS.518..539M}
{Mo}, J.-F., {Zhu}, W., {Wang}, Y., {Tang}, L., \& {Feng}, L.-L. 2023, \mnras,
  518, 539, \dodoi{10.1093/mnras/stac3104}

\bibitem[{{Murray}(2014)}]{2014ascl.soft12006M}
{Murray}, S. 2014, {HMF: Halo Mass Function calculator}, Astrophysics Source
  Code Library, record ascl:1412.006.
\newblock \doeprint{1412.006}

\bibitem[{{Murray} {et~al.}(2013){Murray}, {Power}, \&
  {Robotham}}]{2013A&C.....3...23M}
{Murray}, S.~G., {Power}, C., \& {Robotham}, A.~S.~G. 2013, Astronomy and
  Computing, 3, 23, \dodoi{10.1016/j.ascom.2013.11.001}

\bibitem[{{Nelson} {et~al.}(2018){Nelson}, {Pillepich}, {Springel},
  {Weinberger}, {Hernquist}, {Pakmor}, {Genel}, {Torrey}, {Vogelsberger},
  {Kauffmann}, {Marinacci}, \& {Naiman}}]{2018MNRAS.475..624N}
{Nelson}, D., {Pillepich}, A., {Springel}, V., {et~al.} 2018, \mnras, 475, 624,
  \dodoi{10.1093/mnras/stx3040}

\bibitem[{{Niu} {et~al.}(2021){Niu}, {Li}, {Luo}, {Wang}, {Yao}, {Zhang},
  {Zhu}, {Wang}, {Ye}, {Zhang}, {Niu}, {Tang}, {Duan}, {Krco}, {Dai}, {Feng},
  {Miao}, {Pan}, {Qian}, {Xue}, {Yuan}, {Yue}, {Zhang}, \&
  {Zhang}}]{2021ApJ...909L...8N}
{Niu}, C.-H., {Li}, D., {Luo}, R., {et~al.} 2021, \apjl, 909, L8,
  \dodoi{10.3847/2041-8213/abe7f0}

\bibitem[{{Niu} {et~al.}(2022){Niu}, {Aggarwal}, {Li}, {Zhang}, {Chatterjee},
  {Tsai}, {Yu}, {Law}, {Burke-Spolaor}, {Cordes}, {Zhang}, {Ocker}, {Yao},
  {Wang}, {Feng}, {Niino}, {Bochenek}, {Cruces}, {Connor}, {Jiang}, {Dai},
  {Luo}, {Li}, {Miao}, {Niu}, {Anna-Thomas}, {Sydnor}, {Stern}, {Wang}, {Yuan},
  {Yue}, {Zhou}, {Yan}, {Zhu}, \& {Zhang}}]{2022Natur.606..873N}
{Niu}, C.~H., {Aggarwal}, K., {Li}, D., {et~al.} 2022, \nat, 606, 873,
  \dodoi{10.1038/s41586-022-04755-5}

\bibitem[{{Ocker} {et~al.}(2022{\natexlab{a}}){Ocker}, {Cordes}, {Chatterjee},
  \& {Gorsuch}}]{2022ApJ...934...71O}
{Ocker}, S.~K., {Cordes}, J.~M., {Chatterjee}, S., \& {Gorsuch}, M.~R.
  2022{\natexlab{a}}, \apj, 934, 71, \dodoi{10.3847/1538-4357/ac75ba}

\bibitem[{{Ocker} {et~al.}(2022{\natexlab{b}}){Ocker}, {Cordes}, {Chatterjee},
  {Niu}, {Li}, {McKee}, {Law}, {Tsai}, {Anna-Thomas}, {Yao}, \&
  {Cruces}}]{2022ApJ...931...87O}
{Ocker}, S.~K., {Cordes}, J.~M., {Chatterjee}, S., {et~al.} 2022{\natexlab{b}},
  \apj, 931, 87, \dodoi{10.3847/1538-4357/ac6504}

\bibitem[{Padmanabhan(2000)}]{padmanabhan2000theoretical}
Padmanabhan, T. 2000, Theoretical astrophysics: volume 3, galaxies and
  cosmology, Vol.~3 (Cambridge University Press)

\bibitem[{{Petroff} {et~al.}(2019){Petroff}, {Hessels}, \&
  {Lorimer}}]{2019A&ARv..27....4P}
{Petroff}, E., {Hessels}, J.~W.~T., \& {Lorimer}, D.~R. 2019, \aapr, 27, 4,
  \dodoi{10.1007/s00159-019-0116-6}

\bibitem[{{Pillepich} {et~al.}(2018){Pillepich}, {Nelson}, {Hernquist},
  {Springel}, {Pakmor}, {Torrey}, {Weinberger}, {Genel}, {Naiman}, {Marinacci},
  \& {Vogelsberger}}]{2018MNRAS.475..648P}
{Pillepich}, A., {Nelson}, D., {Hernquist}, L., {et~al.} 2018, \mnras, 475,
  648, \dodoi{10.1093/mnras/stx3112}

\bibitem[{{Planck Collaboration} {et~al.}(2016){Planck Collaboration}, {Ade},
  {Aghanim}, {Arnaud}, {Ashdown}, {Aumont}, {Baccigalupi}, {Banday},
  {Barreiro}, {Bartlett}, {Bartolo}, {Battaner}, {Battye}, {Benabed},
  {Beno{\^\i}t}, {Benoit-L{\'e}vy}, {Bernard}, {Bersanelli}, {Bielewicz},
  {Bock}, {Bonaldi}, {Bonavera}, {Bond}, {Borrill}, {Bouchet}, {Boulanger},
  {Bucher}, {Burigana}, {Butler}, {Calabrese}, {Cardoso}, {Catalano},
  {Challinor}, {Chamballu}, {Chary}, {Chiang}, {Chluba}, {Christensen},
  {Church}, {Clements}, {Colombi}, {Colombo}, {Combet}, {Coulais}, {Crill},
  {Curto}, {Cuttaia}, {Danese}, {Davies}, {Davis}, {de Bernardis}, {de Rosa},
  {de Zotti}, {Delabrouille}, {D{\'e}sert}, {Di Valentino}, {Dickinson},
  {Diego}, {Dolag}, {Dole}, {Donzelli}, {Dor{\'e}}, {Douspis}, {Ducout},
  {Dunkley}, {Dupac}, {Efstathiou}, {Elsner}, {En{\ss}lin}, {Eriksen},
  {Farhang}, {Fergusson}, {Finelli}, {Forni}, {Frailis}, {Fraisse},
  {Franceschi}, {Frejsel}, {Galeotta}, {Galli}, {Ganga}, {Gauthier}, {Gerbino},
  {Ghosh}, {Giard}, {Giraud-H{\'e}raud}, {Giusarma}, {Gjerl{\o}w},
  {Gonz{\'a}lez-Nuevo}, {G{\'o}rski}, {Gratton}, {Gregorio}, {Gruppuso},
  {Gudmundsson}, {Hamann}, {Hansen}, {Hanson}, {Harrison}, {Helou},
  {Henrot-Versill{\'e}}, {Hern{\'a}ndez-Monteagudo}, {Herranz}, {Hildebrandt},
  {Hivon}, {Hobson}, {Holmes}, {Hornstrup}, {Hovest}, {Huang}, {Huffenberger},
  {Hurier}, {Jaffe}, {Jaffe}, {Jones}, {Juvela}, {Keih{\"a}nen}, {Keskitalo},
  {Kisner}, {Kneissl}, {Knoche}, {Knox}, {Kunz}, {Kurki-Suonio}, {Lagache},
  {L{\"a}hteenm{\"a}ki}, {Lamarre}, {Lasenby}, {Lattanzi}, {Lawrence}, {Leahy},
  {Leonardi}, {Lesgourgues}, {Levrier}, {Lewis}, {Liguori}, {Lilje},
  {Linden-V{\o}rnle}, {L{\'o}pez-Caniego}, {Lubin}, {Mac{\'\i}as-P{\'e}rez},
  {Maggio}, {Maino}, {Mandolesi}, {Mangilli}, {Marchini}, {Maris}, {Martin},
  {Martinelli}, {Mart{\'\i}nez-Gonz{\'a}lez}, {Masi}, {Matarrese}, {McGehee},
  {Meinhold}, {Melchiorri}, {Melin}, {Mendes}, {Mennella}, {Migliaccio},
  {Millea}, {Mitra}, {Miville-Desch{\^e}nes}, {Moneti}, {Montier}, {Morgante},
  {Mortlock}, {Moss}, {Munshi}, {Murphy}, {Naselsky}, {Nati}, {Natoli},
  {Netterfield}, {N{\o}rgaard-Nielsen}, {Noviello}, {Novikov}, {Novikov},
  {Oxborrow}, {Paci}, {Pagano}, {Pajot}, {Paladini}, {Paoletti}, {Partridge},
  {Pasian}, {Patanchon}, {Pearson}, {Perdereau}, {Perotto}, {Perrotta},
  {Pettorino}, {Piacentini}, {Piat}, {Pierpaoli}, {Pietrobon}, {Plaszczynski},
  {Pointecouteau}, {Polenta}, {Popa}, {Pratt}, {Pr{\'e}zeau}, {Prunet},
  {Puget}, {Rachen}, {Reach}, {Rebolo}, {Reinecke}, {Remazeilles}, {Renault},
  {Renzi}, {Ristorcelli}, {Rocha}, {Rosset}, {Rossetti}, {Roudier},
  {Rouill{\'e} d'Orfeuil}, {Rowan-Robinson}, {Rubi{\~n}o-Mart{\'\i}n},
  {Rusholme}, {Said}, {Salvatelli}, {Salvati}, {Sandri}, {Santos},
  {Savelainen}, {Savini}, {Scott}, {Seiffert}, {Serra}, {Shellard}, {Spencer},
  {Spinelli}, {Stolyarov}, {Stompor}, {Sudiwala}, {Sunyaev}, {Sutton},
  {Suur-Uski}, {Sygnet}, {Tauber}, {Terenzi}, {Toffolatti}, {Tomasi},
  {Tristram}, {Trombetti}, {Tucci}, {Tuovinen}, {T{\"u}rler}, {Umana},
  {Valenziano}, {Valiviita}, {Van Tent}, {Vielva}, {Villa}, {Wade}, {Wandelt},
  {Wehus}, {White}, {White}, {Wilkinson}, {Yvon}, {Zacchei}, \&
  {Zonca}}]{2016A&A...594A..13P}
{Planck Collaboration}, {Ade}, P.~A.~R., {Aghanim}, N., {et~al.} 2016, \aap,
  594, A13, \dodoi{10.1051/0004-6361/201525830}

\bibitem[{{Platts} {et~al.}(2019){Platts}, {Weltman}, {Walters}, {Tendulkar},
  {Gordin}, \& {Kandhai}}]{2019PhR...821....1P}
{Platts}, E., {Weltman}, A., {Walters}, A., {et~al.} 2019, \physrep, 821, 1,
  \dodoi{10.1016/j.physrep.2019.06.003}

\bibitem[{{Pol} {et~al.}(2019){Pol}, {Lam}, {McLaughlin}, {Lazio}, \&
  {Cordes}}]{2019ApJ...886..135P}
{Pol}, N., {Lam}, M.~T., {McLaughlin}, M.~A., {Lazio}, T.~J.~W., \& {Cordes},
  J.~M. 2019, \apj, 886, 135, \dodoi{10.3847/1538-4357/ab4c2f}

\bibitem[{{Price} {et~al.}(2019){Price}, {Foster}, {Geyer}, {van Straten},
  {Gajjar}, {Hellbourg}, {Karastergiou}, {Keane}, {Siemion}, {Arcavi}, {Bhat},
  {Caleb}, {Chang}, {Croft}, {DeBoer}, {de Pater}, {Drew}, {Enriquez}, {Farah},
  {Gizani}, {Green}, {Isaacson}, {Hickish}, {Jameson}, {Lebofsky}, {MacMahon},
  {M{\"o}ller}, {Onken}, {Petroff}, {Werthimer}, {Wolf}, {Worden}, \&
  {Zhang}}]{2019MNRAS.486.3636P}
{Price}, D.~C., {Foster}, G., {Geyer}, M., {et~al.} 2019, \mnras, 486, 3636,
  \dodoi{10.1093/mnras/stz958}

\bibitem[{{Prochaska} \& {Zheng}(2019)}]{2019MNRAS.485..648P}
{Prochaska}, J.~X., \& {Zheng}, Y. 2019, \mnras, 485, 648,
  \dodoi{10.1093/mnras/stz261}

\bibitem[{{Prochaska} {et~al.}(2019){Prochaska}, {Macquart}, {McQuinn},
  {Simha}, {Shannon}, {Day}, {Marnoch}, {Ryder}, {Deller}, {Bannister},
  {Bhandari}, {Bordoloi}, {Bunton}, {Cho}, {Flynn}, {Mahony}, {Phillips},
  {Qiu}, \& {Tejos}}]{2019Sci...366..231P}
{Prochaska}, J.~X., {Macquart}, J.-P., {McQuinn}, M., {et~al.} 2019, Science,
  366, 231, \dodoi{10.1126/science.aay0073}

\bibitem[{{Qiang} {et~al.}(2022){Qiang}, {Li}, \& {Wei}}]{2022JCAP...01..040Q}
{Qiang}, D.-C., {Li}, S.-L., \& {Wei}, H. 2022, \jcap, 2022, 040,
  \dodoi{10.1088/1475-7516/2022/01/040}

\bibitem[{{Qiu} {et~al.}(2020){Qiu}, {Shannon}, {Farah}, {Macquart}, {Deller},
  {Bannister}, {James}, {Flynn}, {Day}, {Bhandari}, \&
  {Murphy}}]{2020MNRAS.497.1382Q}
{Qiu}, H., {Shannon}, R.~M., {Farah}, W., {et~al.} 2020, \mnras, 497, 1382,
  \dodoi{10.1093/mnras/staa1916}

\bibitem[{{Ravi} {et~al.}(2019){Ravi}, {Catha}, {D'Addario}, {Djorgovski},
  {Hallinan}, {Hobbs}, {Kocz}, {Kulkarni}, {Shi}, {Vedantham}, {Weinreb}, \&
  {Woody}}]{2019Natur.572..352R}
{Ravi}, V., {Catha}, M., {D'Addario}, L., {et~al.} 2019, \nat, 572, 352,
  \dodoi{10.1038/s41586-019-1389-7}

\bibitem[{{Ravi} {et~al.}(2023{\natexlab{a}}){Ravi}, {Catha}, {Chen}, {Connor},
  {Cordes}, {Faber}, {Lamb}, {Hallinan}, {Harnach}, {Hellbourg}, {Hobbs},
  {Hodge}, {Hodges}, {Law}, {Rasmussen}, {Sharma}, {Sherman}, {Shi}, {Simard},
  {Somalwar}, {Squillace}, {Weinreb}, {Woody}, \&
  {Yadlapalli}}]{2023arXiv230101000R}
{Ravi}, V., {Catha}, M., {Chen}, G., {et~al.} 2023{\natexlab{a}}, arXiv
  e-prints, arXiv:2301.01000, \dodoi{10.48550/arXiv.2301.01000}

\bibitem[{{Ravi} {et~al.}(2023{\natexlab{b}}){Ravi}, {Catha}, {Chen}, {Connor},
  {Faber}, {Lamb}, {Hallinan}, {Harnach}, {Hellbourg}, {Hobbs}, {Hodge},
  {Hodges}, {Law}, {Rasmussen}, {Sharma}, {Sherman}, {Shi}, {Simard},
  {Squillace}, {Weinreb}, {Woody}, {Yadlapalli}, {Ahumada}, {Dong}, {Fremling},
  {Huang}, {Karambelkar}, \& {Miller}}]{2023ApJ...949L...3R}
---. 2023{\natexlab{b}}, \apjl, 949, L3, \dodoi{10.3847/2041-8213/acc4b6}

\bibitem[{{Renzini} \& {Peng}(2015)}]{2015ApJ...801L..29R}
{Renzini}, A., \& {Peng}, Y.-j. 2015, \apjl, 801, L29,
  \dodoi{10.1088/2041-8205/801/2/L29}

\bibitem[{{Rodriguez-Gomez} {et~al.}(2019){Rodriguez-Gomez}, {Snyder}, {Lotz},
  {Nelson}, {Pillepich}, {Springel}, {Genel}, {Weinberger}, {Tacchella},
  {Pakmor}, {Torrey}, {Marinacci}, {Vogelsberger}, {Hernquist}, \&
  {Thilker}}]{2019MNRAS.483.4140R}
{Rodriguez-Gomez}, V., {Snyder}, G.~F., {Lotz}, J.~M., {et~al.} 2019, \mnras,
  483, 4140, \dodoi{10.1093/mnras/sty3345}

\bibitem[{{Ryder} {et~al.}(2023){Ryder}, {Bannister}, {Bhandari}, {Deller},
  {Ekers}, {Glowacki}, {Gordon}, {Gourdji}, {James}, {Kilpatrick}, {Lu},
  {Marnoch}, {Moss}, {Prochaska}, {Qiu}, {Sadler}, {Simha}, {Sammons}, {Scott},
  {Tejos}, \& {Shannon}}]{2023Sci...382..294R}
{Ryder}, S.~D., {Bannister}, K.~W., {Bhandari}, S., {et~al.} 2023, Science,
  382, 294, \dodoi{10.1126/science.adf2678}

\bibitem[{{Schechter}(1976)}]{1976ApJ...203..297S}
{Schechter}, P. 1976, \apj, 203, 297, \dodoi{10.1086/154079}

\bibitem[{{Shannon} {et~al.}(2018){Shannon}, {Macquart}, {Bannister}, {Ekers},
  {James}, {Os{\l}owski}, {Qiu}, {Sammons}, {Hotan}, {Voronkov}, {Beresford},
  {Brothers}, {Brown}, {Bunton}, {Chippendale}, {Haskins}, {Leach},
  {Marquarding}, {McConnell}, {Pilawa}, {Sadler}, {Troup}, {Tuthill},
  {Whiting}, {Allison}, {Anderson}, {Bell}, {Collier}, {G{\"u}rkan}, {Heald},
  \& {Riseley}}]{2018Natur.562..386S}
{Shannon}, R.~M., {Macquart}, J.~P., {Bannister}, K.~W., {et~al.} 2018, \nat,
  562, 386, \dodoi{10.1038/s41586-018-0588-y}

\bibitem[{{Sharma} {et~al.}(2023){Sharma}, {Somalwar}, {Law}, {Ravi}, {Catha},
  {Chen}, {Connor}, {Faber}, {Hallinan}, {Harnach}, {Hellbourg}, {Hobbs},
  {Hodge}, {Hodges}, {Lamb}, {Rasmussen}, {Sherman}, {Shi}, {Simard},
  {Squillace}, {Weinreb}, {Woody}, {Yadlapalli}, \& {Deep Synoptic Array
  Team}}]{2023ApJ...950..175S}
{Sharma}, K., {Somalwar}, J., {Law}, C., {et~al.} 2023, \apj, 950, 175,
  \dodoi{10.3847/1538-4357/accf1d}

\bibitem[{{Sharma} {et~al.}(2024){Sharma}, {Ravi}, {Connor}, {Law}, {Ocker},
  {Sherman}, {Kosogorov}, {Faber}, {Hallinan}, {Harnach}, {Hellbourg}, {Hobbs},
  {Hodge}, {Hodges}, {Lamb}, {Rasmussen}, {Somalwar}, {Weinreb}, {Woody},
  {Leja}, {Anand}, {Das}, {Qin}, {Rose}, {Dong}, {Miller}, \&
  {Yao}}]{2024Natur.635...61S}
{Sharma}, K., {Ravi}, V., {Connor}, L., {et~al.} 2024, \nat, 635, 61,
  \dodoi{10.1038/s41586-024-08074-9}

\bibitem[{{Sheikh} {et~al.}(2024){Sheikh}, {Farah}, {Pollak}, {Siemion},
  {Chamma}, {Cruz}, {Davis}, {DeBoer}, {Gajjar}, {Karn}, {Kittling}, {Lu},
  {Masters}, {Premnath}, {Schoultz}, {Shumaker}, {Singh}, \&
  {Snodgrass}}]{2024MNRAS.52710425S}
{Sheikh}, S.~Z., {Farah}, W., {Pollak}, A.~W., {et~al.} 2024, \mnras, 527,
  10425, \dodoi{10.1093/mnras/stad3630}

\bibitem[{{Shin} {et~al.}(2023){Shin}, {Masui}, {Bhardwaj}, {Cassanelli},
  {Chawla}, {Dobbs}, {Dong}, {Fonseca}, {Gaensler}, {Herrera-Mart{\'\i}n},
  {Kaczmarek}, {Kaspi}, {Leung}, {Merryfield}, {Michilli}, {M{\"u}nchmeyer},
  {Pearlman}, {Rafiei-Ravandi}, {Smith}, {Stairs}, \&
  {Tendulkar}}]{2023ApJ...944..105S}
{Shin}, K., {Masui}, K.~W., {Bhardwaj}, M., {et~al.} 2023, \apj, 944, 105,
  \dodoi{10.3847/1538-4357/acaf06}

\bibitem[{{Spitler} {et~al.}(2014){Spitler}, {Cordes}, {Hessels}, {Lorimer},
  {McLaughlin}, {Chatterjee}, {Crawford}, {Deneva}, {Kaspi}, {Wharton},
  {Allen}, {Bogdanov}, {Brazier}, {Camilo}, {Freire}, {Jenet},
  {Karako-Argaman}, {Knispel}, {Lazarus}, {Lee}, {van Leeuwen}, {Lynch},
  {Ransom}, {Scholz}, {Siemens}, {Stairs}, {Stovall}, {Swiggum},
  {Venkataraman}, {Zhu}, {Aulbert}, \& {Fehrmann}}]{2014ApJ...790..101S}
{Spitler}, L.~G., {Cordes}, J.~M., {Hessels}, J.~W.~T., {et~al.} 2014, \apj,
  790, 101, \dodoi{10.1088/0004-637X/790/2/101}

\bibitem[{{Springel}(2010)}]{2010MNRAS.401..791S}
{Springel}, V. 2010, \mnras, 401, 791, \dodoi{10.1111/j.1365-2966.2009.15715.x}

\bibitem[{{Tendulkar} {et~al.}(2017){Tendulkar}, {Bassa}, {Cordes}, {Bower},
  {Law}, {Chatterjee}, {Adams}, {Bogdanov}, {Burke-Spolaor}, {Butler},
  {Demorest}, {Hessels}, {Kaspi}, {Lazio}, {Maddox}, {Marcote}, {McLaughlin},
  {Paragi}, {Ransom}, {Scholz}, {Seymour}, {Spitler}, {van Langevelde}, \&
  {Wharton}}]{2017ApJ...834L...7T}
{Tendulkar}, S.~P., {Bassa}, C.~G., {Cordes}, J.~M., {et~al.} 2017, \apjl, 834,
  L7, \dodoi{10.3847/2041-8213/834/2/L7}

\bibitem[{{Tendulkar} {et~al.}(2021){Tendulkar}, {Gil de Paz}, {Kirichenko},
  {Hessels}, {Bhardwaj}, {{\'A}vila}, {Bassa}, {Chawla}, {Fonseca}, {Kaspi},
  {Keimpema}, {Kirsten}, {Lazio}, {Marcote}, {Masui}, {Nimmo}, {Paragi},
  {Rahman}, {Pay{\'a}}, {Scholz}, \& {Stairs}}]{2021ApJ...908L..12T}
{Tendulkar}, S.~P., {Gil de Paz}, A., {Kirichenko}, A.~Y., {et~al.} 2021,
  \apjl, 908, L12, \dodoi{10.3847/2041-8213/abdb38}

\bibitem[{{Thornton} {et~al.}(2013){Thornton}, {Stappers}, {Bailes},
  {Barsdell}, {Bates}, {Bhat}, {Burgay}, {Burke-Spolaor}, {Champion}, {Coster},
  {D'Amico}, {Jameson}, {Johnston}, {Keith}, {Kramer}, {Levin}, {Milia}, {Ng},
  {Possenti}, \& {van Straten}}]{2013Sci...341...53T}
{Thornton}, D., {Stappers}, B., {Bailes}, M., {et~al.} 2013, Science, 341, 53,
  \dodoi{10.1126/science.1236789}

\bibitem[{{Tumlinson} {et~al.}(2017){Tumlinson}, {Peeples}, \&
  {Werk}}]{2017ARA&A..55..389T}
{Tumlinson}, J., {Peeples}, M.~S., \& {Werk}, J.~K. 2017, \araa, 55, 389,
  \dodoi{10.1146/annurev-astro-091916-055240}

\bibitem[{{Vedantham} \& {Phinney}(2019)}]{2019MNRAS.483..971V}
{Vedantham}, H.~K., \& {Phinney}, E.~S. 2019, \mnras, 483, 971,
  \dodoi{10.1093/mnras/sty2948}

\bibitem[{{Walker} {et~al.}(2020){Walker}, {Ma}, \&
  {Breton}}]{2020A&A...638A..37W}
{Walker}, C. R.~H., {Ma}, Y.-Z., \& {Breton}, R.~P. 2020, \aap, 638, A37,
  \dodoi{10.1051/0004-6361/201833157}

\bibitem[{{Xu} {et~al.}(2022){Xu}, {Niu}, {Chen}, {Lee}, {Zhu}, {Dong},
  {Zhang}, {Jiang}, {Wang}, {Xu}, {Zhang}, {Fu}, {Filippenko}, {Peng}, {Zhou},
  {Zhang}, {Wang}, {Feng}, {Li}, {Brink}, {Li}, {Lu}, {Yang}, {Caballero},
  {Cai}, {Chen}, {Dai}, {Djorgovski}, {Esamdin}, {Gan}, {Guhathakurta}, {Han},
  {Hao}, {Huang}, {Jiang}, {Li}, {Li}, {Li}, {Li}, {Li}, {Liu}, {Luo}, {Men},
  {Niu}, {Peng}, {Qian}, {Song}, {Stern}, {Stockton}, {Sun}, {Wang}, {Wang},
  {Wang}, {Wang}, {Wu}, {Xiao}, {Xiong}, {Xu}, {Xu}, {Yang}, {Yang}, {Yao},
  {Yi}, {Yue}, {Yu}, {Yu}, {Yuan}, {Zhang}, {Zhang}, {Zhang}, {Zhao}, {Zheng},
  {Zhu}, \& {Zou}}]{2022Natur.609..685X}
{Xu}, H., {Niu}, J.~R., {Chen}, P., {et~al.} 2022, \nat, 609, 685,
  \dodoi{10.1038/s41586-022-05071-8}

\bibitem[{{Xu} \& {Han}(2015)}]{2015RAA....15.1629X}
{Xu}, J., \& {Han}, J.~L. 2015, Research in Astronomy and Astrophysics, 15,
  1629, \dodoi{10.1088/1674-4527/15/10/002}

\bibitem[{{Yamasaki} \& {Totani}(2020)}]{2020ApJ...888..105Y}
{Yamasaki}, S., \& {Totani}, T. 2020, \apj, 888, 105,
  \dodoi{10.3847/1538-4357/ab58c4}

\bibitem[{{Yao} {et~al.}(2017){Yao}, {Manchester}, \&
  {Wang}}]{2017ApJ...835...29Y}
{Yao}, J.~M., {Manchester}, R.~N., \& {Wang}, N. 2017, \apj, 835, 29,
  \dodoi{10.3847/1538-4357/835/1/29}

\bibitem[{{Zhang}(2018)}]{2018ApJ...867L..21Z}
{Zhang}, B. 2018, \apjl, 867, L21, \dodoi{10.3847/2041-8213/aae8e3}

\bibitem[{{Zhang}(2020)}]{2020Natur.587...45Z}
---. 2020, \nat, 587, 45, \dodoi{10.1038/s41586-020-2828-1}

\bibitem[{{Zhang} {et~al.}(2020){Zhang}, {Yu}, {He}, \&
  {Wang}}]{2020ApJ...900..170Z}
{Zhang}, G.~Q., {Yu}, H., {He}, J.~H., \& {Wang}, F.~Y. 2020, \apj, 900, 170,
  \dodoi{10.3847/1538-4357/abaa4a}

\bibitem[{{Zhang} {et~al.}(2023){Zhang}, {Li}, {Zhang}, {Cao}, {Feng}, {Wang},
  {Qu}, {Niu}, {Zhu}, {Han}, {Jiang}, {Lee}, {Li}, {Luo}, {Niu}, {Tsai},
  {Wang}, {Wang}, {Wu}, {Xu}, {Yang}, {Zhang}, {Zhou}, \&
  {Zhu}}]{2023ApJ...955..142Z}
{Zhang}, Y.-K., {Li}, D., {Zhang}, B., {et~al.} 2023, \apj, 955, 142,
  \dodoi{10.3847/1538-4357/aced0b}

\bibitem[{{Zhang} {et~al.}(2021){Zhang}, {Yan}, {Li}, {Zhang}, \&
  {Wang}}]{2021ApJ...906...49Z}
{Zhang}, Z.~J., {Yan}, K., {Li}, C.~M., {Zhang}, G.~Q., \& {Wang}, F.~Y. 2021,
  \apj, 906, 49, \dodoi{10.3847/1538-4357/abceb9}

\bibitem[{{Zhou} {et~al.}(2020){Zhou}, {Zhu}, {Wang}, \&
  {Feng}}]{2020ApJ...895...92Z}
{Zhou}, Z.-B., {Zhu}, W., {Wang}, Y., \& {Feng}, L.-L. 2020, \apj, 895, 92,
  \dodoi{10.3847/1538-4357/ab8d32}

\bibitem[{{Zhu} \& {Feng}(2021)}]{2021ApJ...906...95Z}
{Zhu}, W., \& {Feng}, L.-L. 2021, \apj, 906, 95,
  \dodoi{10.3847/1538-4357/abcb90}

\bibitem[{{Zhu} {et~al.}(2018){Zhu}, {Feng}, \& {Zhang}}]{2018ApJ...865..147Z}
{Zhu}, W., {Feng}, L.-L., \& {Zhang}, F. 2018, \apj, 865, 147,
  \dodoi{10.3847/1538-4357/aadbb0}

\end{thebibliography}
\bibliographystyle{aasjournal}



\end{document}